# Entering voltage hysteresis in phase separating materials: revealing the thermodynamic origin of a newly discovered phenomenon and its impact on the electric response of a battery

Tomaž Katrašnik[1,*], Jože Moškon[2], Klemen Zelič[1], Igor Mele[1], Francisco Ruiz-Zepeda[2] and Miran Gaberšček[2,3,**]

[1]University of Ljubljana, Faculty of Mechanical Engineering, Aškerčeva 6, SI-1000 Ljubljana, Slovenia

[2]National Institute of Chemistry, Hajdrihova 19, SI-1000 Ljubljana, Slovenia

[3]University of Ljubljana, Faculty of Chemistry and Chemical Technology, Večna pot 113, SI-1000 Ljubljana, Slovenia

Corresponding authors:

* tomaz.katrasnik@fs.uni-lj.si

** miran.gaberscek@ki.si

**SUMMARY**

**Hysteresis is a general phenomenon regularly observed in measurements of various materials properties such as magnetism, elasticity, capillary pressure, adsorption, battery voltage etc. Usually, the hysteretic behaviour is an intrinsic property that cannot be avoided or circumvented in dynamic operation of the system. Here we show, however, that at least as regards the hysteretic behaviour of phase-separating battery materials, one can enter (deeply) into the hysteretic loop in specific, yet realistic, transient operating conditions. Within the hysteretic loop a (significant) portion of particle population resides in an intraparticle phase separated state. Interestingly, the transition to the more conventional interparticle phase separation state found outside the hysteretic loop is very slow. Further, we establish a direct interrelation between the intraparticle phase separated electrode state and altered electric response of the electrode, which significantly impacts DC and AC characteristics of the battery. The experimental evidence of entering the hysteretic loop and the resulting altered response of the battery are explained based on thermodynamic reasoning, advanced modelling and insightful experiments. We believe that the understanding of this phenomenon will help optimise the diagnostics and monitoring of batteries, while also providing pertinent motivation for the enhancement of battery design and performance.**

## INTRODUCTION

Lithium iron phosphate (LFP) based batteries are regaining on importance due to their high safety, high power density, acceptable energy density, durability, absence of resource-constrained critical metals and, in particular, very good price-performance ratio.[1,2] This opens their way into mass market electromobility and large-scale energy storage applications, in addition to other areas of applications. Short- and mid-term energy storage and, in particular, mobility applications are characterized by highly transient operating regimes, while these applications also require precise monitoring and control of a large number of cells to ensure their proper operation, longevity and safety.[3–6] Likewise, batteries comprising other phase separating materials as lithium titanate oxide (LTO)[7], lithium manganese phosphate (LMP) and others are also utilized.

Despite a significant progress in understanding phenomena in phase separating materials[7–10], phase-separating reactions in real electrodes have remained poorly understood. To a large extent this insufficient understanding is due to the high complexity of battery electrodes which consist of numerous particles connected by binders and conductive additives.[7,11] This is particularly true for electrochemical signatures of various inhomogeneous multi-particle intercalation phenomena, which call for an explanation of the correlation between the macroscopic electrochemistry and microscopic processes. One of the seminal works in this area was the experimental demonstration of the existence of voltage hysteresis and its quantitative explanation based on the inhomogeneous multi-particle intercalation phenomena in phase-separating batteries a decade ago.[9] The findings were further elaborated in many follow up papers.[12–15] Other known phenomena interrelating macroscopic electrochemistry and microscopic processes are the so-called "memory effect"[15] and also the ''group-by-group'' multi-particle Li intercalation in a battery system that undergoes Li phase separation yielding electrochemical oscillations.[7] Although different mechanisms have been identified, the model by Dreyer et al.[9], along with the tacit assumption that hysteresis cannot be avoided in phase separating materials, has remained unchallenged. In this paper, however, we do challenge this basic assumption by trying to find an answer to the following fundamental question: Is it possible to enter the hysteretic loop under dynamic operating conditions?

This fundamental dilemma opens several intriguing, yet unresolved, questions when applied to batteries containing electrodes comprised of phase separating materials (LFP, LTO, LPM, etc.): Why can nominally identical cells feature significantly different voltage outputs while being at the same depth-of-discharge (DOD) and temperature or after being exposed to the same low current over a long time? Why do parallelly connected cells of the same type tend to feature significantly different currents when being at the same DOD and temperature? Why can cells at the same DOD and temperature feature very different EIS spectra after being at rest for a long time?

The present paper offers answers to these intriguing questions by explaining the correlation between the microscopic processes occurring during the operation of batteries with phase-separating electrodes and their electrochemical output. In this context we show that using different excitation protocols (different history of battery operation) one can reach different states of electrode and thus observe different electrical outputs. The latter include states (voltages) within the hysteresis loop which until now seemed to be impossible under dynamic operating conditions. The possibility of entering in the hysteresis loop during battery operation, however, has far reaching consequences for all the essential questions raised above, as well as for any imaginable application, so this aspect deserves particularly careful consideration.

In general, hysteresis is a consequence of irreversible energy loss in a system during transitions between different points in a phase diagram. If after transition driving forces are exactly inverted, the system generally does not follow the exact same path in the reverse direction across the phase diagram. Only in special cases it is possible to approach the equality in the second law of thermodynamics ($dS \geq dQ/T$) during the transition, and the system limits towards the exactly inverted path. However, driving the system out of thermodynamic equilibrium results in energy dissipation ($dS > dQ/T$), irreversibility and consequently hysteresis. In electrochemistry, extremely small currents are intuitively considered as a near-thermodynamic equilibrium case. This is not true for phase separating materials, which exhibit a hysteretic behaviour even at extremely small currents.[9] This is a consequence of the potential barrier of the chemical potential in the phase separating electrode that imposes states of the system far from the thermodynamic equilibrium even in the case of infinitesimally small currents.[9] The explanation of such a barrier is offered in reference[9] and briefly revisited in this paper. Considering the fact that hysteresis is a non-equilibrium phenomenon, questions from the second and the third paragraph can be reformulated to a single fundamental dilemma: Is there any known (or unknown) electrode state that would allow entrance into the intrinsic hysteretic loop (close to the equilibrium state) of a battery under dynamic operating conditions?

In order to find answers to such fundamental questions one might want to consider other, more familiar similar systems from everyday life. As regards hysteresis, the underlying mechanism occurring in a battery was shown to be analogous to the one appearing in an ensemble of rubber balloons.[16,17] Both the active particle (in this work "active particle" denotes either a single primary storage particle or an aggregate[18]) in a phase separating electrode and a rubber balloon are characterised by a bi-stable nature. The extensive and intensive variables from the conjugated variable pair that determines the particle (balloon) free energy, are connected by a non-monotone functional dependency.[19,20] Therefore, it seems that an ensemble of particles sharing lithium and an ensemble of rubber balloons sharing air, also share a mathematical analogy.

Considering this direct analogy with rubber balloons, one might tend to conclude that there is no appropriate state of matter (or balloon ensemble configuration etc.) that would allow such entrance. Similarly, looking into other systems exhibiting a hysteretic behaviour, one finds that hysteresis in dynamic conditions can only be circumvented in very rare cases.[21–24] However, to the best of our knowledge the entrance into hysteresis loop has not been demonstrated for multi-particle systems described by bi-stability (i.e. ensemble of rubber balloons and analogous systems [25–28]). A large number of battery papers clearly demonstrates the ubiquity of the hysteretic behaviour of phase-transitioning battery materials.[29–33] However, only sporadically the published measurements indicate voltages that are apparently inside the hysteretic loop.[9,34] In fact, an example can already be found in the original paper by Dreyer et al.[9] where Fig. 2d shows a point deeply inside the loop. However, this state was reached after a very prolonged relaxation period of time during which the system had relaxed to a thermodynamic equilibrium, which by definition lies in the centre of hysteresis.[34] Recently, entering into the hysteretic loop was reported for the case of silicon electrode.[35,36] It needs to be stressed, however, that a lithiated silicon is not a typical phase separating material[36,37] so the thermodynamic background – which is still not entirely clear – might be different than in the present case. Still, we believe that the present findings will offer some inspiration also for a better understanding of hysteresis in lithiated silicon and other similar systems.

Based on known behaviour of many phase separating materials under different conditions and understanding the main underlying mechanisms, we have been able to identify transient operating conditions that indeed result in a clear and deep entrance in the hysteresis loop under dynamic operating conditions. Even more, the system can remain within the hysteresis almost to the end of the discharge if subjected to small currents (Fig. 1a and Supplementary Section 1). Although several other materials have shown the same effect (e.g. LMP, see Supplementary Section 1.3), we here focus on a LFP – due to abundance of previous results on this system which allows unambiguous identification of the new phenomenon. As can be seen from Fig. 1a, one can reach voltages that are very close to the middle of hysteresis loop if a large current stimulus – denoting either a longer period of large current or a shorter large current burst – is applied which is then reduced to a (very) small current. Interestingly, the voltage remains inside the usual hysteretic loop for a very long time, more or less almost to the end of the ongoing half-cycle if subjected to small currents. This phenomenon, which is based on fundamentally different underlying microscopic processes compared to the previously reported phenomena associated with the so-called "memory effect"[15] and ''group-by-group'' multi-particle Li battery intercalation[7,11] (Supplementary Section 3), features a large monotonous deviation of the cell voltage. Thus, in addition to shedding new light into the fundamental mechanisms occurring in contemporary battery electrodes, the possibility of entering into hysteretic loop during battery operation may have a crucial impact on various battery applications, as discussed later on.

To demonstrate the generality of the newly discovered phenomenon, we devised a number of transient protocols (Supplementary Section 1), which make it possible to enter in the hysteresis loop (i) at different lithiation levels of the electrode and different applied currents, (ii) during charge and discharge, (iii) at various electrode thicknesses/loadings, (iv) for current stimuli of different lengths or transferred change, (v) for different dilution ratios of active material in the electrodes and (vi) for different materials and, finally, (vii) the new phenomenon was also demonstrated on commercial cells. On one side, this additionally confirms the generality of the identified phenomenon, while on the other, it also demonstrates that during highly transient operating conditions comprising low load periods (e.g. when powering only auxiliary components), it is (very) likely that the electrode and thus battery will enter into a voltage hysteresis.

In order to explain the observed phenomenon, let us first briefly revisit the original particle-by-particle model that explains the existence of hysteresis, postulated by Dreyer et al.[9] The model explained successfully the generally of the observed phenomenon of non-vanishing voltage gap in phase separating materials when charge-discharge currents were extrapolated to zero. The model connects a non-monotone chemical potential of these materials and the resulting particle-by-particle charging/discharging mechanism at small overpotentials.[9] Multiple subsequent publications further confirmed the existence of this inherent voltage hysteresis.[12,14,33,38–43] Although different mechanisms have been identified for higher rates[14,32,33,44], the model by Dreyer et al.[9] along with the tacit assumption that hysteresis cannot be avoided in phase separating materials has remained unchallenged, as exposed above. Thus, at first sight the result shown in Fig. 1a seems to be in contradiction with this model. However, as shown in continuation, the reason for the observed behaviour is not a flaw or deficiency of the model by Dreyer et al.[9], but a result of simultaneous entrance of a large share of active particle population into the intraparticle phase separated state. Active particle population denotes the share of active particles that features lithiation levels, which, for discharge, spans between spinodal point A in Fig. 2a and the intersection of the intraparticle phase separated electrode state potential with the right most branch of the chemical potential beyond the right spinodal. The phenomenon of interparticle phase separation was clearly observed experimentally in a recent study[10], however the corresponding intraparticle phase separated state comprising a large share of active particle population has not yet been correlated with battery performance. The latter represents a fundamental contribution of this article, that is, establishment of a new bridge between the macroscopic electrochemistry and microscopic processes occurring in a phase separating battery electrode.

**THEORY**

In the idealised case, the intraparticle phase separated state represents an ensemble of particles all of which are split uniformly into two phases occupying equal volumes within each particle - if the DOD – is 0.5. In a more realistic situation, where a distribution of particle properties and their heterogeneous ion and electron wiring is taken into account, one however may expect a distribution of the degree of particle lithiation such as schematically depicted in Fig. 1a (central scheme). Indeed, this scheme is well supported by experimental observations using STEM-EELS (Fig. 1c) where a lot of the particles are clearly split into two phases, albeit each particle has a slightly different degree of lithiation (for details on this experiment see Supplementary section 2). In any case, the intraparticle phase separation differs clearly from the interparticle phase separation which is experimentally shown in Fig. 1b and occurs at small currents without applying current stimuli. The intraparticle phase separated state exhibits a lower chemical potential, which leads to a higher electrode potential with respect to the other known electrode states depicted in Fig. 1a. Using state-of-the art materials, a precondition to reach this state is to apply a current stimulus that ends after the first spinodal point (different variants are presented in Fig. 1a and Supplementary Section 1). Briefly, significantly shorter current stimuli than those presented in Fig. 1a also enable entering into the hysteresis (Supplementary Section 1.1.3).

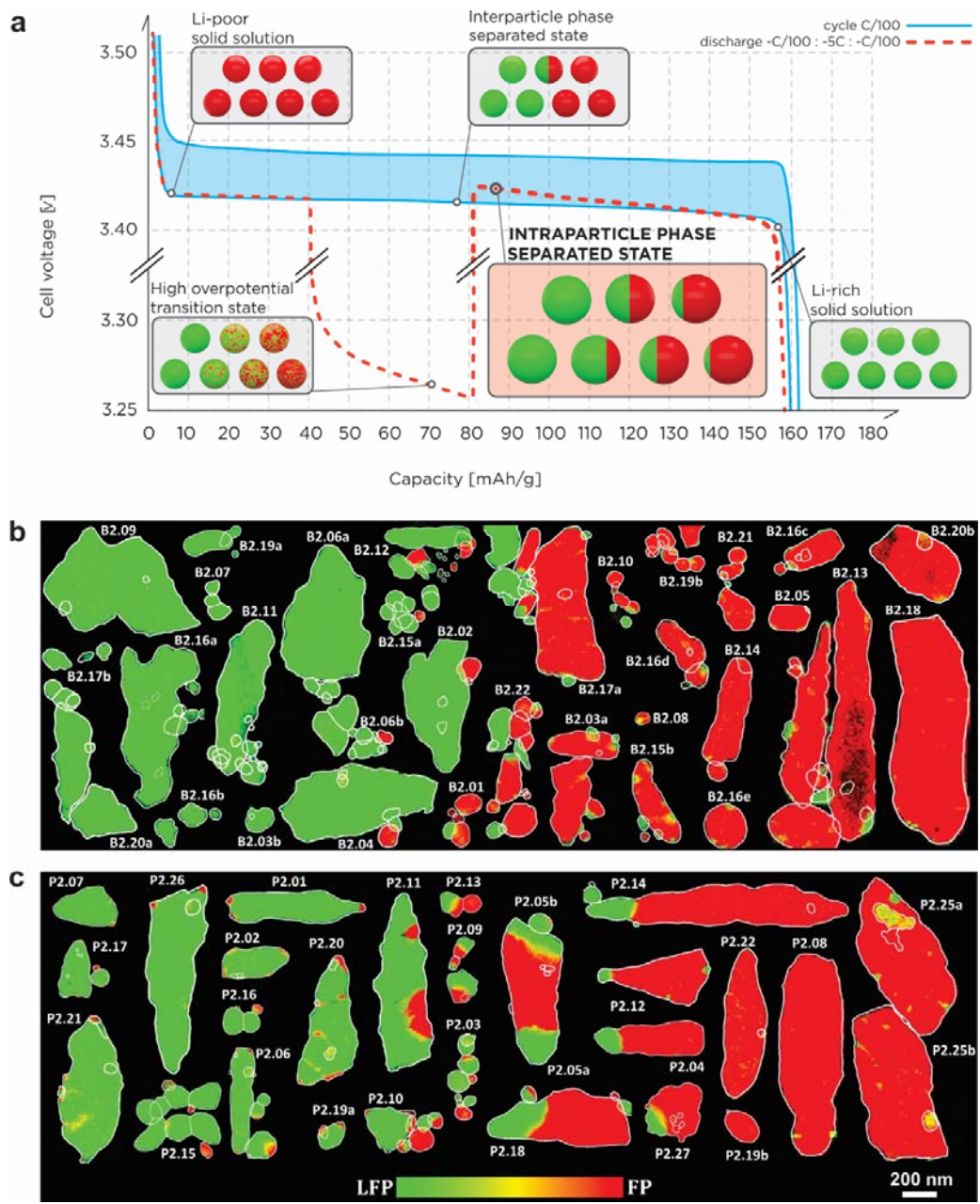

**Fig. 1 a** Path dependent entrance into the voltage hysteresis in phase separating battery materials. The blue curve shows the conventionally observed voltage hysteresis of LFP at small currents (measured on cell #1). The red curve demonstrates the entrance into hysteresis loop after reducing the current from the preceding large current stimulus (starting at 25 % and ending at 50 % DOD) applied during a battery discharge. The entrance into the voltage hysteresis loop is explained by an intraparticle phase separated state depicted by the central scheme. This state is fundamentally different compared to the other frequently encountered electrode states which comprise: solid solution state at low and high lithiation levels[14,33,45], high overpotential transient state within the miscibility gap [14,33,45] (more details provided in Supplementary Section 4.2) under application of a large current and interparticle phase separated state[10] or mosaic instability[13] being characterized by a particle-by-particle mode of lithiation[9] under application of a small current. **b** and **c** STEM-EELS colour map demonstrating the interparticle (**b**) and intraparticle (**c**) phase separation states (LFP (green) and FP (red)) that occur at small currents and after current stimulus, obtained from cells #2 and #3 respectively.

The thermodynamic basis for present explanations relies on a postulate that the lithiation level of the electrode is equal to the sum of Li mole fractions of the many-particle system $q$[9]

$$q = \frac{1}{N} \sum_{l=1}^{N} x_l, \qquad (1)$$

where $x_l$ is the Li mole fraction of an individual particle (indexed by $l$) and $N$ is number of (de)lithiating particles inside electrode. This constraint imposes that in a many-particle system it is only possible to control the total amount of lithium in all particles, thus the amount of Li inside individual particles is not uniquely determined. Another driving force for this behaviour arises from free energy minimisation of the many-particle system

$$F_{TOT} = \sum_{l=1}^{N} \left[ \int_{V_l} f_l(x_l) dV \right], \qquad (2)$$

where $F_{TOT}$ denotes the total free energy of many-particle system, $V_l$ and $f_l$ are volume and free energy density of an individual particle. The amount of lithium inside individual particle ($x_l$) and exchange of lithium among active particles is, therefore, driven by minimisation of energy of the whole ensemble.

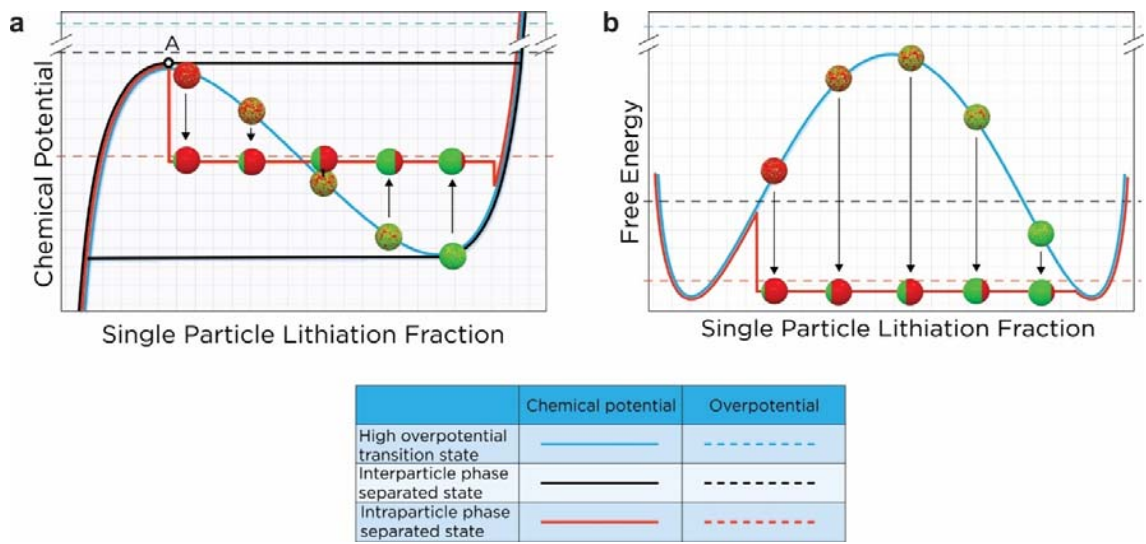

**Fig. 2** Schematic presentation of **a** chemical potential and **b** free energy during a discharge process. Under application of a large current, the majority of active particle population undergoes a non-

equilibrium transition following the spinodal potential, which is, as elaborated in Supplementary Section 4.2, denoted high overpotential transition state. After a sudden reduction of current and thus overpotential, the majority of particles undergo intraparticle phase separation (i.e. they internally split into two phases) yielding the intraparticle phase separated electrode state. Chemical potential of the particle in the intraparticle phase separated state is close to the equilibrium state and it is characterised by a minor offset of the free energy associated with intraparticle strain and interfacial energy due to the internal phase boundary.[46] Corresponding intraparticle phase separated electrode state, where the majority of particle population is intraparticle phase separated, is thus characterised by a lower energy level and, therefore, by a lower intraparticle phase separated chemical potential compared to the interparticle phase separated chemical potential imposed by point A in Fig. **a**. The latter is characteristic for particle-by-particle mode of lithiation of multi-particle systems. At small currents, overpotentials corresponding to intraparticle phase separated and interparticle phase separated state are just slightly above the corresponding potentials with slightly larger overpotential corresponding to interparticle phase separated state due to lower share of active particle population.

This thermodynamic basis offers a direct explanation of all states depicted in Fig. 1 (see also Supplementary Section 4). At small overpotentials that barely exceed the initial energy barrier imposed by the convex part of the free energy density of a particle (Fig. 2a and 2b) the particles undergo interparticle phase separation, that is, individual particle splits internally into two phases.[38] However, in contrast to the intraparticle phase separated state, in this particle-by-particle lithiation regime (Supplementary material Fig. S32c and S32d) only a small fraction of particles is active simultaneously (ref[9] and Fig. 3b) which forms a mosaic pattern of particles[9] (see also Fig. 1b). The resulting interparticle phase separated state inherently yields a voltage hysteresis.[9] Despite small currents this hysteresis is characterised by non-negligible entropy generation. To a large extent this is a consequence of the chemical potential barrier that imposes states of the system far from the thermodynamic equilibrium and Li redistribution between the particles.[9] Another limiting (dis)charging regime is observed at large currents, when the system is far from the thermodynamic equilibrium and thus characterised by high overpotentials. In that regime the overpotential is sufficiently large that the majority of particles lithiate simultaneously via a pathway involving poorly defined, non-equilibrium disordered structure denoted as high-overpotential transient state (Fig. 2a and 2b and Supplementary Section 4.2), as confirmed by multiple simulation and experimental studies.[14,32,33,44,45] This high overpotential transient state is characterised by a large share of active particle population.[10,38,45]

The thermodynamic background that explains transient protocols for entering into the hysteretic loop under dynamic operating conditions is graphically explained in Fig. 2. One can identify the following key conditions that need to be satisfied in order for this to occur:

- Application of a sufficiently large current stimulus that ends inside the spinodal region; this ensures that a large share of active particle population undergoes a transition through high overpotential transient state (Figs. 2a and 2b).
- Sufficiently fast reduction of the current to a very low value so that local overpotentials are appropriately reduced.
- Simultaneously, minimisation of energy of the whole ensemble (eq. (2)) drives the intraparticle and interparticle redistribution of lithium.
- Preferential occurrence of intraparticle phase separation (Fig. 2a and 2b) over the interparticle phase separation due to longer characteristic times for the latter[10]; this results in a higher potential of the electrode.

The existence and interplay between intraparticle and interparticle phase separations were investigated in several experimental studies.[39,44,47–49] Detailed reasoning on different characteristic times for intraparticle and interparticle phase separation is provided in Supplementary Section 5, which addresses a causal chain of phenomena that covers the transition from the high overpotential transient state to the long term-lived intraparticle phase separated state, and subsequent relaxation to the lower energy - interparticle phase separated state with long characteristic times.

The proposed scenario can also be explained phenomenologically. After a large current stimulus, a phase-separating electrode should be at a higher potential level compared to the case of a constant small current discharge. Thus, it can be concluded that most of the voltage difference presented in Fig. 1a, i.e. entering into the hysteresis, arises from the difference between the maximum spinodal level (Fig. 2a – point A) and the intraparticle phase separated potential level of a large share of particle population, as depicted in Fig. 2. This statement is further supported by the following simulation and experimental results.

**RESULTS**

**Simulation results**

The schematic explanation proposed above is further supported by simulation results obtained with the extended continuum level porous electrode model[50] shown in Fig. 3 and discussed in detail in Supplementary sections 6.1 and 6.2. Obviously, the model is able to replicate the C/100 hysteresis very accurately. Fig. 3b and 3c show that in the small current regime the active particle population is significantly larger at the same level of electrode lithiation, i.e. after 50 % DOD, if a large current stimulus is applied. The combination of significantly increased active particle population and a higher cell voltage (Fig. 3a), therefore, provides a unique proof of the hypothesis on simultaneous entrance of a larger share of active particle population into the intraparticle phase separated state after a large current stimulus (more details are provided in Supplementary Section 6.2). Namely, both parameters can only be achieved simultaneously if the electrode is in the intraparticle phase separated state. This is because a higher battery voltage can only be realised if a large share of particles features lower chemical potentials such as inherently found in particles that are in the intraparticle phase separated state (c.f. Fig. 2).

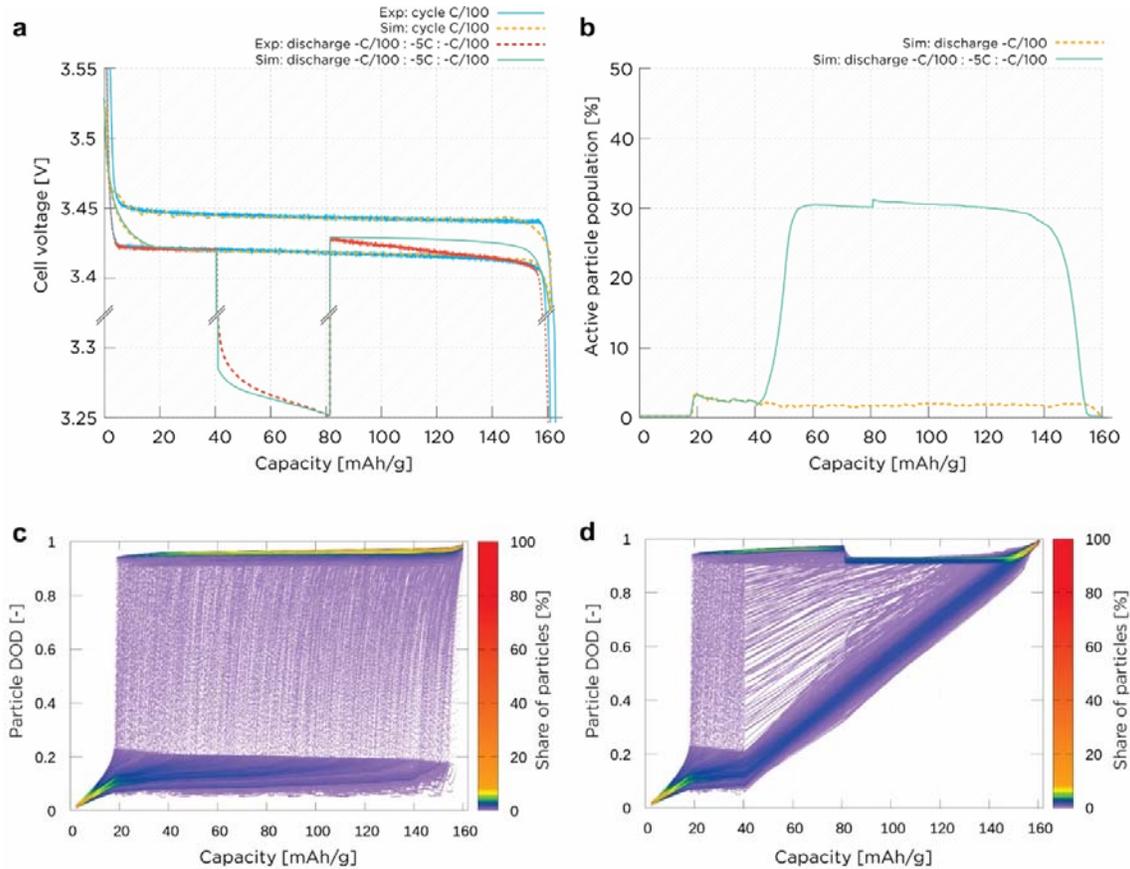

**Fig. 3 a** Comparison between experimental (cell #1) and simulation results for hysteresis obtained at C/100 charge and discharge cycle as well as for -C/100 discharge with a -5C discharge stimulus between 25 % and 50 % DOD (discharge -C/100 – -5C – -C/100). **b** simulated active particle population for -C/100 discharge and -C/100 discharge with a -5C discharge stimulus between 25 % and 50 % DOD. **c** and **d** Evolution of the particle's DODs as a function of the electrode level or cell capacity during discharge. Fig. **c** shows initial solid solution lithiation until the spinodal point followed by particle-by-particle lithiation lasting nearly to the end of the cell capacity range. Fig. **d** clearly shows initial solid solution lithiation until the spinodal point followed by particle-by-particle lithiation until 25 % DOD and subsequent lithiation in the high overpotential transient state increasing active particle population also shown in **b**. This is a prerequisite for entering the intraparticle phase separated state upon reduction of the current. This intraparticle phase separated state is characterised by a larger share of active particle population at high potentials further supporting key conditions that need to be satisfied in order to enter into the hysteretic loop under dynamic operating conditions.

**Experimental results**

Figs. 4c and 4d offer a direct experimental proof that after applying a current stimulus of -3C the cells exhibit markedly different electrochemical characteristics compared to a nominally identical cell discharged at -C/30 (Fig. 4b) within the same initial DOD range (from 0 to 0.1). The significantly different current response (Fig. 4c) as well as the different ratio of apparent cell-level total (DC) resistances (Fig. 4c) of both cells clearly confirm that after a current stimulus the electrodes enter a different, i.e. intraparticle phase separating, state. The macroscopically observed significant variation in apparent DC cell resistance after a current stimulus is directly related to the higher potential level of the intraparticle phase separated electrode state (Fig. 1a and Fig. 2a) and the lower overpotential for lithiation due to a larger share of active particle population. Importantly, after the cells had been connected in parallel (inset of Fig. 4b), the voltage of both cells increased, which is consistent with the findings in Figs. 1 and 3 as well as with the thermodynamic reasoning presented in Fig. 2. Expectedly, the trends of the current and the resistance ratio are reversed towards the end of the discharge period (Fig. 4c and 4d) as the present cells feature nominally identical capacities.

The connection between a lower overpotential for lithiation and a larger share of active particle population can be explained thermodynamically, as depicted in Figs. 5a and 5b. During a small current discharge, a large share of particles is at the lithiation level that corresponds to the first spinodal point (Fig. 5a). This is characterised by a large derivative of free energy, $\Delta F_1$ (i.e. potential), as a function of variation of the single particle lithiation fraction ($\Delta c$) and consequently current (Fig. 5a). In contrast, the intraparticle phase separated state is characterised by a much smaller gradient of free energy, $\Delta F_2$, as a function of variation of the single particle lithiation fraction, $\Delta c$ (Fig. 5a) and thus impedance (including DC resistance). One could also say, from a different perspective, that phase boundaries which are inherently present in intraparticle phase separated particles, are moved without a large energy penalty.[46]

Finally, the proposed mechanism is additionally confirmed by an independent experimental evidence employing the electrochemical impedance spectroscopy (EIS) measurements after galvanostatic excitation with different rates (C/100 and 5C), terminated at a DOD of 0.1 (Fig. 5c). The experiments show a pronounced effect of preceding current stimulus on the low-frequency part of impedance response in which solid state processes strongly prevail. Specifically, after a large current stimulus (red curve) this part reveals impedance values that are by 20-40 % lower compared to the case with small current (blue curve). This different behaviour is a further proof that intraparticle phase separated state is characterised by a lower overpotential for lithiation due to a larger share of active particle population being closer to the equilibrium state.

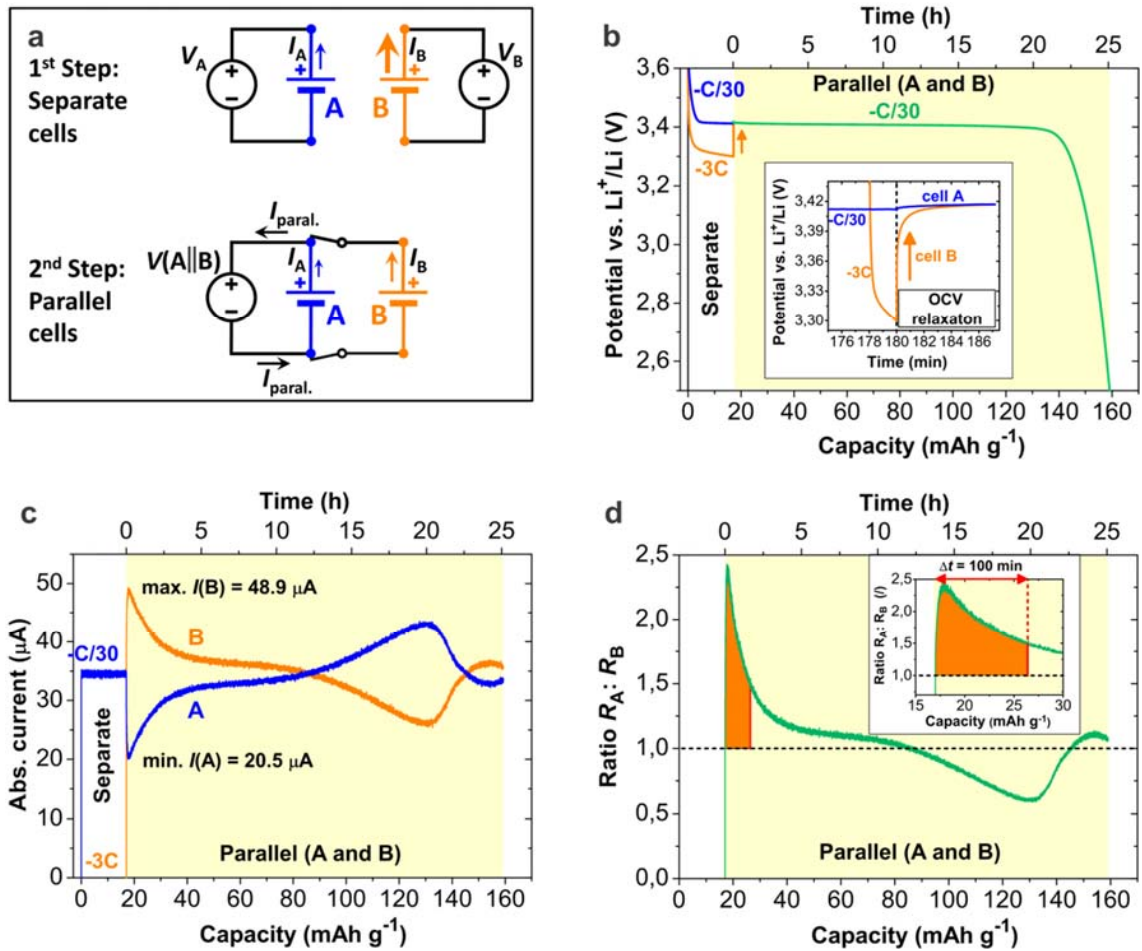

**Fig. 4 a** Schematic presentation of a parallel battery experiment performed on two nominally identical LFP-Li cells (A = cell #4 and B = cell #5). In the 1$^{st}$ Step each cell was connected to its own current (voltage) source with the corresponding terminal voltages $V_A$ and $V_B$. During the following 2$^{nd}$ Step the cells were connected in parallel to a single current (voltage) source with a terminal voltage $V(A \| B)$. **b** During the 1$^{st}$ Step the cells were partially galvanostatically discharged whereby their global DOD was changed from 0 to 0.1; however, the magnitude of discharge current was different: -C/30 for cell A (blue curve) and -3C for cell B (orange curve). The discharge was followed by a short relaxation period at open circuit (inset in Fig. **b**). In the 2$^{nd}$ Step the cells were connected in parallel (A∥B, Fig. **b**, green curve). While the pair was discharged using a total current of -C/30, the individual currents through the cells ($I_A$ and $I_B$) were also monitored. **c** A strong deviation of the individual cell currents from the total current value -C/30 is observed. Importantly, in the initial part of the 2$^{nd}$ Step a markedly larger current is observed in cell B which in the 1$^{st}$ step was subjected to a larger current (-3C). **d** The observed effect can be expressed in terms of the ratio of the apparent DC cell resistances $R_A : R_B$, where the maximal value of the ratio was found to be as large as 2.4. The zero value on the time scale corresponds to the end of the current stimulus.

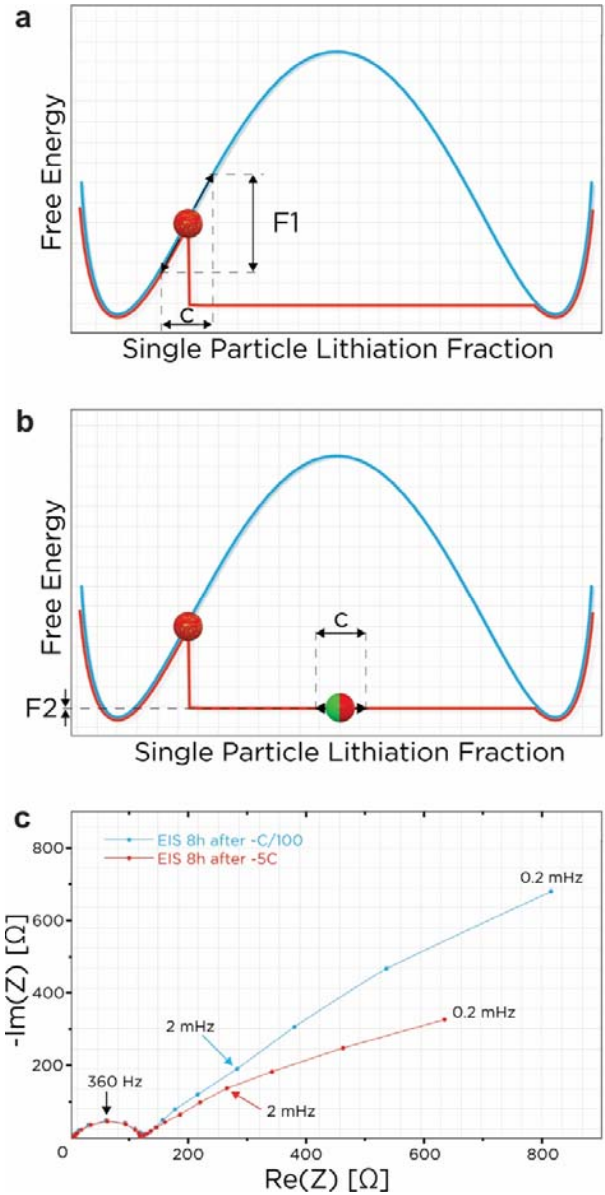

**Fig. 5 a** Schematic explanation of larger impedance of the interparticle phase separated state compared to the **b** intraparticle phase separated state indicating smaller gradient of free energy ($f$) and thus potential as a function of variation of the single particle lithiation fraction ($c$); colour code in figure **a** and **b** is identical as in Fig. 2. **c** EIS measurements of the experimental cell #6 measured 8 hours after termination of 5C discharge at DOD 0.1 (red curve) and after termination of C/100 discharge at DOD 0.1 (blue curve).

**DISCUSSION**

On a broader scale, entering the hysteresis loop has already been reported for systems other than bi-stable multi-particles.[25–28] For example, such states have been reached by changing the temperature of a multi-phase system and tracing the temperature path dependence or shifting the system vertically across the phase lines of phase diagram. Unlike those previous studies, we have demonstrated and explained the existence of long-living states inside the hysteresis of phase separating Li-ion battery by shifting the multi-phase system (i.e. phase separating active particle) horizontally across the phase diagram in the direction of variation in fraction of lithiation. Moreover, we have shown that the system remains within hysteresis also under dynamic operating conditions.

This finding is of an overarching importance not only in terms of better understanding the fundamental processes occurring in state-of-the-art battery systems but also for their application, which will be addressed in some detail in the following paragraphs. First of all, we provide clear experimental and theoretical evidence that using (highly) transient charge/discharge protocols it is possible to enter (deeply) into the hysteretic loop while the battery is under (normal) operation. Secondly, additional application of very small currents enables approaching the equilibrium state. It is exactly by this demonstration, which is clearly connected with the existence of long-living intraparticle phase separated electrode state, that allows for the unprecedented possibility of revealing physics-based answers to all introductory questions.

In terms of scientific progress, this paper gives a comprehensive resolution of the initial fundamental dilemma about whether it is possible to enter the intrinsic hysteretic battery loop under dynamic operating conditions. It provides a clear experimentally and theoretically supported interrelation between the microscopic intraparticle phase separated state (Fig 1c), which is associated with a lower chemical potential (Fig 2) and thus a higher potential of an electrode, and the resulting macroscopic electrochemical output of the cell being within the voltage hysteresis (Fig 1a). Furthermore, the existence of long-living intraparticle phase separated electrode state that occurs upon entrance into hysteresis loop (Fig 1a) explains why cells being at the same DOD and temperature and being exposed to the same low current over a long time can feature significantly different voltage outputs. Namely, the magnitude of this phenomenon is directly correlated with the share of active particle population in the intraparticle phase separated state, which is dependent on the current stimulus or transient current trace, but also on micro and nano structure of the active and inactive material (Supplementary Section 1). In addition, a higher share of active particle population in the intraparticle phase separated state results in a higher potential level of the electrode (Fig. 1a and Fig. 2a), which also clearly answers the question why parallelly connected cells of the same type feature significantly different currents despite being at the same DOD and temperature (Fig. 4). Additionally, intraparticle phase separated state is characterised by a much smaller gradient of a free energy as a function of variation of the single particle lithiation fraction (Fig. 5a) and a larger share of active particle population. These particular characteristics of the intraparticle phase separated state answer the question why cells at the same DOD and temperature feature much lower impedance in the low-frequency part of impedance response in which solid state processes strongly prevail (Fig. 5c).

Therefore, the present findings may have a broad impact on the analysis, diagnostics and monitoring of such batteries during their dynamic operation. The state-of-charge (SOC) estimation of a given battery, which is correlated with DOD, is one of the most important issues in a battery management system (BMS).[4,5,51,52] However, in cells comprising LFP or LTO electrodes, it is very challenging to measure SOC/DOD directly - due to the flatness and hysteresis of the open circuit voltage (OCV) curve[5,51–54], whereas the classical approach of current integration (Coulomb counting) is also associated with challenges of accumulative error and inaccurate initial values.[5,51,54] In the light of findings revealed in this study, Coulomb counting, which is a common approach in larger systems, can face further difficulties during transient operation, particularly when applied on multiple cells. Due to the inherent cell-to-cell variation and due to the further amplification of this variation during degradation, cells will tend to enter the intraparticle phase separated electrode states featuring (slightly) different share of active particle population and thus cell voltages when subjected to adequate transient operation. As demonstrated on a limiting case in Fig. 5, such a scenario will lead to differences in inter-cell currents and thus differences in the SOC/DOD level, which cannot be detected if the current output of multiple cells is measured.

These challenges further aggravate when trying to determine the state-of-energy (SOE), which is not only the integral of the current but also depends on the voltage which, in turn, includes the nonlinearity of Li-ion batteries (LIB).[55] SOC is also strongly correlated to state-of-health (SOH), which is crucial to keep the LIBs working within safe limits and to maximize their performance.[56,57] SOH is normally obtained from estimating the OCV and the internal resistance[56] being an important health indicator in addition to the cell capacity.[3,58] However, these two health indicators are not directly measurable with commercially available sensors, and they tend to be indicated and estimated through other measured variables such as the voltage, current and temperature.[3] In the light of findings revealed in this study, this can impose further challenges as apparent DC cell resistances changes significantly with increasing share of active particle population in the intraparticle phase separated state (Fig 4d).

It is well known[59], that the relaxation phenomena also influence the electric response of the cells. From this perspective the present findings can be interpreted also as introduction of an additional characteristic time scale that is associated with the relaxation of the intraparticle phase separated state to the interparticle phase separated state, which in turn impacts cell voltage and impedance. This is of particular importance for the correct usage and interpretation of electrochemical impedance spectroscopy (EIS) results. EIS has been widely applied in the laboratories to obtain a profound insight into the phenomenology of the intra-cell phenomena[60,61] and to assess battery SOH[62] or associated remaining useful life (RUL)[63] and SOC[54,64], where models can also be used together with EIS measurements for a better SOC estimation.[65,66] In addition, it was shown in[64,67] that EIS measurements on a vehicle could be made through cell excitation driven by the motor controller, bringing the intriguing EIS phenomena associated with interparticle phase separated state (Fig 5c) closer to future analysis, diagnostics and monitoring applications. Namely, our findings clearly indicate that the low-frequency part of impedance response changes drastically when cells enter the long-living intraparticle phase separated electrode state (Fig 5c).

Another challenging aspect in multi cell system is also cell balancing of serially and parallelly connected cells, where balancing algorithms require inputs of at least voltage, current and temperature.[62,68] During fast charging and, consequently, low current balancing procedure cells can enter the intraparticle phase separated states characterized by different shares of active particle population between the cells - due to inherent cell-to-cell variation. Resultantly, cells can yield different cell voltages at the same SOC or vice versa, which impairs the balancing procedure.

Although reliable measurements are key for adequate monitoring and control of batteries, BMSs, in general, also incorporate various types of models, model-based state observers and/or data driven methods to refine SOX (i.e. SOC, SOE, SOH) and RUL estimates.[3,5] Furthermore, machine learning methods and artificial intelligence is being widely applied to further refine these estimates.[57,58,63,69] However, applied models, model-based observers and/or data driven methods including machine learning methods and artificial intelligence utilize the experimental values of voltage, current, temperature and potentially also capacity and cycling numbers for calibrating and training data, while in observer applications these experimental data are continuously used in observer algorithms.[70,71] Our results show that entering long-living intraparticle phase separated electrode state results in voltage deviations that are an order of magnitude larger compared to the expected accuracy threshold of the measurement equipment, while intraparticle phase separated electrode state also features larger monotonous deviation of the cell voltage than the previously reported memory effect[15,72] and electrochemical oscillations in Li-Ion batteries.[7]

Finally, it can be envisaged that the results shown in this paper will ultimately help developing enhanced functionalities of the next-generation Battery Management Systems (BMS) and, in particular, enable development of systems for much more accurate determination of SOC, SOE and SOH of batteries utilising phase separating materials that are operated in transient operating regimes.

Ultimately, inspired by the newly revealed causal chain of events in phase separating materials, the battery community may try to search for new ways of minimisation of entropy generation in future electrode designs and/or with elaboration of tailored control strategies of batteries and battery systems by exploring interactions between material properties, particle size dependent effects[73,74], impact of particle connectivity[50,75] and current profiles. In particular, if a battery could operate predominantly in the intraparticle phase separated state, it would exhibit very low losses and thus a minimised heat generation.

## Experimental procedures

**Active Cathode Materials**

Three types of LiFePO$_4$ (LFP) battery-grade powder materials were used as active materials for LFP cathodes. a) Commercially available powder provided by Targray (SLFP02002): according to the specifications, this LFP material has a specific surface area of 11 ± 2 m$^2$/g, an agglomerate size of 3 ± 1 μm by the D50 criterion, and a native carbon content of 2 ± 1 wt.%. b) The second type of LFP material, was prepared by a novel Pulse Combustion Reactor method (PCR-method) in a slightly reductive environment, as described in detail in our previous paper.[76] Note however, that extensive experiments did not reveal any meaningful effect of the synthesis procedure on the present phenomenon. Briefly, the second material was synthesized in a reactor setup consisting of a Helmholtz-type pulse combustor with a natural frequency of 280 Hz (at a temperature of around 1250 K) and a 4 m long stainless steel reactor pipe. Air is supplied to the combustor by way of a blower through an aerodynamic valve. The method allows very precise control of atmosphere in the reactor, as well as the frequency and amplitude of pulses, all of which have a pronounced effect on the reaction outcome.[77] In the present synthesis the spraying gas was 99.9 % nitrogen with a pressure of 1.5 bar and a flow of 45 mL min$^{-1}$. The precursor was composed of 13.5 g of LiNO$_3$, 76.1 g of Fe(NO$_3$)$_3$·9H$_2$O, 36.4 g of triethyl phosphate, 78.5 g of glycine, 30.9 g of NH$_4$NO$_3$, everything dissolved in 400 g of deionized water. We used a 4 mole % excess of lithium due to losses in the reactor and annealing oven. The temperature at 0.5 m after the spray nozzle was maintained at (700 ± 5) °C with the amount of precursor sprayed (20 ± 1) mL min$^{-1}$. The frequency of combustion was maintained at 240 Hz. The prepared material was collected in an electrostatic precipitator and annealed in an electrical oven under a constant argon flow and in presence of carbon at 700 °C for 6 hours. The annealed material had a tap density of approximately 1 g cm$^{-3}$ and was used as prepared for the preparation of cathodes. c) The third type of LFP material was used specifically for the experiments where the STEM-EELS analysis was done (Figs. 1b and 1c). The material is essentially the commercial LFP powder provided by Targray and was not exposed to any further chemical modification or heat treatment. In order to be able to perform STEM-EELS measurements we selectively removed large agglomerates and largest LFP aggregates from the starting Targray LFP powder. We performed the separation by implementing centrifugation of dispersed Targray LFP powder in inert organic solvent. 9 g of the starting Targray LFP was dispersed in 300 mL of isopropanol (propan-2-ol) where beforehand 70 mg of Triton X-100 (Sigma-Aldrich) non-ionic surfactant was dissolved; the latter was shown to be completely compatible with LFP cathode.[3]

Ultra-Turrax® T25 (IKA®) homogeniser/disperser was used at 13000 rpm for 30 mins to break apart large agglomerates and followed by 20 mins of sonication with ultrasonic liquid horn disperser (Sonics, Vibra-Cell™) by applying sequence of 8 s pulse + 2s pause steps in cooling ice bath was performed to promote effective dispersion of particles. We applied first centrifugation step (5 mins at 2300 rpm) to remove large aggregates (settled down at the bottom of the centrifuge tube), and second centrifugation step (30 mins at 10000 rpm) to remove the desired fraction of LFP particles from the excess Triton X-100 in the residual solution and the smallest (nano-) particles of LFP and potential impurities that remained in the dispersion. The sediment of LFP particles was re-dispersed in fresh isopropanol and the anticipated powder sample of LFP particles with selected particle size was obtained (150 mg) after final drying under reduced pressure (rotary evaporator, 3 mbar, 40 °C).

**Electrode Preparation**

Conventional porous electrodes were prepared by applying standard procedure for preparation of laboratory-scale cathodes. Starting from a selected active material (AM) powder and carbon black (CB) conductive additive we prepared a homogeneous slurry (dispersion) of the solid components in beforehand prepared 15 mg mL$^{-1}$ solution of PVdF binder (182702 Aldrich) in NMP solvent (99.5 % pure, 8.06072.2500 Merck) to obtain the final standard (dry) electrode composite composition AM : CB : PVdF = 90 : 5 : 5 (wt.%). Special (non-standard) composite formulations used in this work are provided in Table 1. In case of LMP-based cathodes the composition was AM : CB : PVdF = 80 : 10 : 10 (wt.%). The slurry was homogenized in a planetary mill for 30 min at 300 rpm and after that it was applied to the surface of carbon coated Al foil by using automated doctor-blade applicator. The distance between the blade and the surface of Al foil was set to 200 µm. The Al foil with coating was in the first step dried for 3 h at 90°C at reduced pressure (10 mbar). After that we cut out circular electrodes with 16 mm diameter (geometric area $A$ = 2 cm$^2$). We pressed the electrodes by applying 5 t (2.5 tons per 1 cm$^2$) for 1 min in a hydraulic press. Before transferring in an argon-filled glove box the electrodes were additionally dried under vacuum overnight at 90°C in vacuum chamber. This means the prepared electrodes had typical LFP mass loadings of about 3 – 4 mg per cm$^2$ of the electrode geometric area (composite loading 3.3 – 4.4 mg per cm$^2$) and the obtained composite thickness was in the range 18 to 19 µm. Electrode porosity was calculated by taking the values of electrode composite mass, thickness and the bulk densities of the composite components to be 41 – 44 vol. %. Porosity of the LFP cathodes is comparable with the typical practical range found for LFP cathodes with good electrochemical performance[78,79]. Special (dedicated) electrode composite formulations and electrode parameters are presented in Table 1.

**Preparation of Electrochemical Cells and Electrochemical Measurements**

All the electrochemical experiments were conducted by using vacuum sealed pouch-type electrochemical cells. The conventional 2-electrode LFP-Li cells were assembled in an Ar-filled glove-box by using pre-prepared circular cathodes as a working electrode and circular metallic lithium foil (geometric area of 2.5 cm$^2$) as a counter electrode. The lithium foil used for preparation of lithium anodes was immediately before that scratched and manually rolled in glove-box in order to remove surface layer(s) thus yielding freshly exposed shiny surface. As a separator we used a glass-fiber filter paper (Whatman, GF/A glass microfiber) with geometric area of 3.5 cm$^2$ and thickness of 260 µm (non-compressed) and about 200 µm (when squeezed in assembled cell), respectively. The electrolyte used was commercial "LP-40" 1 M solution of LiPF$_6$ in ethylene carbonate/diethyl carbonate (EC:DEC = 1:1 wt/wt, Merck).

Prior to all the current stimulus experiments the LFP-Li cells with "pristine" LFP electrodes were galvanostatically pre-cycled with C/10 (5 cycles) in voltage window 2.7 – 4.1 V, followed by a +C/10 charge up to 3.9 V vs. Li, potential hold for 6 h at 3.9 V, and relaxation at OCV for 6 h. This means we ensured the cells were free of the initial cycling effects and the global DOD of the LFP electrodes was driven to be outside the 2-phase regime to be close to 0 ($x \approx 0$ in Li$_x$FePO$_4$). Similar pre-cycling routine was applied for LMP-Li cells (3 cycles at C/20 in voltage window 2.7 – 4.5 V). All the galvanostatic and impedance electrochemical experiments were performed using a "VPM3" (Bio-Logic) potentiostat/galvanostat running with EC-Lab® software. For the dedicated experiment with LFP-Li cell where we have measured impedance response after partial discharge into plateau region of LFP we have used small amplitude (5 mV) sinusoidal perturbation in frequency span from 20 kHz down to 0.2 mHz. Intentionally we have used very low number of measured points (3 points per decade of frequency sweep) in order to reduce the measurement time to minimal and thus minimise the effect of further cell relaxation. In order to diminish the effects of surrounding temperature variation, all the measurements were performed by keeping battery cell temperature constant in a thermostatic bath (dipped in silicone oil with temperature 25 ± 0.1 °C). Specifications of all the electrochemical cells used in the present study are provided in Table 1.

**Table 1.** Specifications of the testing cells used in the present study. Provided is information about: type of active insertion material (AM), (dry) electrode composite composition (CB = carbon black, PVdF = polyvinylidene difluoride binder), electrode parameters and type of electrochemical measurement. "LFP by PCR-method" denotes LFP material synthesized in house by PCR-method. *Cathode with "diluted LFP" contains extremely low mass fraction of LFP and very high mass fraction of CB – effectively providing very good electronic and ionic wiring of the AM. All the cells presented in Table 1 included LFP based cathode and Lithium metal anode.

| Cell No. | Active material (AM) | Electrode composition AM : CB : PVdF (wt. %) | Electrode parameters | | Measurement, DOD at stopping |
|---|---|---|---|---|---|
| | | | AM mass loading (mg/cm$^2$) | Porosity (vol. %) | |
| #1 | Targray LFP | 90 : 5 : 5 | 4.5 | 41 ± 1 | Baseline, current stimuli, complete C/100 cycle |
| #2 | Targray LFP with selected particle size/aggregation | 90 : 5 : 5 | 0.6 | ~50 | Baseline, DOD = 0.5 |
| #3 | Targray LFP with selected particle size/aggregation | 90 : 5 : 5 | 0.6 | ~50 | Current stimuli, DOD = 0.51 |
| #4 | LFP by PCR-method | 90 : 5 : 5 | 4.1 | 43 ± 1 | Parallel battery experiment |
| #5 | LFP by PCR-method | 90 : 5 : 5 | 4.1 | 43 ± 1 | Parallel battery experiment |
| #6 | LFP by PCR-method | 90 : 5 : 5 | 2.1 | 43 ± 1 | EIS measurements, |

| | | | | | DOD = 0.9 |
|---|---|---|---|---|---|
| #8 | LFP by PCR-method | 90 : 5 : 5 | 3.1 | 43 ± 1 | Current stimuli of base -C/5 |
| #9 | LFP by PCR-method | 90 : 5 : 5 | 3.4 | 43 ± 1 | Current stimuli in both directions (effect of electrode thickness) |
| #10 | LFP by PCR-method | 90 : 5 : 5 | 0.3 | ~43 | Current stimuli in both directions (effect of electrode thickness) |
| #11 | Targray LFP | 5 : 73 : 22 | 0.018* | 75 ± 2 | Baseline, current stimuli (variation of stimuli duration) |
| #12 | Targray LFP | 90 : 5 : 5 | 4.5 | 41 ± 1 | Baseline, current stimuli, complete C/100 cycle |
| #13 | Targray LFP | 90 : 5 : 5 | 3.1 | 41 ± 1 | Baseline, current stimuli |
| #14 | LMP | 80 : 10 : 10 | 0.4 | ~54 | Baseline, current stimuli (variation of baseline current magnitude) |
| #15 | LFP by PCR-method | 90 : 5 : 5 | 3.3 | 43 ± 1 | Current stimuli (variation of baseline current magnitude) |

**Sample preparation for Scanning Transmission Electron Microscopy - Electron Energy-Loss Spectroscopy (STEM-EELS) experiments**

Particles of Li$_x$FePO$_4$ electrode material were obtained from LFP cathodes which were beforehand subjected to selected electrochemical protocols in the corresponding LFP-Li cells as follows. We conducted two main electrochemical experiments (baseline and with a current stimuli) and two additional ones for the Fe$^{2+}$ and Fe$^{3+}$ references. Initially the LFP-Li cells were pre-cycled (3 cycles at C/10 in voltage range 2.5 – 4.1 V), followed by slow +C/20 charge and 20 h voltage hold at 3.8 V with intention to do complete de-lithiation and thus drive all the active particles in equilibrium Li-poor solid solution (FP) corresponding to the thermodynamic state at 3.8 V vs. Li. The cell that we used for obtaining material for the Fe$^{3+}$ reference was stopped in this condition. For the baseline experiment the cell was discharged by -C/100 down to DOD = 0.5; for the experiment with electrode activation the cell was first discharged by -C/100 down to DOD = 0.25, followed by -5C current stimuli until DOD = 0.5, and finally additionally discharged for 50 mins with -C/100 exactly to the point corresponding to the local maximum of the voltage curve (see Fig. 1a). The cell that was used to obtain material for the Fe$^{2+}$ reference was discharged by slow -C/20 discharge, followed by 20 h voltage hold at 2.7 V with intention to do complete lithiation and thus drive all the active particles in equilibrium Li-rich solid solution (LFP) corresponding to the thermodynamic state at 2.7 V vs. Li.

All the samples of LFP/FP (Li$_x$FePO$_4$) particles were obtained from the corresponding cells by transferring them quickly into Ar-filled glove-box, rapid disassembling, and washing of the cathode in 40 mL of high-purity anhydrous diethyl carbonate (DEC, Sigma-Aldrich) for 5 mins (with stirring). We let the washed electrode to dry in glove-box (about 2 mins). This means we removed the electrolyte from cathode and further relaxation of the local Li$_x$FePO$_4$ composition within particles was strongly diminished (inhibited) [6]. The inspected electrode was dipped in 3 mL of Ar-purged Ethyl acetate, sealed, and transferred in ultrasound (US) bath. After 20 mins of electrode exposure to US agitation we first let dispersion to settle down (10 mins), followed by pipetting out 10 µL of dispersion of particles and drop casting them onto a Cu lacey TEM grid. After 2 mins of drying the grid was placed in an Ar-filled container and sealed for transfer to TEM room (within 3 mins). The grid was then mounted to the TEM holder and inserted into the TEM within 1 minute. The total exposure of the sample(s) to air atmosphere was less than 3 mins.

**STEM-EELS measurements**

STEM-EELS observations were performed with a cold-field emission gun Cs corrected JEOL ARM microscope equipped with a GIF Quantum (Gatan) Dual EELS spectrometer. The acceleration voltage was set to 200 kV and the estimated electron-beam current was ~48.9 pA. The convergence and EELS collection semiangles were 18 and 60 mrad, respectively. Maps and standards spectra were collected using an energy dispersion of 0.1 and 0.25 eV per channel in the spectrometer. The exposure time during mapping acquisition was 0.5 s per pixel using sub-pixel scan. Different pixel sizes were used ranging from 5 to 14 nm depending on the size of the particles. The chemical maps for the determination of local $Fe^{2+}/Fe^{3+}$ composition (LFP/FP phases) in individual LFP particles were analyzed by multiple linear least square (MLLS) fitting. More technical details concerning the EELS data collection and description of the analysis regarding the measured EELS mappings are presented in Supplementary Section 2.

# Brief description of the simulation model

The main aim of the simulation results is to provide a bridge between the thermodynamic reasoning presented in Fig. 2 and experimental evidence presented in Fig. 1. The specific goal was the explanation of the first order experimental phenomena found in this paper, i.e. the pulse effect itself. For clarity and generality, various details than might further improve the fitting of secondary characteristics of measured curves but are non-essential for the explanation of the pulse effect are not necessarily covered as this would be beyond the scope of current paper. The simulation model is described in more detail in Supplementary section 6.1, whereas in the following paragraphs we merely outline its general characteristics.

Simulations were conducted using an extended continuum level porous electrode model inspired by porous electrode theory[80,81], which is upgraded to more consistently virtually represent the electrode topology and material characteristics. It is based on the governing equations for concentration of ionic species in the electrolyte and liquid as well as solid phase potential. These equations are solved under assumption of electro-neutrality, while considering widely adopted Butler-Volmer equation, with exchange current density derived from the regular solution theory[82], to model charge transfer molar flux coupling transport between electrolyte and solid domain.

Another merit of the applied modelling framework is a plausible determination of the chemical potential of active particles via a multi-scaling approach that preserves a high level of consistency with lower scales and thus increases the modelling fidelity of the cathode (elaborated in references[50,83,84]). Due to its relatively small size and short characteristic times for diffusion[14], the particles were simulated as 0D particles[84] with chemical potential derived through a consistent reduction of the detailed spatially resolved model relying on the regular solution theory for phase separating materials.[83] This 0D virtual representation of particles is inevitable to ensure finite computational times when performing electrode level simulations that are needed to virtually represent the electric performance response of the battery, which comprises a large number of particles.

The model considers that equilibrium chemical potential is inherently linked to the lithiation level of a particle and thus to the mass conservation in the active particle, which is very important to plausibly model redistribution of Li between active particles via electrolyte or via direct contact by considering the real electrode topology and materials properties[50]. For the simulations of the half-cell, a 2D unstructured mesh with total of 273 control volumes was used.

The applied approach thus opens possibilities for long-term transient simulations of the entire electrochemical cell or half-cell while considering specific topological characteristics of materials, which are crucial for credible analyses of path dependent entrance into the voltage hysteresis in phase separating battery materials.

## Supplemental information

Supplemental Information includes supplemental experimental procedures and detailed description of the simulation model including results presented in 39 figures and can be found with this article online at *TBC*.

## Acknowledgements

The authors acknowledge financial support from the Slovenian Research Agency (research core funding No. P2-0401 and P2-0393 and project J7-8270). This project has received funding from the European Union's Horizon 2020 research and innovation programme under grant agreement No. 769506. The authors acknowledge help of Andrej Debenjak for providing of a dedicated in-house build amplifier and technical support during implementation of the parallel battery experiment. The authors greatly acknowledge the help of Marjan Bele for performing SEM analysis and sharing of many suggestions during the preparation of material for STEM-EELS experiments.

## Author contributions

The concept for this work was defined by T.K., K.Z. and M.G.; T.K. and K.Z. elaborated and performed the thermodynamic analysis; J.M. synthesised materials and electrodes and performed measurements; F.R.-Z. and J.M. conducted STEM-EELS experiments and analysis. T.K., K.Z. and I.M. conceived the simulation model, I.M. performed simulations; T.K., K.Z. and M.G. wrote the draft manuscript with input from all authors; all authors reviewed & edited the final manuscript.

## Declaration of interests

The authors declare no competing interests.

Incoherent Nanoscale Domains on the Sequence of Lithiation in LiFePO4 Porous Electrodes. Advanced Materials *27*, 6591–6597.

49. Yu, Y.-S., Kim, C., Shapiro, D.A., Farmand, M., Qian, D., Tyliszczak, T., Kilcoyne, A.L.D., Celestre, R., Marchesini, S., Joseph, J., et al. (2015). Dependence on crystal size of the nanoscale chemical phase distribution and fracture in Li x FePO4. Nano letters *15*, 4282–4288.

50. Mele, I., Pačnik, I., Zelič, K., Moškon, J., and Katrašnik, T. (2020). Advanced porous electrode modelling framework based on more consistent virtual representation of the electrode topology. Journal of The Electrochemical Society *167*, 60531.

51. Tang, X., Wang, Y., and Chen, Z. (2015). A method for state-of-charge estimation of LiFePO4 batteries based on a dual-circuit state observer. Journal of Power Sources *296*, 23–29.

52. Ko, Y., and Choi, W. (2021). A new soc estimation for lfp batteries: Application in a 10 ah cell (HW 38120 L/S) as a hysteresis case study. Electronics *10*, 705.

53. Dong, G., Wei, J., Zhang, C., and Chen, Z. (2016). Online state of charge estimation and open circuit voltage hysteresis modeling of LiFePO4 battery using invariant imbedding method. Applied Energy *162*, 163–171.

54. la Rue, A., Weddle, P.J., Kee, R.J., and Vincent, T.L. (2020). Feature selection for state-of-charge estimation of LiFePO 4-Li 4 Ti 5 O 12 batteries via electrochemical impedance. In 2020 American Control Conference (ACC), pp. 231–236.

55. Wang, Y., Zhang, C., and Chen, Z. (2014). A method for joint estimation of state-of-charge and available energy of LiFePO4 batteries. Applied energy *135*, 81–87.

56. Duong, V.-H., Bastawrous, H.A., Lim, K., See, K.W., Zhang, P., and Dou, S.X. (2015). Online state of charge and model parameters estimation of the LiFePO4 battery in electric vehicles using multiple adaptive forgetting factors recursive least-squares. Journal of Power Sources *296*, 215–224.

57. Roman, D., Saxena, S., Robu, V., Pecht, M., and Flynn, D. (2021). Machine learning pipeline for battery state-of-health estimation. Nature Machine Intelligence *3*, 447–456.

58. Hu, X., Xu, L., Lin, X., and Pecht, M. (2020). Battery lifetime prognostics. Joule *4*, 310–346.

59. Zhang, Y., Song, W., Lin, S., and Feng, Z. (2014). A novel model of the initial state of charge estimation for LiFePO4 batteries. Journal of Power Sources *248*, 1028–1033.

60. Gaberšček, M. (2021). Understanding Li-based battery materials via electrochemical impedance spectroscopy. Nature Communications *12*, 1–4.

61. Zelič, K., Katrašnik, T., and Gaberšček, M. (2021). Derivation of Transmission Line Model from the Concentrated Solution Theory (CST) for Porous Electrodes. Journal of The Electrochemical Society *168*, 70543.

62. Lipu, M.S.H., Hannan, M.A., Karim, T.F., Hussain, A., Saad, M.H.M., Ayob, A., Miah, M.S., and Mahlia, T.M.I. (2021). Intelligent algorithms and control strategies for battery management system in electric vehicles: Progress, challenges and future outlook. Journal of Cleaner Production, 126044.
24

**Entering voltage hysteresis in phase separating materials: revealing the thermodynamic origin of a newly discovered phenomenon and its impact on the electric response of a battery**

**SUPPLEMENTARY MATERIAL**

Tomaž Katrašnik[1,*], Jože Moškon[2], Klemen Zelič[1], Igor Mele[1], Francisco Ruiz-Zepeda[2] and Miran Gaberšček[2,3,**]

[1]University of Ljubljana, Faculty of Mechanical Engineering, Aškerčeva 6, SI-1000 Ljubljana, Slovenia

[2]National Institute of Chemistry, Hajdrihova 19, SI-1000 Ljubljana, Slovenia

[3]University of Ljubljana, Faculty of Chemistry and Chemical Technology, Večna pot 113, SI-1000 Ljubljana, Slovenia

Corresponding authors:

* tomaz.katrasnik@fs.uni-lj.si

** miran.gaberscek@ki.si

# 1. Electrochemical experiments at different conditions and on different active materials

## 1.1 Laboratory cells with Lithium iron phosphate (LFP) based cathode

**Table S1.** Specifications of the testing cells that were used for obtaining LFP related data presented in the Supplementary Information. Provided is information about: type of active insertion material (AM), (dry) electrode composite composition (CB = carbon black, PVdF = polyvinylidene difluoride binder), electrode parameters and type of electrochemical measurement. "LFP by PCR-method" denotes LFP material synthesised in house by PCR-method. *Cathode with "diluted LFP" contains extremely low mass fraction of LFP and very high mass fraction of CB – effectively providing very good electronic and ionic wiring of the AM. All the cells presented in Table S1 included LFP based cathode and Lithium metal anode (for further information about cell preparation see Experimental procedures in the main text).

| Cell No. | Active material (AM) | Electrode composition AM : CB : PVdF (wt. %) | Electrode parameters | | Measurement, DOD at stopping |
|---|---|---|---|---|---|
| | | | AM mass loading (mg/cm$^2$) | Porosity (vol. %) | |
| #8 | LFP by PCR-method | 90 : 5 : 5 | 3.1 | 43 $\pm$ 1 | Current stimuli of base -C/5 |
| #9 | LFP by PCR-method | 90 : 5 : 5 | 3.4 | 43 $\pm$ 1 | Current stimuli in both directions (effect of electrode thickness) |
| #10 | LFP by PCR-method | 90 : 5 : 5 | 0.3 | ~43 | Current stimuli in both directions (effect of electrode thickness) |
| #11 | Targray LFP | 5 : 73 : 22 | 0.018* | 75 $\pm$ 2 | Baseline, current stimuli (variation of stimuli duration) |
| #12 | Targray LFP | 90 : 5 : 5 | 4.5 | 41 $\pm$ 1 | Baseline, current stimuli, complete C/100 cycle |
| #13 | Targray LFP | 90 : 5 : 5 | 3.1 | 41 $\pm$ 1 | Baseline, current stimuli |
| #14 | LMP | 80 : 10 : 10 | 0.4 | ~54 | Baseline, current stimuli (variation of baseline current magnitude) |
| #15 | LFP by PCR-method | 90 : 5 : 5 | 3.3 | 43 $\pm$ 1 | Current stimuli (variation of baseline current magnitude) |

### 1.1.1 Experiments with different magnitudes of current stimuli and base currents

We performed a series of tests with a systematic variation of the current stimuli - within a range that can occur in a real battery operation - at two different magnitudes of base currents (Fig. S1a, S1b, Fig. S2). The results of the experiments with both selected base current magnitudes, i.e. -C/100 (Fig. S1a, S1b) and -C/5 (Fig. S2), clearly show entrance into the voltage hysteresis in all of the cases whereby the tests were conducted by applying a wide range of magnitude of current stimuli. In addition, we performed a series of experiments with a systematic variation of base current magnitude while applying the same -1C current stimulus (Fig. S1c, S1d), which also clearly reveal a reproducible entrance into the voltage hysteresis. Due to the fact that the magnitudes of both the current stimuli as well as of the base current are similar to those typically found in everyday battery transient operation the presented sets of data confirm the generality of the observed phenomenon in transient operating conditions. Namely, the magnitudes of the current stimuli fall within the range of power peaks of high power operation, while the magnitudes of base currents fall within the range of low current discharges that can represent low load periods when powering e.g. auxiliary systems in a vehicle or in an energy storage system.

These results further confirm that the phenomenon shown in Fig. 1 of the main text is reproducible also in other operating conditions (at other magnitudes of base current and current stimulus) as well as at different levels of lithiation (DOD). Moreover, supplementing the basic experiment (Fig. 1) by the results shown in Figs. S1 and S2 and additionally in Fig. S4 it is definitely demonstrated that the phenomenon of entering into voltage hysteresis of LFP may be associated with a very large span of quantities of the charge transferred during the current stimulus. The only necessary condition for entering voltage hysteresis is that a current stimulus ends within the spinodal region of the phase-separating material. All of the described experimental observations are confirming the relevance of the discovered phenomenon in real battery operation.

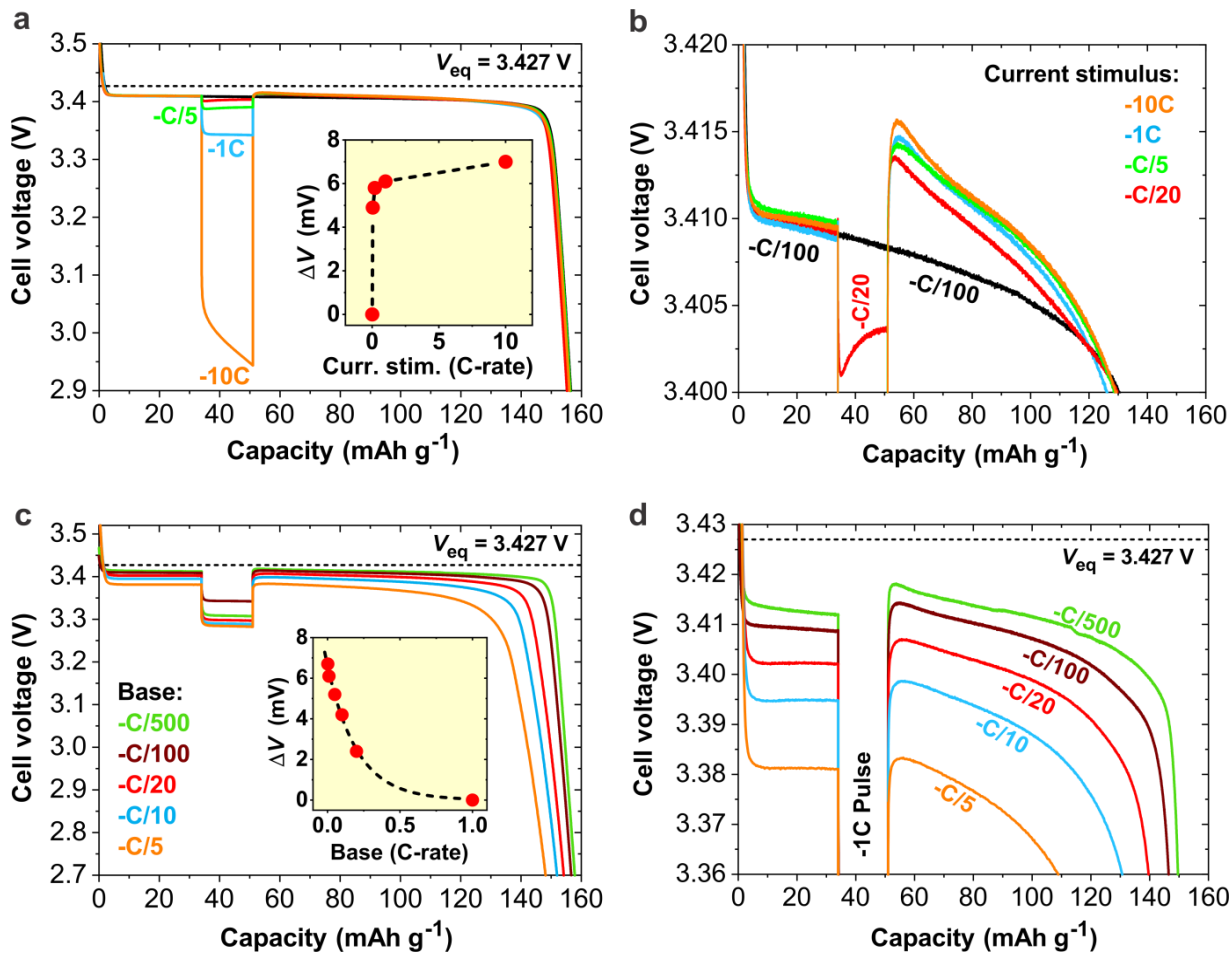

**Fig. S1** Set of experiments performed on the LFP-Li cell #8 with the LFP material synthesised by combustion reactor method which gives very reproducible batches of active composite. **a-b** Set of experiments with base discharge current -C/100 and different magnitudes of the current stimuli in range from C/20 up to 10C. **c-d** Set of pulse experiments with the same magnitude of the current stimuli (1C) and different magnitudes of the base current in range from C/500 up to C/5.

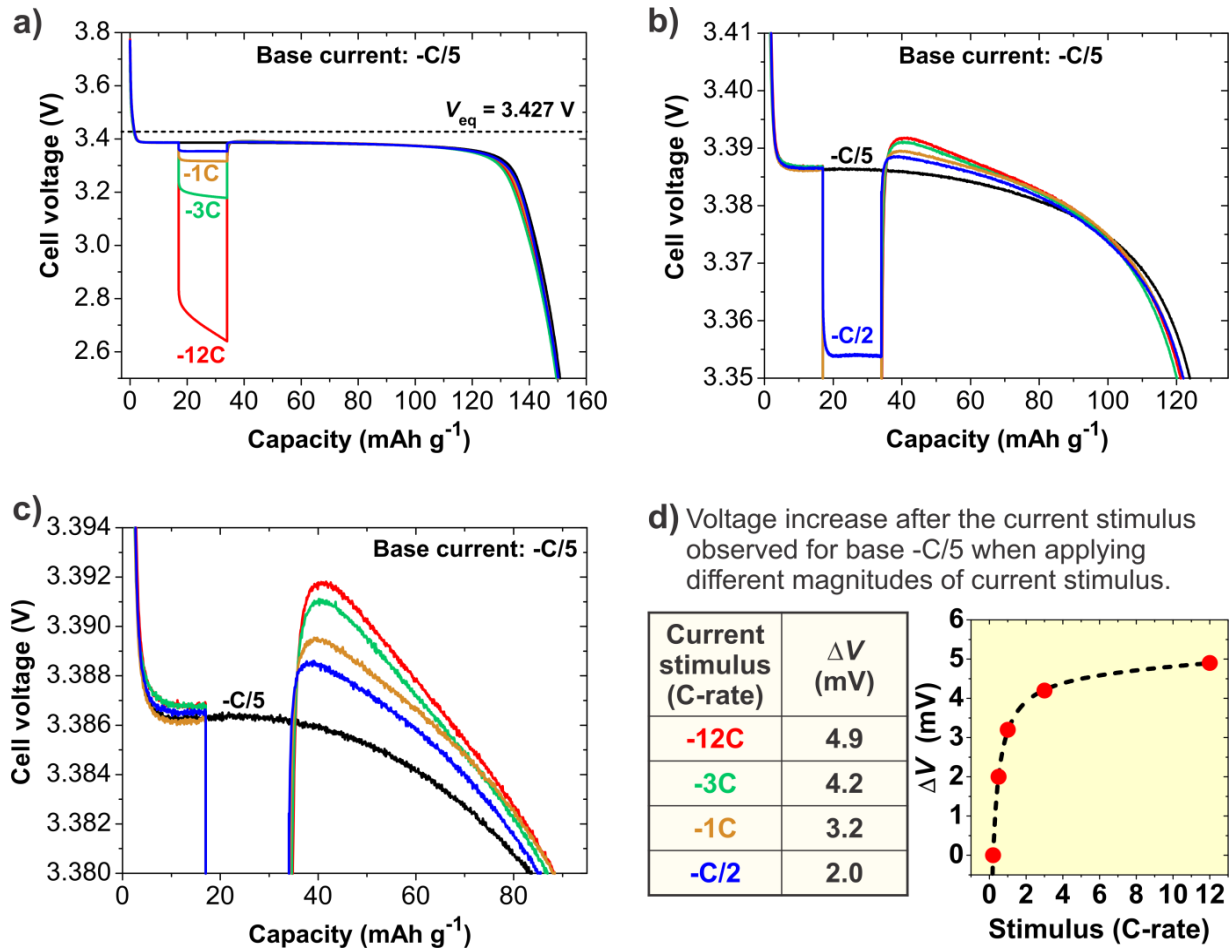

**Fig. S2** Set of experiments with -C/5 base current and different magnitudes of the current stimuli (-12C, -3C, -1C, and -C/2) performed on the LFP-Li cell #8 with the LFP material synthesised by combustion reactor method which gives very reproducible batches of active composite.

### 1.1.2 Experiments with current stimulus both during charge and discharge, performed on LFP cells with different electrode thicknesses

The results shown in Fig. S3 show that the phenomenon presented in Fig. 1 of the main text is observed both during charge and discharge of a LFP-Li cell. Using the Li-LFP cells (#9 and #10), we also checked the effect of cathode material (LFP) loading on the observed phenomenon. Besides the LFP cathode of regular thickness ($L_e$ = 22 µm corresponding to a mass loading of 3.4 mg LFP per 1 cm$^2$, cell #9), we also prepared a very thin cathode ($L_e \approx$ 2 µm corresponding to approximately 0.3 mg cm$^{-2}$, cell #10). The results of Fig. S3 demonstrate that the present phenomenon is not influenced by the amount of the active material (or its thickness). Among others, this rules out the possible impact of the intra-electrode gradients and their relaxation on the key phenomena leading to the entrance in the hysteresis, although loading does have a small impact on its magnitude. Moreover, this experiment rules out the possible impact of the electron and ion wiring properties of the material on the key phenomena yielding the entrance in the hysteresis. We believe this is a very important piece of information as there are many electrode and materials parameters (for example: electrode composition, porosity, tortuosity, electronic conductivity of solid phases, electrolyte transport properties, morphological features of conductive additive(s) and active material particles, binder distribution, wettability properties, etc.) that are known to affect the quality of ionic and electronic wiring. Additionally, the voltage loss associated with the ionic accessibility gradually develops during the course of dis(charge) [1, 2]. Despite all these possible complications, it is generally expected that in cases where the same type of electrode composite (same composition, porosity, etc.) is used and the same electronic connectivity between the composite and adjacent current collector is ensured, the related voltage loss at defined DOD is proportional to $L_e$ [3]. In this respect the result of Fig. S3 which confirms that the electronic and ionic wiring and associated overpotentials do not influence the key phenomena yielding entrance in the hysteresis is of crucial importance.

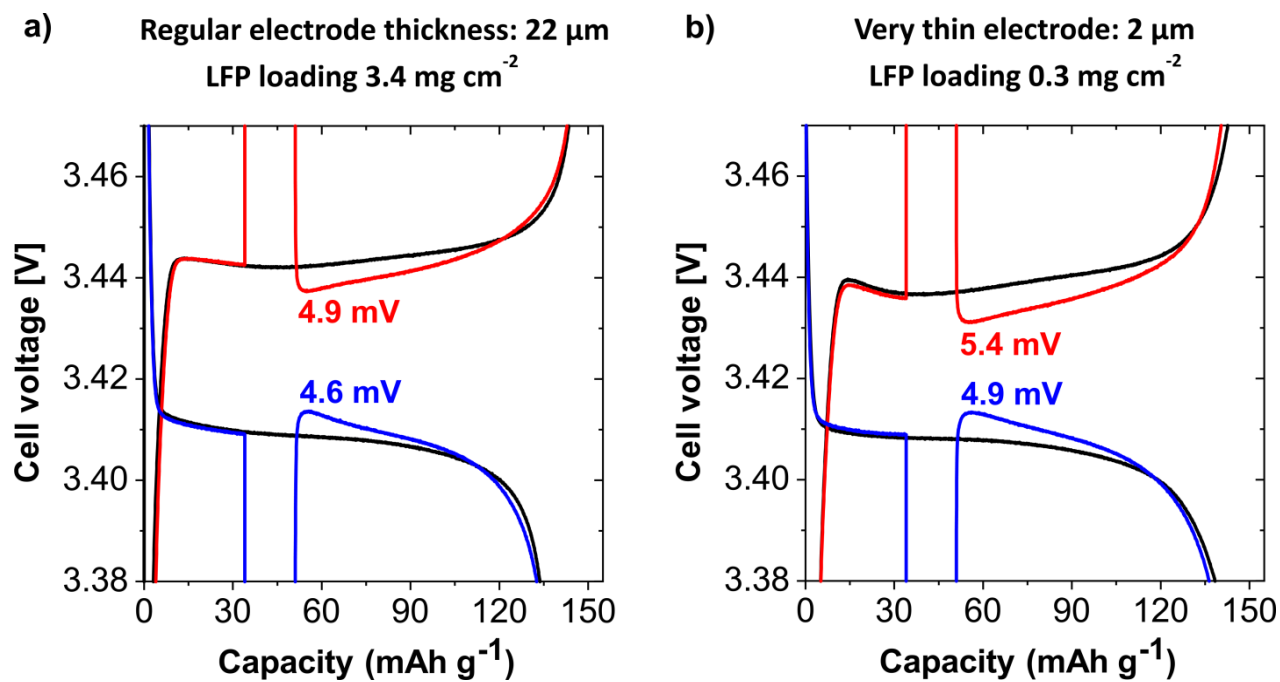

**Fig. S3** Experiment performed with a base current of C/20 and a current stimulus of 3C in both directions (discharge and charge) obtained on a LFP-Li cell with: a) LPF electrode of regular thickness (mass loading 3.4 mg cm$^{-2}$, cell #9), and b) very thin LFP electrode (mass loading 0.3 mg cm$^{-2}$, cell #10).

### 1.1.3 Variation of the length (duration) of current stimulus

Fig. S4 shows the measured voltage curves obtained in a systematic series of discharge tests on LFP-Li cell where the duration of the current stimuli of -5C was varied from extremely long (ΔDOD = 0.5) down to very short current stimuli (ΔDOD = 0.008 in Li$_x$FePO$_4$). The measurements were conducted on the cell #11 with the electrode composite with diluted distribution of LFP material (denoted as diluted electrode) containing an extremely low mass (35 μg) of active material. Electrode composition and electrode parameters are provided in Table S1; basic electrochemical performance is shown in Fig. S5a where it is clearly seen that the measured C/100 voltage hysteresis loop is very close to the one measured on the LFP-cell with regular mass loading (cell #12). This diluted LFP cathode exhibits entrances into hysteresis loop that are very similar to those found in LFP cathodes with the regular loading of active material, as presented in the main text and in the Supplementary Section 1.1.

Despite the similarity of the main effects, there is one important difference inherent to the diluted electrode. Fig. S4 and in particular Fig. S5 reveals that after a longer current stimulus the voltage increases closely to the equilibrium value. The diluted cathode exhibits not only a very good electronic but also a very much improved ionic wiring of particles of the active material, which significantly reduces the corresponding overpotentials. This means that the material specific phenomena are more clearly revealed in such a material. This result further supports the thermodynamic reasoning proposed in the main text, where Fig. 2 depicts that the intraparticle phase separated chemical potential is just slightly above the equilibrium potential. Furthermore, these results further indicate that the ionic and electronic wiring do not affect the key phenomena leading to the entrance in the hysteresis.

In addition, Fig. S4 clearly indicates that the voltage hysteresis is entered during a low discharge current for very different lengths (durations) of the current stimuli, i.e. in the range from $\Delta DOD$ = 0.008 to 0.5. Furthermore, it is discernible from Fig. S4 that the voltage approaches closer to the equilibrium value in the case of longer current stimuli. Both observations are very important and both further support the thermodynamic reasoning proposed in the main text. The robustness of the observed phenomena with respect to the change that is transferred during the current stimulus, i.e. $\Delta DOD$, confirms the statement that at least short period of particle exposure to the high overpotential transient regime (introduced in Supplementary Section 4.2) is required to enter the hysteresis and thus in the intraparticle phase separated state. Furthermore, the sensitivity of the observed phenomenon with the respect to the change that is transferred during the current stimulus, i.e. $\Delta DOD$, additionally supports the thermodynamic reasoning proposed in the main text, as during longer current stimuli more particles enter the lithiation range between the spinodal points. After reduction of the current this larger share of active particles enters the intraparticle phase separated state, which yields voltage levels closer to the equilibrium.

In addition to demonstrating the robustness and also the generality of the observed phenomena, this analysis also clearly exposes that electrode parameters (composition, porosity, etc.) and also characteristics of the current stimuli influence the magnitude of the presented phenomenon. The latter fact has a significant implication on electrode/cell analysis, diagnostics and monitoring as exposed in the Discussion section of the main text.

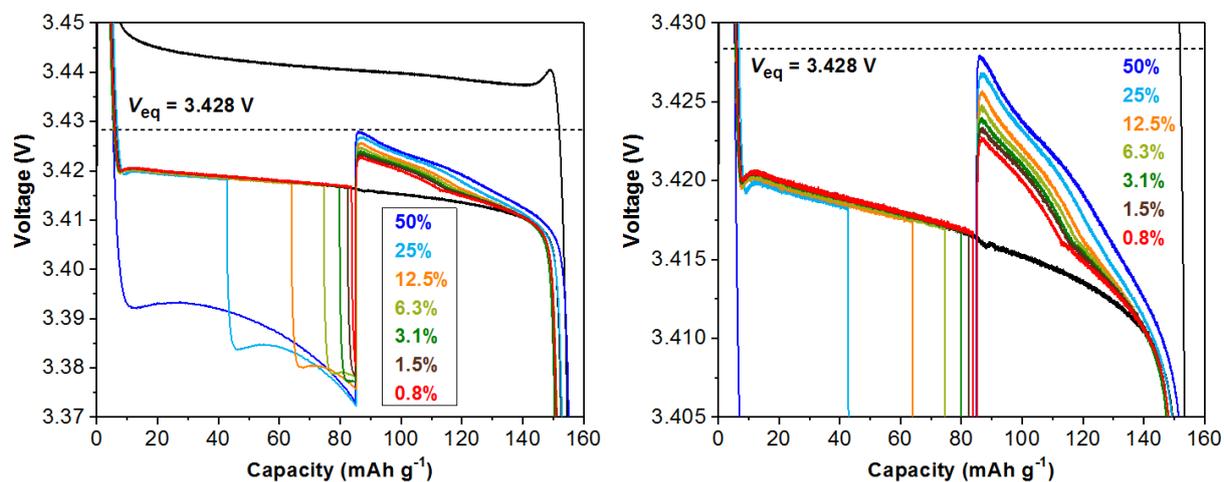

**Fig. S4** The effect of variation of the length (duration) of current stimuli in a wide range of values (a diluted LFP cathode in cell #11 was used for this set of experiments).

### 1.1.4 Impact of dilution of LFP material in electrode composite on electrochemical performance and the effect of entrance into the hysteresis

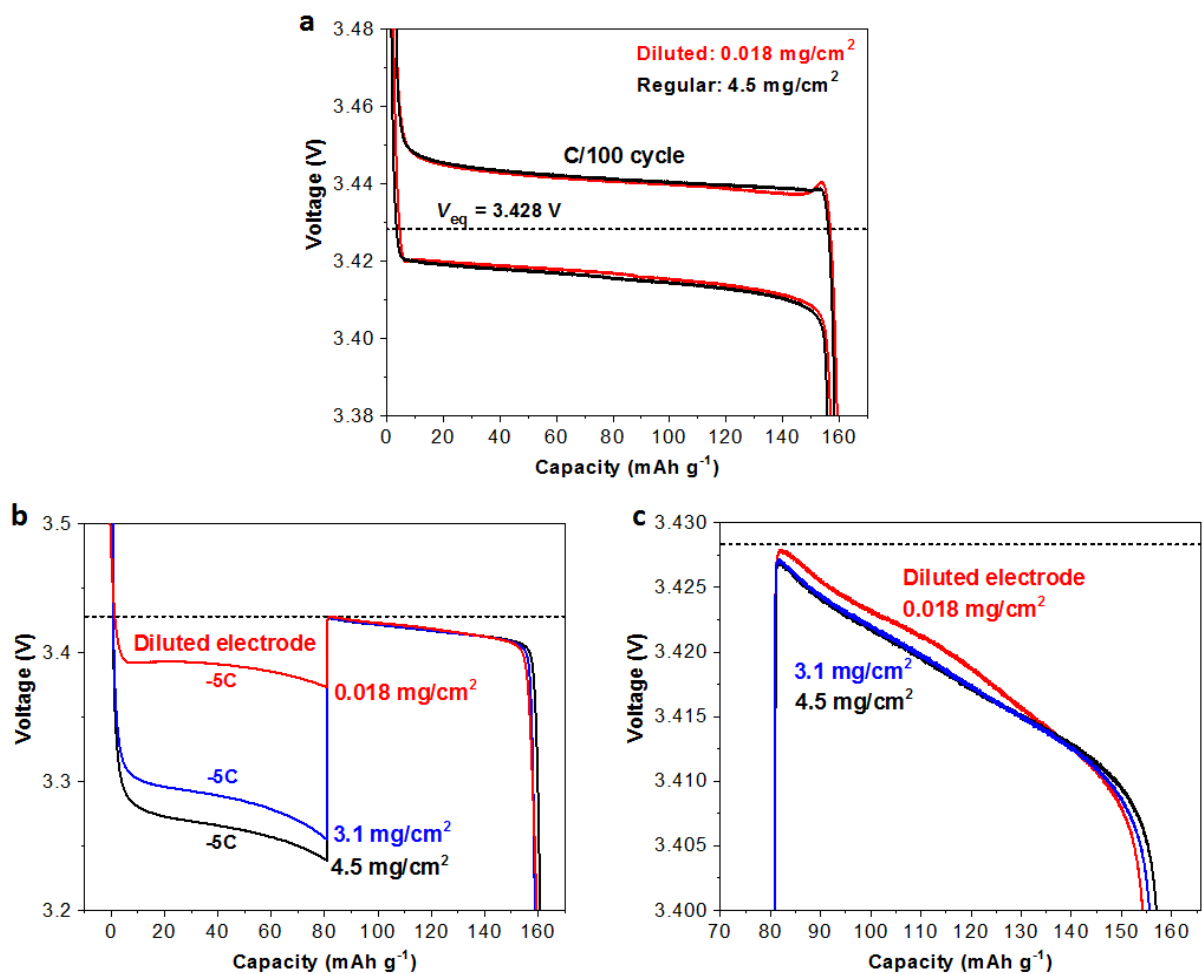

**Fig. S5 a** Comparison of the basic electrochemical performance (C/100 cycle) of the LFP cathode with diluted LFP (cell #11, red line) compared to the cathode with regular mass loading (cell #12). **b-c** Comparison of entrances into the hysteresis for different LFP electrodes characterised by the variation of the LFP mass loading in electrode composite from regular (4.5 mg/cm², black line and 3.1 mg/cm², blue line) down to extremely diluted distribution of the LFP material (0.018 mg/cm², red line).

## 1.2 Commercial LFP-Graphite battery cells

To additionally demonstrate the generality of the observed phenomenon, experiments were conducted also on two types of commercial LFP-Graphite battery cells: cylindrical Optium 32650 (5 Ah) and cylindrical A123 26650 (2.5 Ah). Both of the tested commercial cells (Figs. S6 and Fig. S7) clearly show that after a current stimulus ending after the first spinodal point of the LFP cathode the effect of voltage increase again pulls up the battery voltage into the voltage hysteresis. It should be noted here that for both tested commercial cells voltage hysteresis is larger compared to the hysteresis of laboratory-build LFP-Li cells due to the intrinsic contribution of the graphite anode to the total hysteresis of a LFP-Graphite cell. It is discernible from Figs. S6e and S6f that Optium 32650 features a similar magnitude of the voltage increase after the current stimulus as observed for laboratory cells. On the other hand, A123 26650 features an observable but clearly smaller effect of voltage increase (Fig. S7). The latter can be associated with the high power characteristic of this cell, which might include LFP cathode composed of smaller LFP particles featuring a lower potential barrier of the chemical potential [4, 5]. As mentioned before, these results clearly further confirm generality of the observed phenomenon.

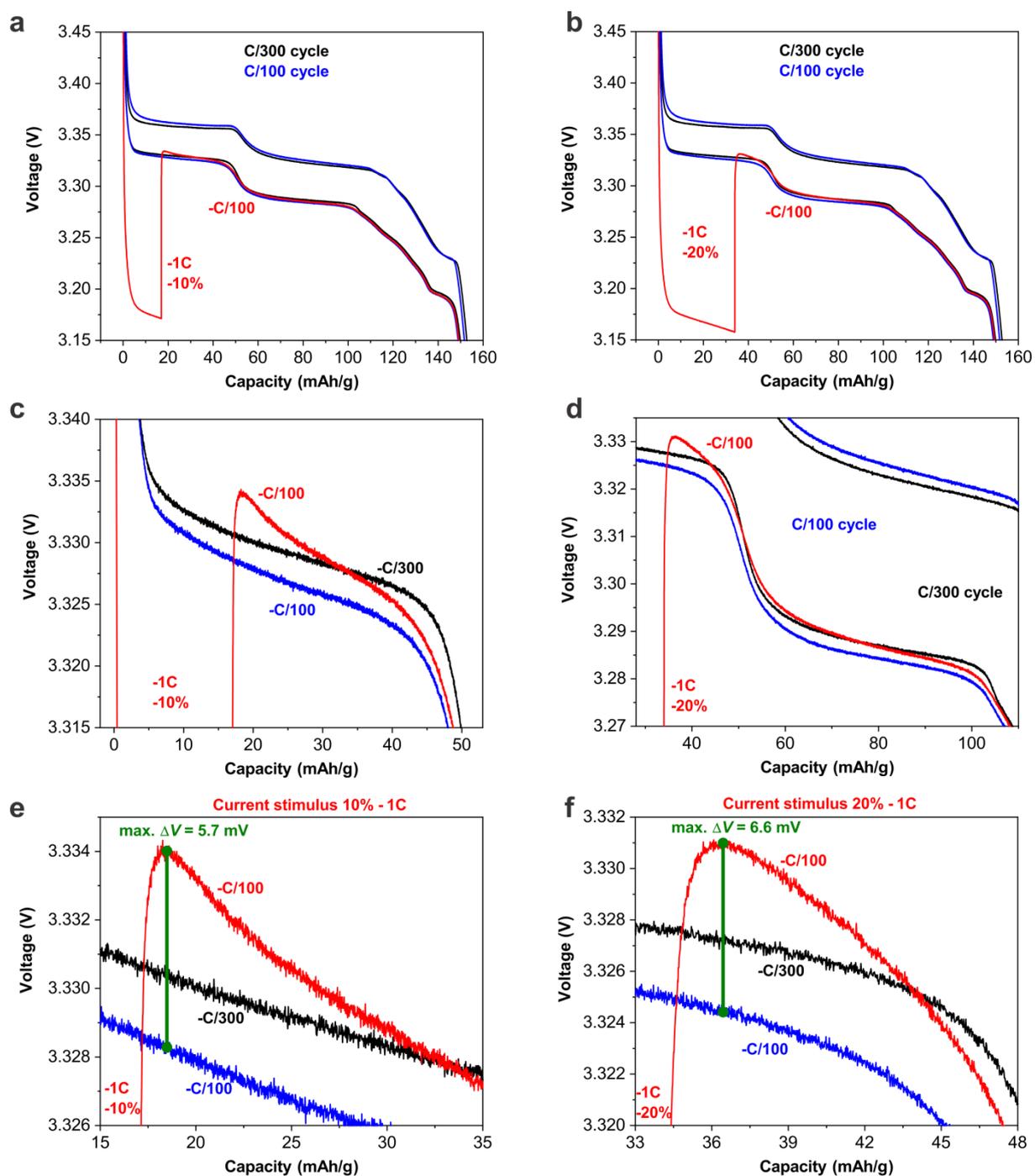

**Fig. S6** Tests on commercial cylindrical Optium 32650 (5 Ah) LFP-Graphite battery cell showing cell voltage for cycles obtained at currents -C/100 (blue) and -C/300 (black) as well as voltage trace for two (discharge) current stimulus experiments. Figures **a**, **c** and **e**: -1C current stimulus for initial 10% of the discharge followed by a -C/100 current (red); and figures **b**, **d** and **f**: -1C current stimulus for initial 20% of the discharge followed by a -C/100 current (red).

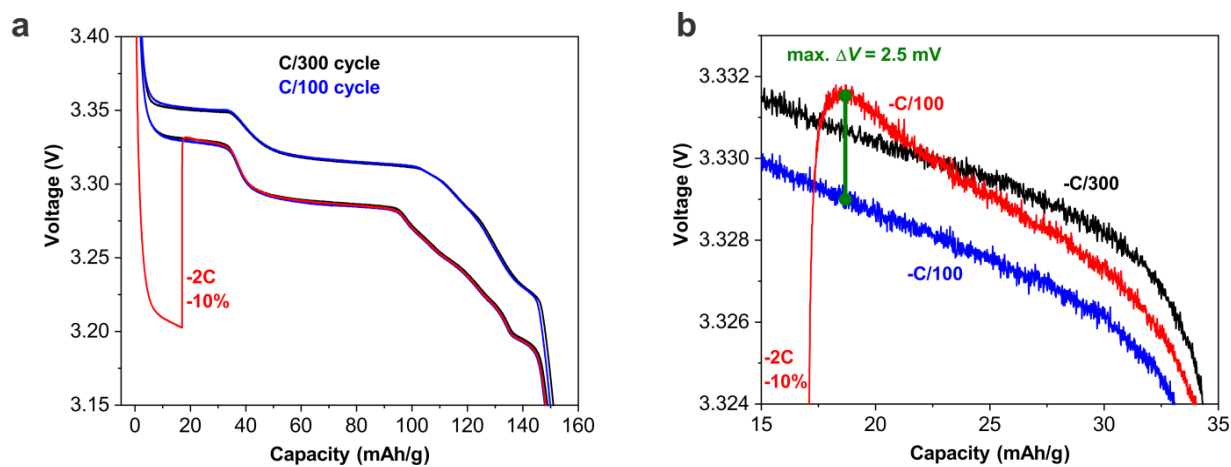

**Fig. S7** Tests on commercial cylindrical A123 26650 (2.5 Ah) LFP-Graphite battery cell showing cell voltage for cycles obtained at currents -C/100 (blue) and -C/300 (black) as well as voltage trace for the following composed (discharge) current stimulus experiment: -2C current stimulus for initial 10% of the discharge followed by a -C/100 current (red).

### 1.3 Lithium Manganese Phosphate (LMP)

#### 1.3.1 Methods

**Synthesis of Active Cathode Materials.** The LiMnPO$_4$ (LMP) active material was synthesised according to the two-step synthesis described in detail elsewhere [6]. In the first step, a homogeneous mixture of reactants without lithium was prepared in a round bottom flask by stirring stoichiometric quantities of manganese acetate (Fluka), citric acid (Sigma Aldrich), and phosphoric acid (Merck) (the molar ratio of Mn : P : citric acid was 1 : 1.1 : 1.5). The pre-dissolved Mn and P precursors and the solution of citric acid (each prepared as separate water solution) were mixed together at RT in a flask. The latter was then transferred to a vacuum rotary evaporator with a bath temperature of 60°C. In the first step of drying the pressure was carefully decreased to 60 mbar whereby most of the water was removed forming a viscous sol. The latter was subjected to a sudden pressure decrease to 10 mbar whereby the sol simultaneously expanded to form a voluminous foamy-like sol that was finally dried at 5 mbar for 2 h. The dried sol was thermally treated at 700°C for 1 h in an Ar atmosphere. In the second step, the composite from the first step was mixed with a 20% excess of LiOH (Aldrich, the Li : Mn molar ratio was 1.2 : 1), using planetary ball milling (Retsch) for 30 min at 300 rpm. The final LiMnPO4-C material was obtained with additional thermal treatment at 700°C in Ar for 12 h.

**Electrode preparation** with the LMP material was the same as for the LFP material presented in the main text with the exception that: a) the distance between the blade and the surface of Al foil was set to 75 µm, and b) electrode porosity determined by calculating the values of electrode composite mass, thickness and the bulk densities of the composite components was ~54 vol. %.

**Preparation of Electrochemical Cells and Electrochemical Measurements** the LMP material was the same as for the LFP material presented in the main text.

### 1.3.2 Entering into the voltage hysteresis of Lithium Manganese Phosphate (LMP)

With LMP material, similar experiments involving application of current stimuli were done as on LFP. The basic results displayed in Fig. S8 reveal a similar qualitative behaviour as shown in Fig. 1 in main text. However, as the intrinsic hysteresis is significantly bigger in LMP than in LFP, the magnitude of the peak after excitation with a current stimulus is also bigger and thus even easier to observe. We can see that the magnitude of the peak following the excitation depends on the base current of the experiment prior and after the current stimulus. Similarly, Fig. S9 reveals that the peak magnitude somewhat depends on the state of charge at the moment when the current stimulus was imposed on the basic discharge curve. All these measured effects support the present explanation of observed phenomena and thus the generality of the observed phenomena in other phase separating materials.

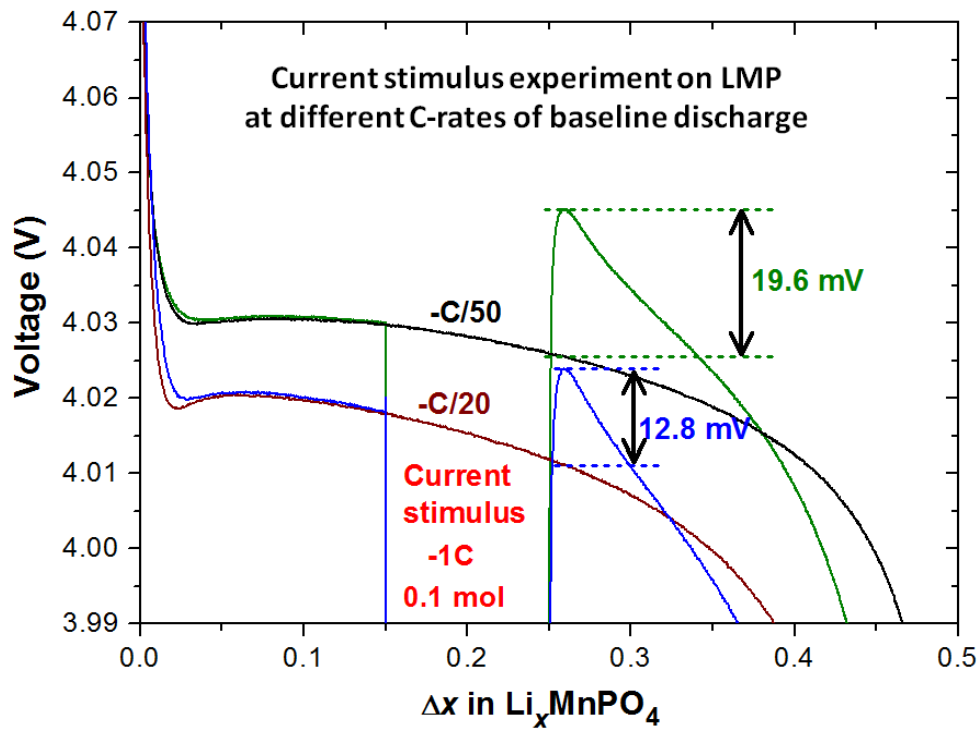

**Fig. S8** Basic experiment employing a current stimulus on a Lithium Manganese Phosphate material (LMP-Li cell #14) at two magnitudes of base current (-C/50 and -C/20).

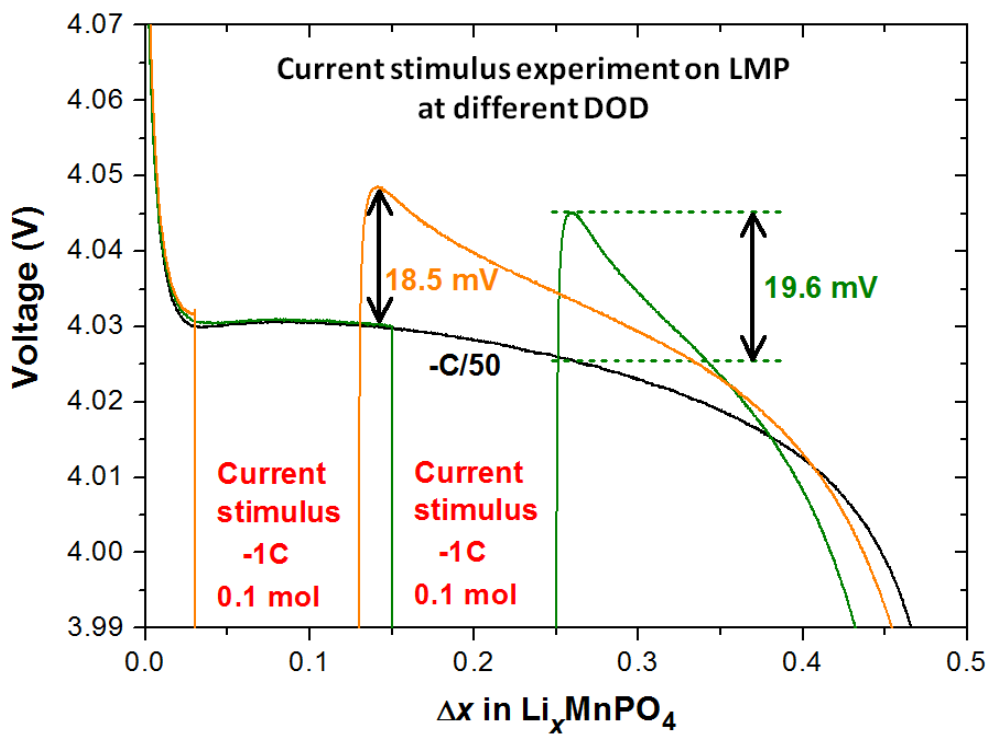

**Fig. S9** Application of current stimuli on a LMP material (LMP-Li cell #14) at different depth of discharge levels (DOD). The base current was -C/50 whereas the magnitude of the current stimulus was -1C.

## 2. Experiments with application of STEM-EELS to determine local $Fe^{2+}/Fe^{3+}$ composition in individual $Li_xFePO_4$ particles

This section presents experimental results with application of STEM-EELS to determine local $Fe^{2+}/Fe^{3+}$ composition in individual $Li_xFePO_4$ particles after applying: i) baseline -C/100 discharge, and ii) 5C current stimuli to the cathode. Ex-situ determination of local $Fe^{2+}/Fe^{3+}$ composition of individual LFP particles was attained by means of Electron Energy-Loss Spectroscopy in Scanning Transmission Electron Microscopy mode (STEM-EELS). The individual $Li_xFePO_4$ particles of electrode material that were used for STEM-EELS LFP/FP phase mappings were obtained from LFP cathodes which had been subjected to selected electrochemical protocols in the corresponding LFP-Li cells. Preparation steps and technical details for obtaining of the samples with $Li_xFePO_4$ particles for the STEM-EELS study are described in the main text (section Methods). Briefly, for the baseline experiment the cell was discharged by -C/100 down to DOD = 0.5; for the experiment confirming existence of the interparticle phase separated state the cell was first discharged by -C/100 down to DOD = 0.25, followed by -5C current stimuli until DOD = 0.5, and finally additionally discharged for 50 mins with -C/100 exactly to the point corresponding to the local maximum of the voltage curve (i.e. at the point with a maximum deviation of voltage from baseline -C/100 discharge curve, see Fig. 1a).

### 2.1 STEM-EELS measurements experimental conditions and data processing
All the EELS data were measured in a JEOL ARM 200 CF operated at 200 kV of acceleration voltage in STEM mode with an estimated beam current of 48.9 pA. Mapping spectra were collected using a GIF

Quantum (Gatan) Dual EELS spectrometer. Spectra were measured simultaneously at each pixel for two energy ranges: the low loss (to capture the zero-loss peak), and the core-loss electrons in an energy range suitable to capture the Fe edge (700 eV to 730 eV), using dispersions of 0.1 eV and 0.25 eV. The energy resolution as measured by the full-width at half-maximum of the zero-loss peak was 0.7 eV. The convergence and EELS collection semiangles were 18 and 60 mrad, respectively. The camera length employed was 3 cm and the energy filter entrance aperture was 5 mm. In order to minimise beam damage, sub-pixel scan was used along with a slightly defocused beam. The pixel size was chosen depending on the particle size, varying from 5 nm for smaller particles (300 nm), and from 8 nm to 14 nm for larger particles. Acquisition time of an individual pixel scan was 0.5 s/pixel. Total time maps ranged from 20 to 50 minutes. All the observed morphologies of LFP particles were mapped while avoiding and skipping thick and overlaying particles that would prevent unambiguous determination of the individual particles' phase maps. Experimental conditions and parameters for EELS acquisition were also consulted on the following references: [7, 8, 9].

After performing the zero-loss energy calibration for each pixel, the corresponding correction of the zero-loss energy was applied at each measured Fe edge EELS map. This feature is possible thanks to the availability of collecting two regions of the spectrum at the same time in the Dual EELS mode of the spectrometer. For the extraction of the valence ratio map of $Fe^{2+}/Fe^{3+}$, references for $Fe^{2+}$ and $Fe^{3+}$ were used (see below for description) in a Multiple Linear Least Squares (MLLS) fitting analysis routine for the Fe-$L_3$ and Fe-$L_2$ edges performed in Digital Micrograph. The output of the MLLS analysis is shown as the $Fe^{2+}/Fe^{3+}$ mapping where the obtained corresponding phase map of LFP/FP is represented as the linear combination of the fitting signals of the two ($Fe^{2+}$ and $Fe^{3+}$) used references where the results are displayed as the superposition of the two colours: red with RGB = [255, 0, 0] for the $Fe^{3+}$ reference, and green with RGB = [0, 255, 0] for the $Fe^{2+}$ reference.

## 2.2 $Fe^{3+}$ (FP) and $Fe^{2+}$ (LFP) References

In order to determine the local distribution of LFP/FP ($Fe^{2+}/Fe^{3+}$) with STEM-EELS maps in the inspected LFP particles, two references were prepared. The *$Fe^{3+}$ reference* was obtained from the cathode of the LFP-Li cell that had been slowly +C/20 charged up to 3.8 V followed by a 30 h voltage hold at 3.8 V with intention to perform a complete de-lithiation and thus drive all the LFP particles in a solid solution state corresponding to the thermodynamic state at 3.8 V vs. Li. The LFP-Li cell that was used to obtain the material for the *$Fe^{2+}$ reference* was discharged by a slow -C/20 discharge, followed by a 30 h voltage hold at 2.8 V with intention to do a complete lithiation and thus drive all the LFP particles in a solid solution state corresponding to the thermodynamic state at 2.8 V vs. Li (Fig. S10).

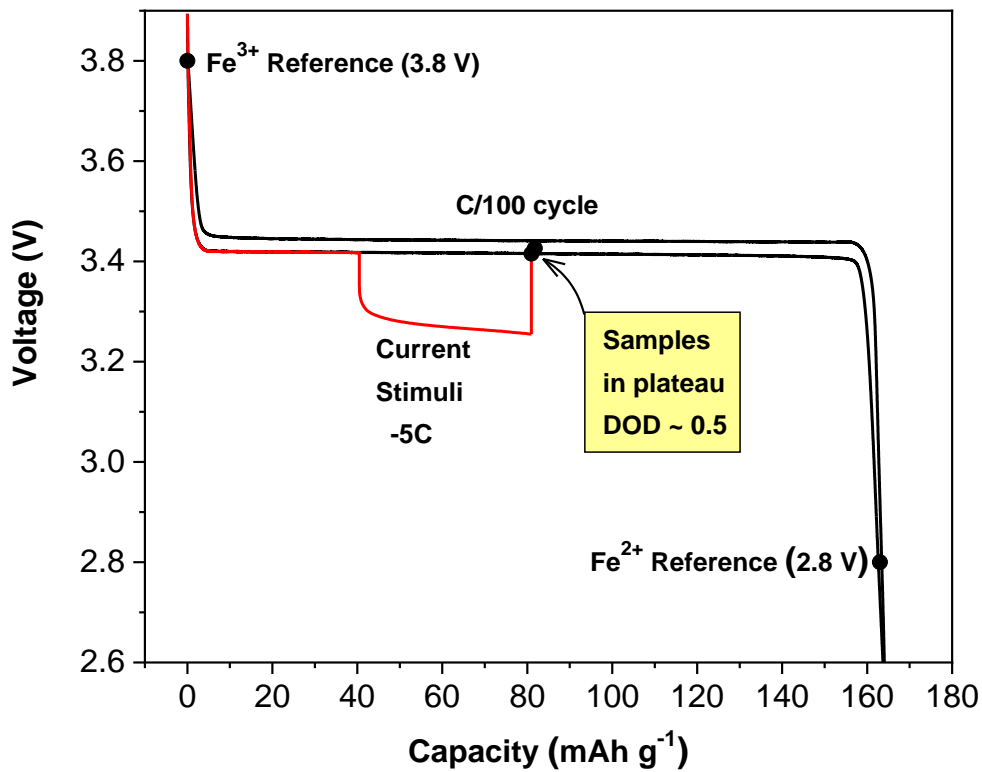

**Fig. S10** Position of the two references ($Fe^{3+}$ and $Fe^{2+}$) relative to the C/100 galvanostatic cycle and 25% ΔDOD -5C current stimuli experiment with LFP cathode based on a commercial Targray LFP material that was similarly used for the ex-situ STEM-EELS analysis.

The obtained energy-loss spectra of Fe-$L_{2,3}$ edge for the two references after zero-loss correction and background removal: $Fe^{3+}$ (red line), and $Fe^{2+}$ (green line) is shown in Fig. S11. The measured counts (after background removal) were normalised by the peak height of the maximum-intensity peak of the Fe-$L_3$ edge.

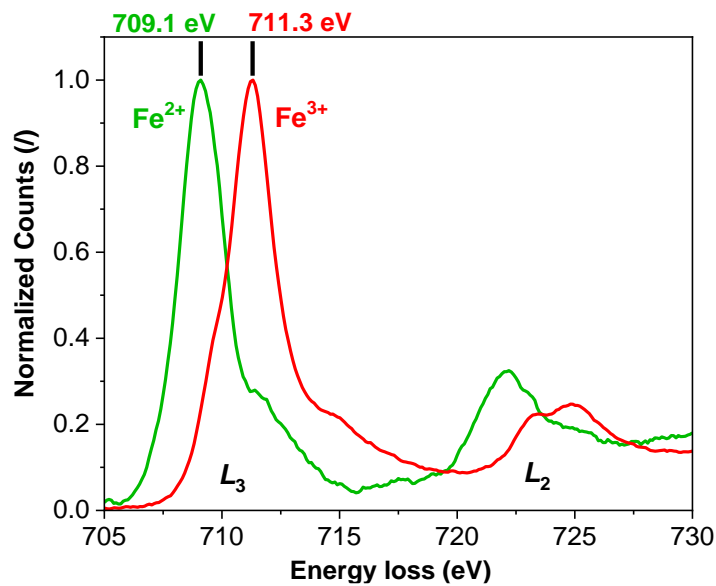

**Fig. S11** Obtained EELS spectra of the two references: $Fe^{3+}$ (red), and $Fe^{2+}$ (green) exhibiting distinct Fe-$L_{2,3}$ edge energy loss spectra. The $Fe^{3+}$ reference corresponds to Li$_x$FePO$_4$ equilibrium solid-solution at 3.8 V vs. Li, and the $Fe^{2+}$ reference corresponds to Li$_x$FePO$_4$ equilibrium solid-solution at 2.8 V vs. Li, respectively.

The general shape of the obtained Fe-$L_{2,3}$ edge profiles of our reference spectra is in good agreement with the reported literature data [7, 8, 9], exhibiting all the main characteristics as well as the detailed features previously observed for fully delithiated (FP) and full lithiated (LFP) LiFePO$_4$ samples. Fe-$L_{2,3}$ edges of LFP olivine materials display a characteristic energy loss near edge structure (ELNES) corresponding to excitations from the $2p^6 3d^n$ Fe ground state toward the $2p^5 3d^{n+1}$ Fe states. The two major features of these edges are two strong white lines ($L_3$ and $L_2$) due to the spin orbit splitting of the 2p core hole and separated by about 12 eV [9]. The edge structure exhibits additional characteristic fine spectral features (Fig. S12). Namely, FePO$_4$ ($Fe^{3+}$ in the octahedral site) is characterised by an $L_3$ edge with a leading shoulder and a peak maximum at about 711 eV, whereas LiFePO$_4$ ($Fe^{2+}$ in the octahedral site) is characterised by an $L_3$ edge peak maximum at about 709 eV, with no leading shoulder and a small plateau feature at about 712 eV [9]. The two detailed features (i.e. the leading shoulder for FePO$_4$ and the plateau feature for LiFePO$_4$) at the Fe-$L_3$ edge are considered to reflect a site dependence (both in the octahedral site) [10] and not a mixing of the two structures.

As shown in Fig. S12, in our case the most intense Fe-$L_3$ peak maxima of the reference spectra exhibit a chemical shift of about 2.2 eV between $Fe^{2+}$ state (~709.1 eV) and $Fe^{3+}$ state (~711.3 eV). The observed shift seems to be slightly larger compared to the reported values that were typically found in ranges 1.8 - 2 eV [7, 8, 9]. Although the small deviation is within the error limit (energy resolution) of our spectra, if it is a real deviation, one possible reason could be the fact that when preparing sample material for our references a complete lithiation and complete de-lithiation of the Li$_x$FePO$_4$ particles was done by performing a long (30 hours) voltage hold far above (3.8 V vs. Li) and below (2.8 V vs. Li) the equilibrium voltage plateau of LFP.

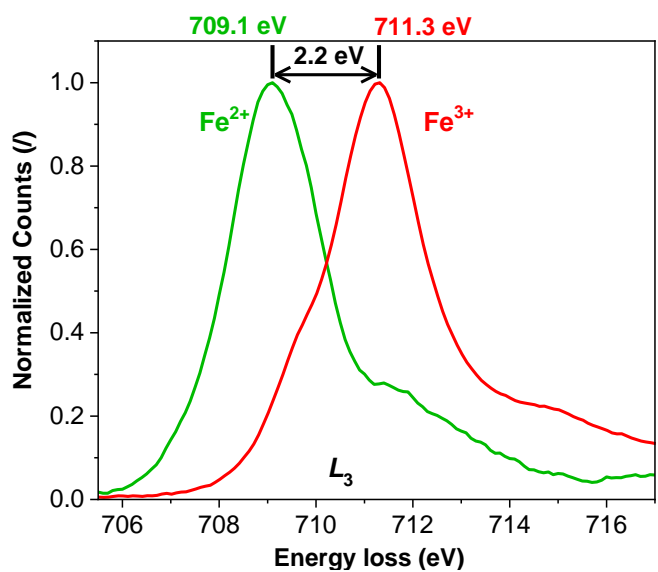

**Fig. S12** Intensity-normalised Fe-$L_3$ edge peaks of the references: $Fe^{2+}$ LFP (green) and $Fe^{3+}$ FP (red) that were used for MLLS phase map analysis. The peak maxima of the two reference spectra exhibit a chemical shift of about 2.2 eV between LFP (~709.1 eV) and FP (~711.3 eV) iron valence states.

The uniformity of our reference samples was checked by collecting STEM-EELS SI spectra and comparing cumulative EELS Fe-$L_{2,3}$ edge spectra of several selected medium-sized areas (roughly 240 nm × 240 nm) with one very large area (roughly 480 nm × 570 nm) of the same analysed particle (Fig. S13). This procedure was repeated on several particles with favourable relative mean thickness (mean free path, mfp ~1) to reduce multiple scattering effects. In the case of the $Fe^{3+}$ reference sample we performed this analysis on a typical FP particle (Fig. S13). We found that the collected spectra from the medium-sized areas exhibit random-noise scattering around the very-large-area spectrum that is in fact the *$Fe^{3+}$ reference* spectrum (Fig. S14).

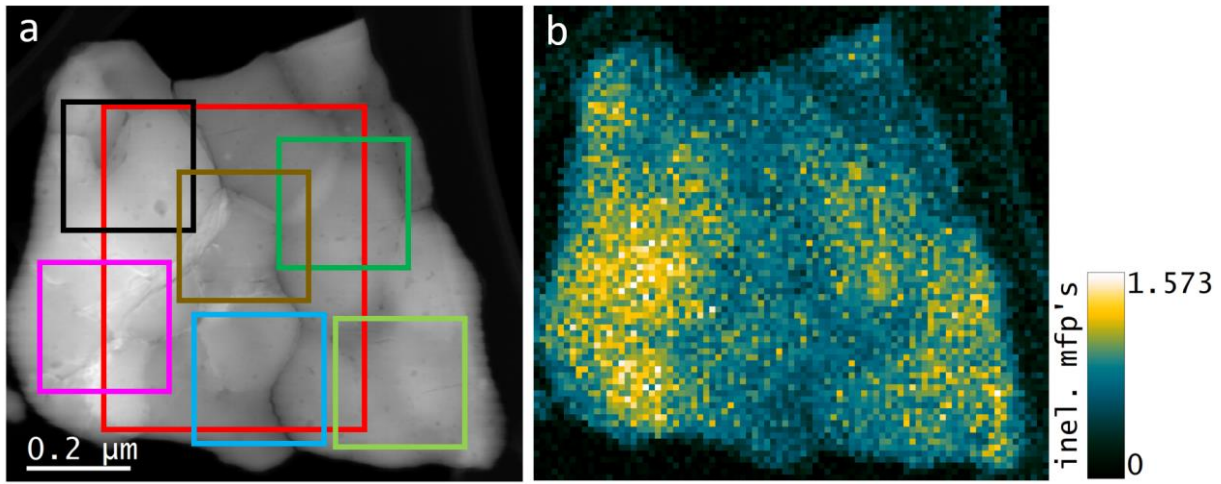

**Fig. S13** Checking uniformity of the $Fe^{3+}$ reference sample on a typical FP particle. **a** Collected EELS Fe-$L_{2,3}$ edge spectra at several selected medium-sized areas (roughly 240 nm × 240 nm) and at one very large area (roughly 480 nm × 570 nm). **b** Corresponding relative thickness map shown in mean free path (mfp) units.

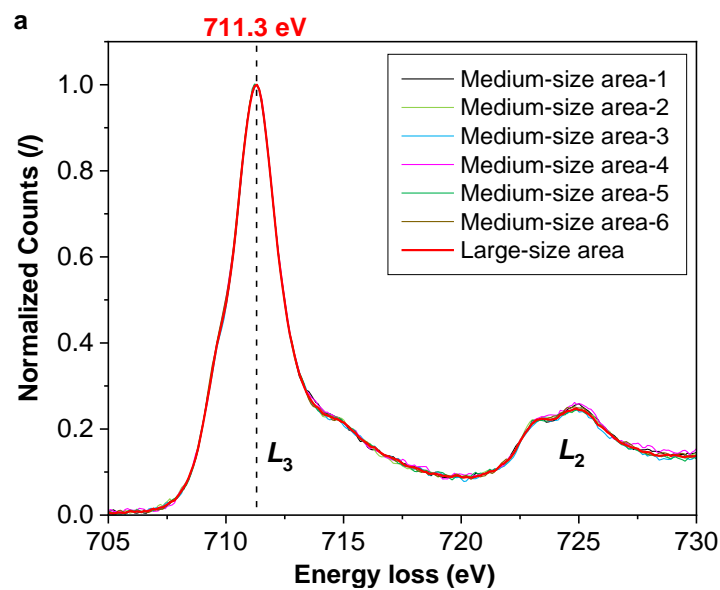

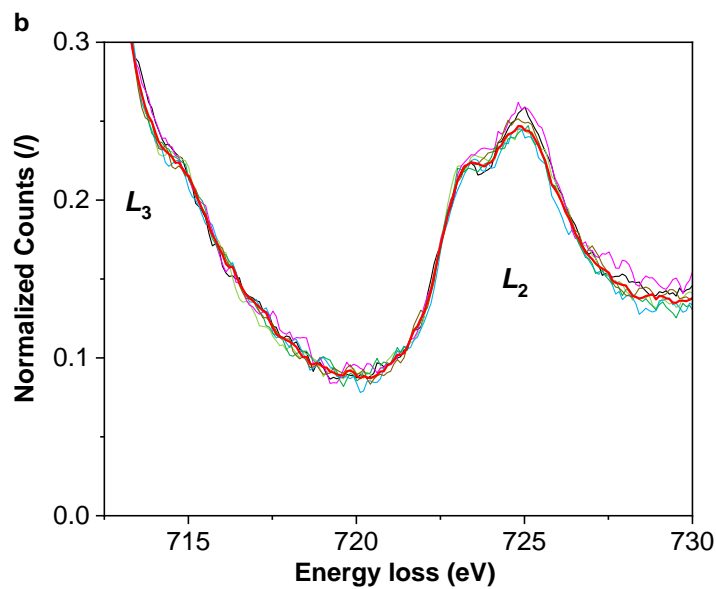

**Fig. S14 a-b** EELS Fe-$L_{2,3}$ edge spectra obtained at several selected areas of the $Fe^{3+}$ reference sample on a typical FP particle shown in Fig. S13. **b** The spectra collected from medium-sized areas (lower counts) are found to exhibit random-noise scattering around the very-large-area spectrum (higher counts) that is in fact the *$Fe^{3+}$ reference* spectrum.

For the case of the $Fe^{2+}$ reference sample we performed this uniformity analysis on a typical LFP particle (Fig. S15). Similarly as in the case of the $Fe^{3+}$ reference we found that the collected spectra from the medium-sized areas exhibited practically random-noise scattering around the very-large-area spectrum that is in fact the *$Fe^{2+}$ reference* spectrum.

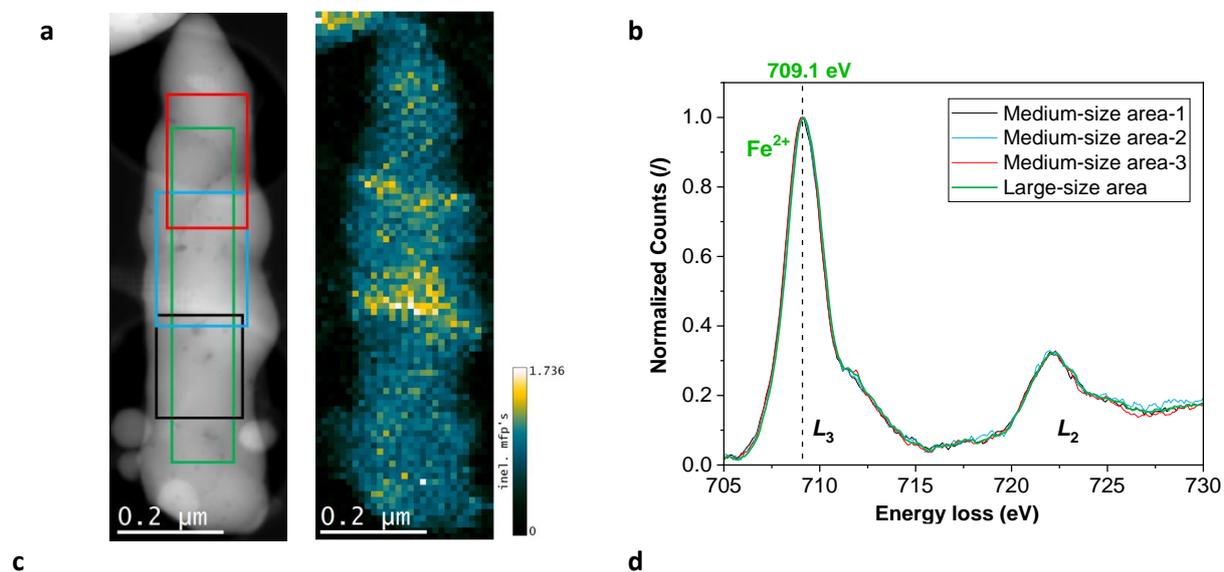

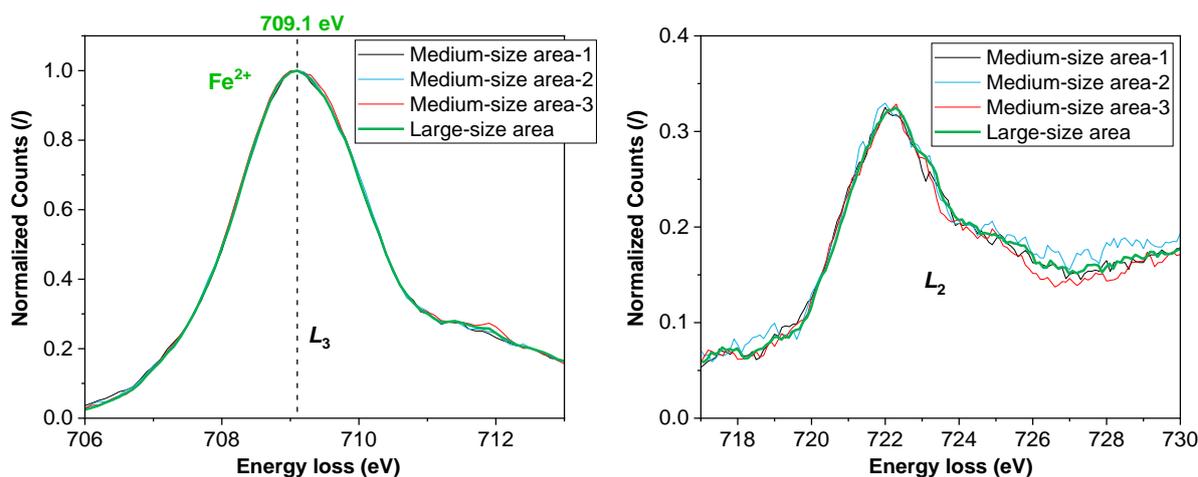

**Fig. S15** Checking uniformity of the $Fe^{2+}$ reference sample on a typical FP particle. **a** Collected EELS Fe-$L_{2,3}$ edge spectra at several selected medium-sized areas (roughly 160 nm × 200 nm, 160 nm × 240 nm, 140 nm × 250 nm) and at one very-large-area (roughly 150 nm × 630 nm). **b-d** The spectra from the medium-sized areas are found to exhibit random-noise scattering around the very-large-area spectrum that is in fact the *$Fe^{2+}$ reference* spectrum.

To summarise, we systematically checked cumulative EELS Fe-$L_{2,3}$ spectra of medium-size-areas and found that they are practically totally matching with the very-large-area spectrum collected at the same particle. Moreover, additional particles of the reference sample were inspected observing similar results. These examples confirmed that our prepared reference samples of FP ($Fe^{3+}$) and LFP ($Fe^{2+}$) were uniform in composition.

### 2.3 MLLS analysis for construction of LFP/FP phase maps

In order to determine the $Fe^{3+}$ or/and $Fe^{2+}$ signals present in each pixel of the measured EELS maps of the analysed particles, multi-linear least square (MLLS) fitting was applied on the Fe edge [7, 8] using Digital Micrograph's routine. The background subtracted signals of Fe-$L_3$ edge along with $L_2$ edge from the *ad-hoc* prepared FP and LFP reference particles (as described previously) were used as standards for the determination of the valence state $Fe^{3+}$ and $Fe^{2+}$, respectively. The measured high-loss signals were background subtracted and fitted in an energy window of 700 eV – 730 eV with acquired references specifically for $Fe^{3+}$ and $Fe^{2+}$ determination. The MLLS routine output corresponded to two maps of the (FP and LFP) fitting coefficients where the brightness of a pixel is proportional to the fit integral of a phase signal. An example of the MLLS fitting output from a scanned area taken from particles of the experiment with applied current stimuli is shown in Fig. S16. A relative thickness map is also shown, obtained by the ratio of the zero-loss peak and the transmitted intensity peak at each pixel, known as the log-ratio method. The relative thickness map ($t/\lambda$) has units of inelastic mean-free-path (mfp). Ideally, to avoid unwanted multiple scattering effects in the collected spectra, observational areas with less than ~1 mfp are considered. To estimate the approximated thickness, we used the MFP Estimator script from David Mitchell [11], yielding values of $\lambda$ = 101.4 for LFP and $\lambda$ = 97.9 for FP.

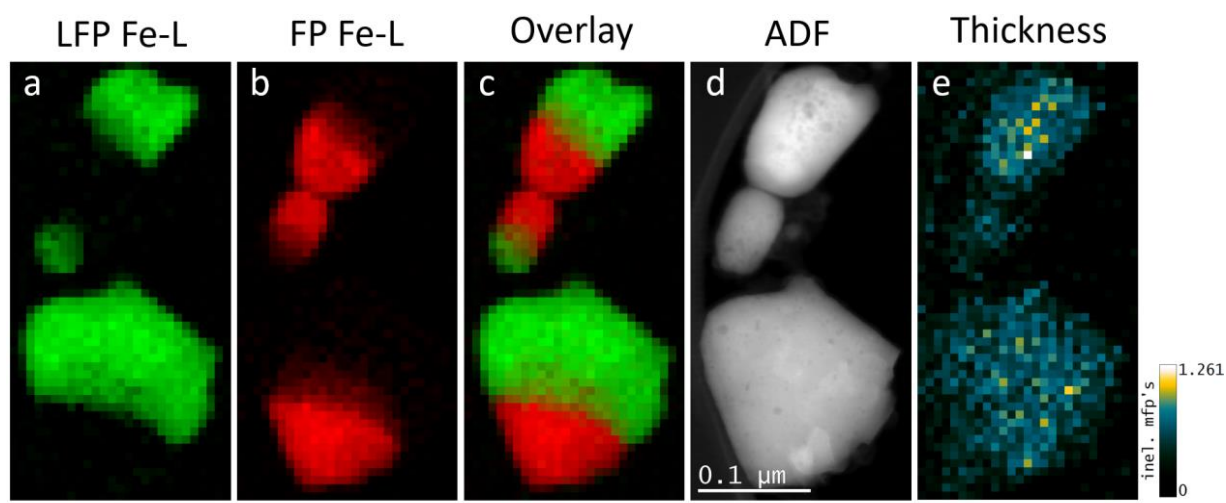

**Fig. S16** STEM-EELS analysed scanned area from particles of the experiment with applied current stimuli (corresponding to particle P2.09 from Fig. 1b in the main text). **a** MLLS fitting coefficient map of LFP Fe-$L_3$ edge, **b** MLLS fitting coefficient map of FP Fe-$L_3$ edge, **c** overlay map of the two determined phases, **d** ADF signal, and **e** the corresponding relative thickness map.

**2.4 Collections of Li$_X$FePO$_4$ particles from the two main STEM-EELS experiments where the LFP/FP phase map analysis was performed (additional information on the data displayed Figs. 1b and 1c in the main paper)**

In Figs. S17 and S18 STEM ADF images of the collection of the Li$_X$FePO$_4$ particles from the two main STEM-EELS experiments and the corresponding LFP/FP phase maps are shown. The particles from the baseline STEM-EELS experiment (Fig. 1b, main text) are displayed in Fig. S17, whereas the particles from the STEM-EELS experiment where we, after -C/100 discharge, applied -5C current stimuli (Fig. 1c, main text) are shown in Fig. S18.

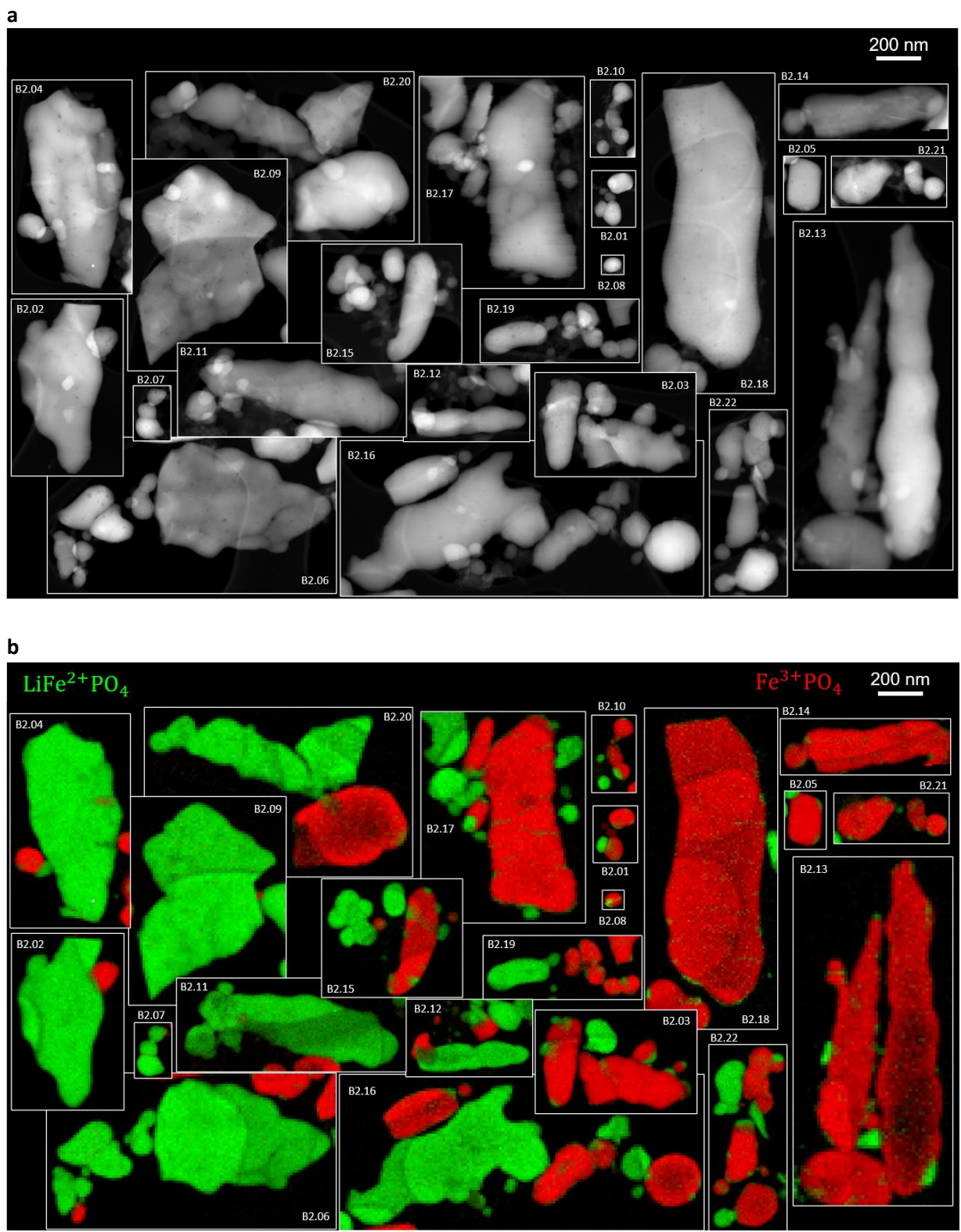

**Fig. S17 a** ADF images of the collection of Li$_X$FePO$_4$ particles and **b** the corresponding STEM-EELS MLLS fit maps from the baseline experiment where the LFP/FP phase map analysis was performed (Fig. 1b, main text).

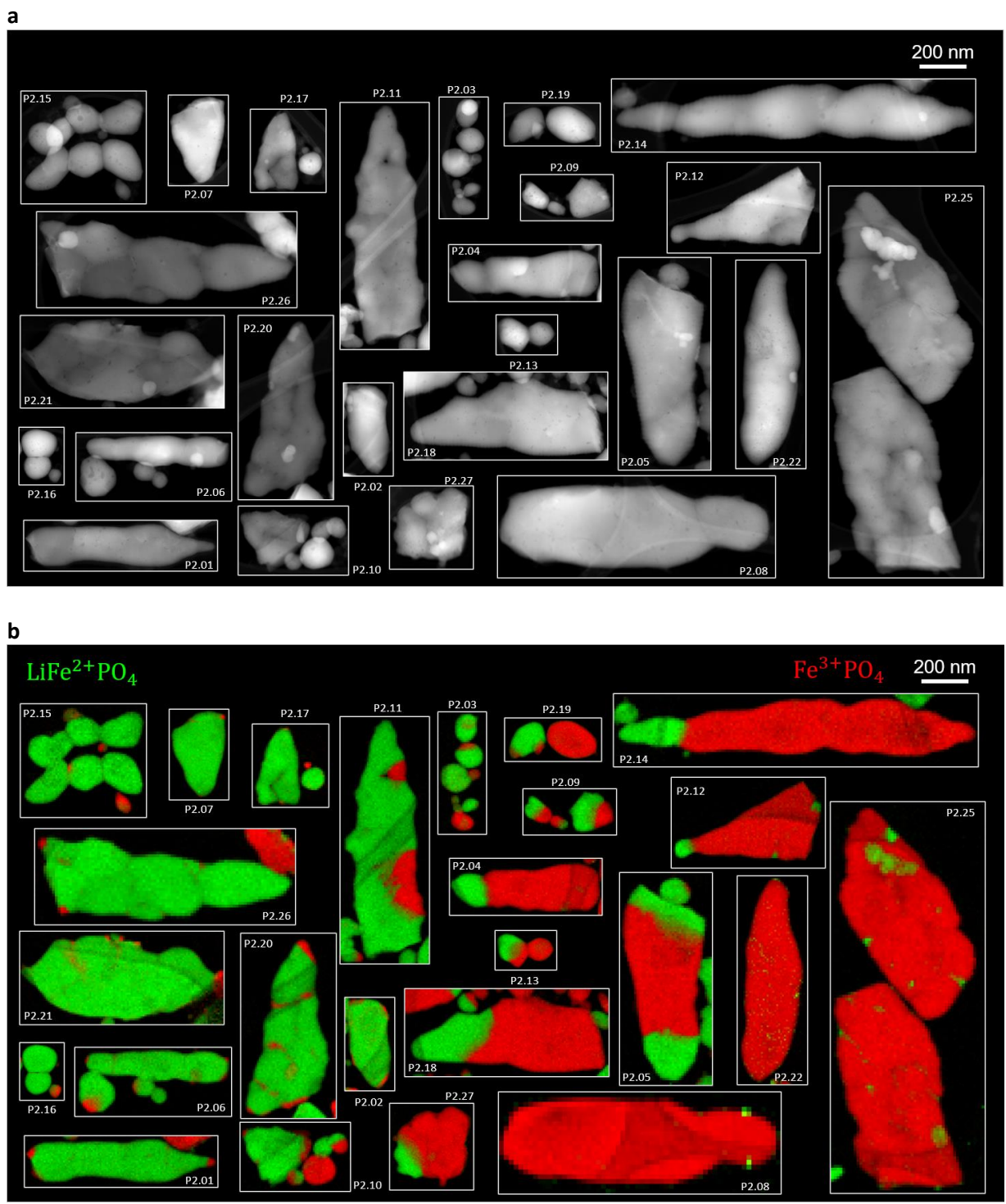

**Fig. S18** ADF images of the collection of Li$_X$FePO$_4$ particles and the corresponding STEM-EELS MLLS fit maps from the experiment with applied current stimuli where the LFP/FP phase map analysis was performed (Fig. 1c, main text).

It can be observed that the analysed particles are of various morphologies and various dimensions: ranging from small rounded ones with typical diameter in the range between ~50 nm up to ~250 nm, thin platelet particles with rather large lateral dimensions (up to 900 nm) and typical thickness span from about 60 nm to ~120 nm, to thicker platelet particles (thickness between ~130 nm and ~200 nm), and finally "carrot-like" rod-shaped particles with one preferential long axis. The smallest particles may correspond to single crystals while the larger ones are aggregates of several primary particles (primary particle size distribution is shown in Fig. S19b). An overview of the analysed samples' particles is shown in Fig. S19a.

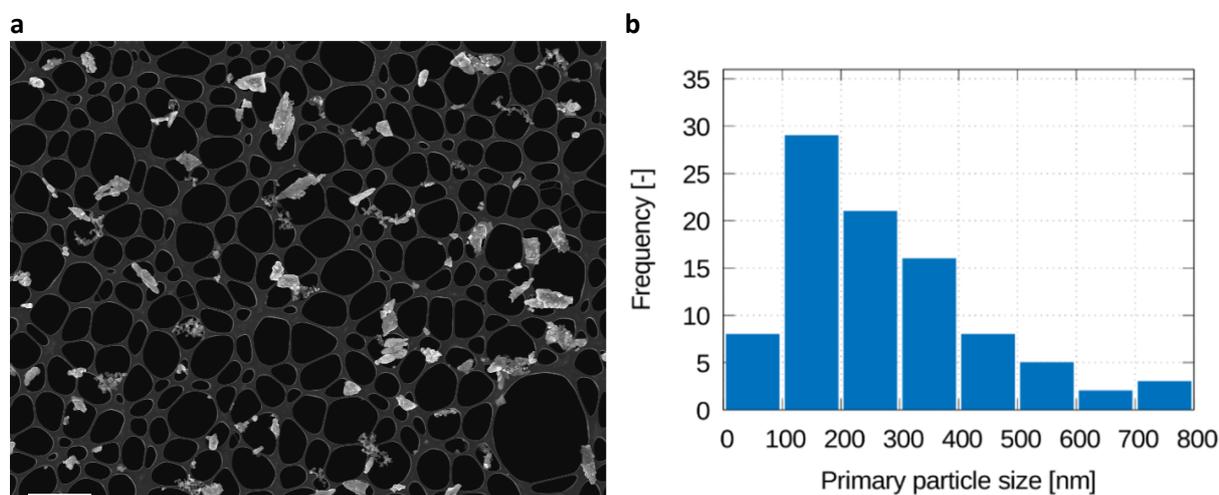

**Fig. S19 a** A set of Li$_x$FePO$_4$ particles imaged by SEM where all types of morphologies can be seen. **b** Size distribution histogram of LFP primary particle (grains) sizes experimentally obtained from TEM measurements with last bin representing the rest of the larger particles (>700 nm) [12].

### 2.5 Analysis of the selected areas in the LFP/FP phase maps

In order to verify the correctness of the LFP/FP phase maps output processed with the help of the MLLS fitting analysis of the Fe-$L_3$ edge (and Fe-$L_2$ edge), we performed a systematic inspection of cumulative energy-loss spectra of selected areas. In Fig. S20 a typical analysed scanned area from baseline experiment (which includes particles B2.06a and B2.06b from Fig. 1 in the main text) is shown. The STEM Annular Dark Field (ADF) image of the selected area suggests that the larger particle B2.06a is composed of several primary particles (grains). The upper part of the particle rests partially on the carbon film (from the TEM grid), this is evidenced by the border line crossing near the central part of particle B2.06a. From the corresponding relative thickness map (Fig. S20d) we can deduce that the relatively large particle B2.06a is in fact quite thin, i.e., having more of a platelet like morphology (varying from regions of 40 nm to 90 nm with maximum of 120 nm) compared to its neighbours (140 nm). In this sense, we have observed that in general, the majority of large particles in the baseline experiment which got completely lithiated (e.g. B2.20a, B2.11, B2.09, B2.16a, B2.02, B2.04) possessed a platelet morphology; see further discussion below about the sequence of lithiation (**section 2.6**). As mentioned before, such types of particles (areas with less than or around ~1 mfp) are preferred for EEL spectra measurements since the effects of plural scattering are less pronounced. In a post-acquisition analysis shown in Fig. S20, we selected a large area within particle B2.06a (avoiding the carbon film from the TEM grid) and obtained the corresponding spectra shown in Fig. S22 (blue line).

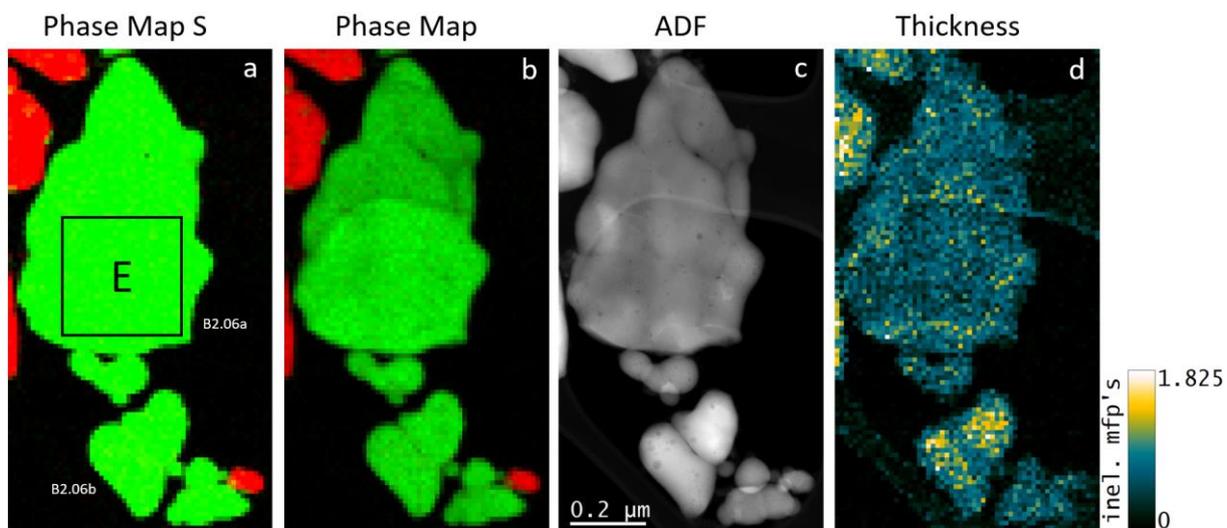

**Fig. S20** Analysed scanned area from baseline experiment (includes particles B2.06a and B2.06b from Fig. 1 in the main text). **a** LFP/FP phase mapping with saturated colours, **b** phase map, **c** ADF image, and **d** corresponding relative thickness map.

Additionally, in the baseline -C/100 lithiation STEM-EELS experiment we selected a second large scanned area containing a large lithiated platelet particle B2.04a with a varying thickness of around 60 nm up to 130 nm (Fig. S21), with a small rounded partially lithiated particle B2.04b at the side. Within the particle B2.04 we selected a large area (250 nm × 350 nm) with LFP phase extracting in post-acquisition analysis the corresponding EELS spectrum of the area as shown in Fig. S22 (black line).

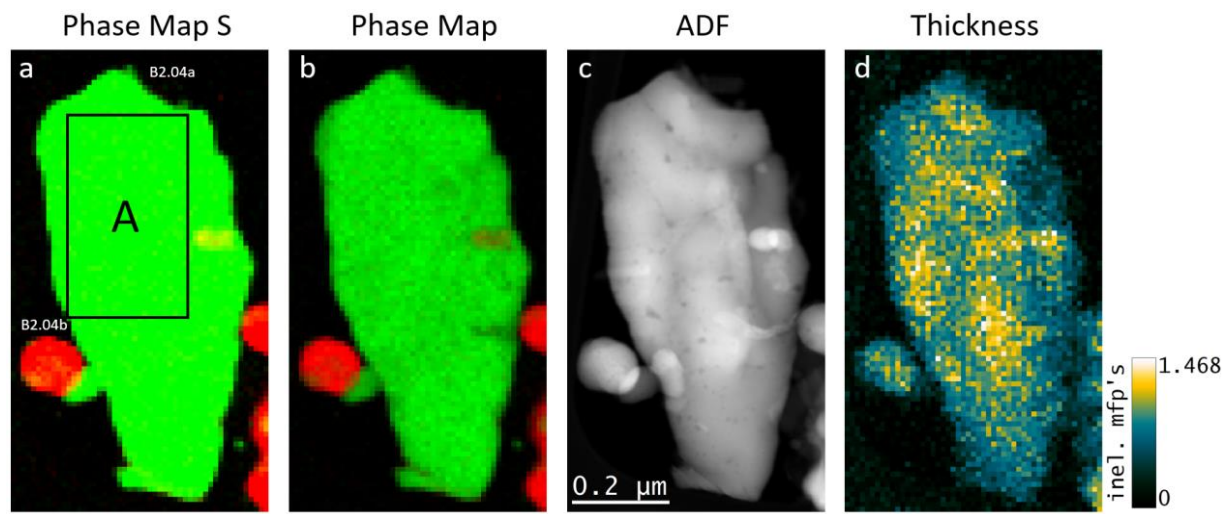

**Fig. S21** Analysed scanned area from baseline experiment that includes a large platelet particle B2.04a and particle B2.04b (from Fig. 1 in the main text). **a** LFP/FP phase mapping with saturated colours, **b** phase map, **c** ADF image, and **d** corresponding relative thickness map.

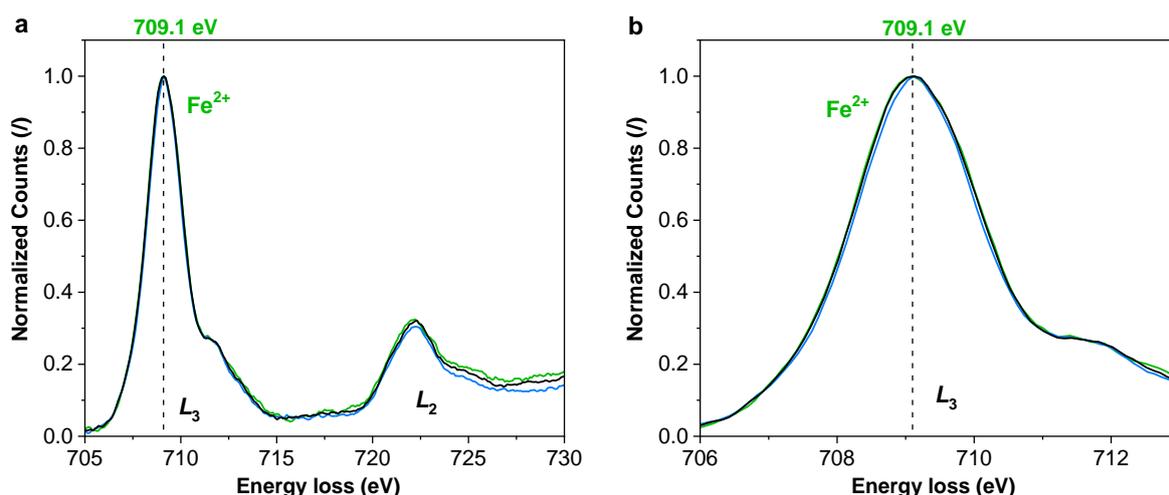

**Fig. S22** Comparison of the Fe-*L* edge EELS spectra obtained from the large area mapping (300 nm × 300 nm) of the particle B2.06a (blue), large area mapping (250 nm × 350 nm) of the particle B2.04a (black) shown together with the $Fe^{2+}$ reference spectrum (green). **a** whole Fe-*L* edge, **b** zoomed Fe-$L_3$ peak.

We find that there is almost no difference (deviation) between the spectrum obtained from the large area of particle B2.06a compared to the $Fe^{2+}$ reference spectrum. Thus we are in fact not able to determine the solubility limit ($\beta$) for the equilibrium Li-rich solid solution ($Li_\beta FePO_4$) corresponding to the 2-phase LFP/FP plateau region. In other words, with EELS mapping we are not able to distinguish between the LFP phase corresponding to the thermodynamic state at 2.8 V vs. Li and the LFP phase found in the large (but thin) platelet-like lithiated $Li_xFePO_4$ particles in the plateau region obtained during low-rate (-C/100) lithiation down to the global electrode composition corresponding to $Li_{0.5}FePO_4$.

In Fig. S23 an analysed scanned area shows a particle from the STEM-EELS experiment with applied current stimuli. The area includes a large platelet particle P2.05a with a thickness that varies from around 90 nm to 130 nm at the central part, while exhibiting two distinct LFP/FP phase boundaries. The second one is a small quasi-spherical particle P2.05b with a larger diameter of around 120 nm that is practically totally lithiated. In this case, we have selected a large area (200 nm × 250 nm) in the central part of particle P2.05a where there is clearly present a de-lithiated FP phase. In Fig. S24 the corresponding EELS spectrum of the selected area of particle P2.05a (black line) is shown in comparison with the $Fe^{3+}$ reference spectrum (red line).

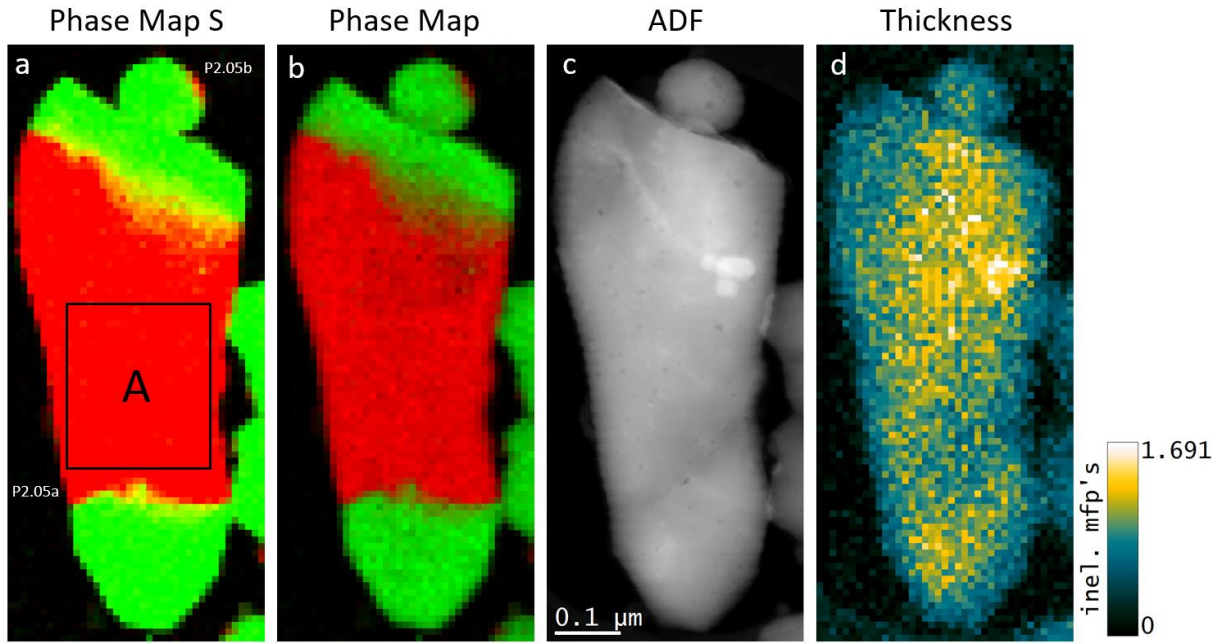

**Fig. S23** Analysed scanned area from the STEM-EELS experiment with applied current stimuli that includes particles P2.05a and P2.05b from **Fig. 1c** in the main text. **a** LFP/FP phase mapping with saturated colours, **b** phase map, **c** ADF image, and **d** corresponding relative thickness map. Selected large area A (200 nm × 250 nm) with non-lithiated FP phase is shown in **a**.

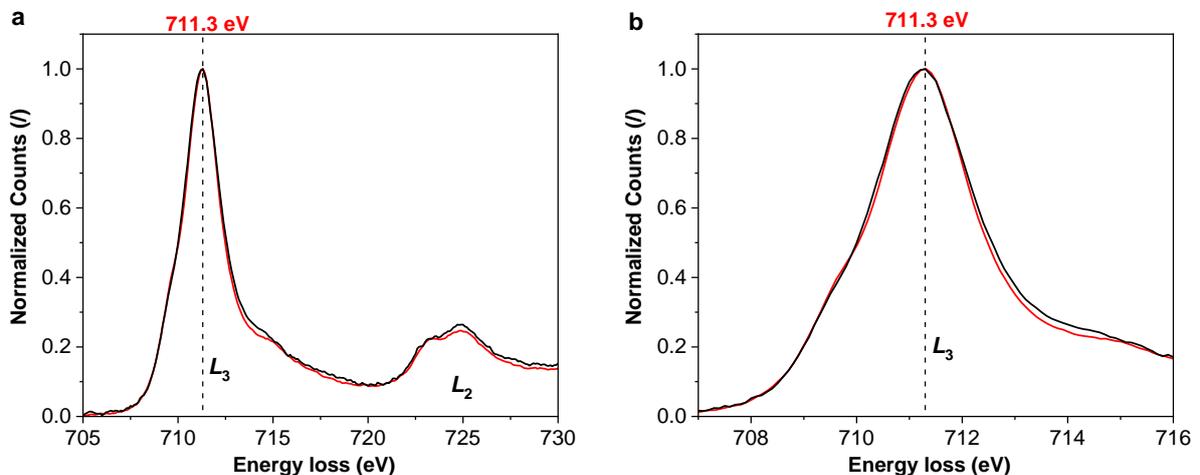

**Fig. S24** Comparison of the Fe-*L* edge EELS spectrum obtained from the large area mapping (200 nm × 250 nm) of the central part of particle P2.05a exhibiting the FP phase (black line) shown together with the $Fe^{3+}$ reference spectrum (red line). **a** whole Fe-*L* edge, **b** zoomed Fe-$L_3$ peak.

Only minor deviations are observed between the spectrum obtained from the selected large area in the central (FP) part of particle P2.5a compared to the $Fe^{3+}$ reference spectrum. Thus in the same way as in the case of LFP phase we are practically not able to determine the solubility limit ($\alpha$) for the equilibrium Li-poor solid solution ($Li_\alpha FePO_4$) corresponding to the 2-phase LFP/FP plateau region. In other words, with EELS mapping we are not able to distinguish between the FP phase that corresponds to the thermodynamic state at 3.8 V vs. Li and the FP phase found in the phase-separated $Li_xFePO_4$ particles in the plateau region obtained during discharge experiment with applied current stimuli and followed by a low-rate (-C/100) continuation of discharge whereby the corresponding LFP-Li cell voltage is within regular C/100 hysteresis.

Therefore, somewhat surprisingly, the results reveal that the LFP and FP phases appearing in the large particles from the -C/100 baseline discharge experiment (Fig. 1b) and in the experiment with applied current stimuli (Fig. 1c) are rather pure phases. A least as regards the Fe-*L* edge EELS signals, these phases are quite close to ones found in the two ($Fe^{2+}$ and $Fe^{3+}$) reference samples.

Particle P2.05a from the experiment with applied current stimuli (Fig. 1c) is a suitable particle to test whether the mathematical linear combination of the Fe-*L* edge signals of the $Fe^{2+}$ and $Fe^{3+}$ references can reproduce the Fe-*L* edge cumulative signal that we obtain from a large analysed area that includes both phases (LFP and FP) simultaneously. We have selected such an area that the maximum intensity of the two Fe-$L_3$ peaks was 1:1 with the signal having large overall counts. Accordingly, the selected large area (250 nm × 350 nm) is shown in Fig. S25 with the corresponding cumulative EELS spectrum shown in Fig. S26.

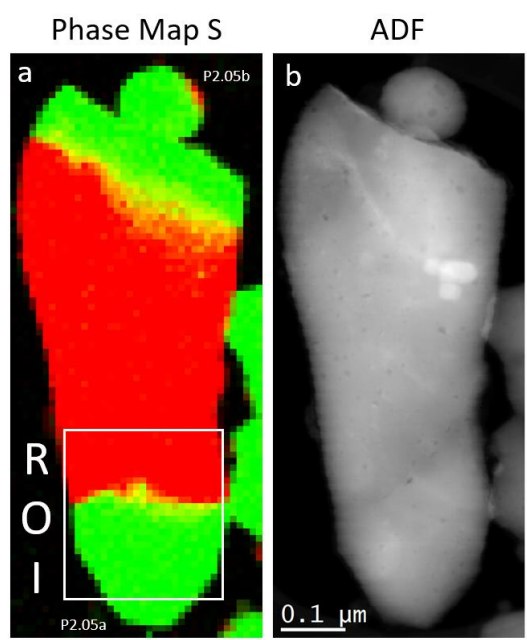

**Fig. S25** Analysed area from the STEM-EELS experiment with applied current stimuli that includes particles P2.05a and P2.05b from **Fig. 1c**. Selected (ROI) large area of the particle P2.05a that includes both LFP and FP phases procuring the maximal intensity of the two Fe-$L_3$ peaks is 1:1 (Fig. S26, black line).

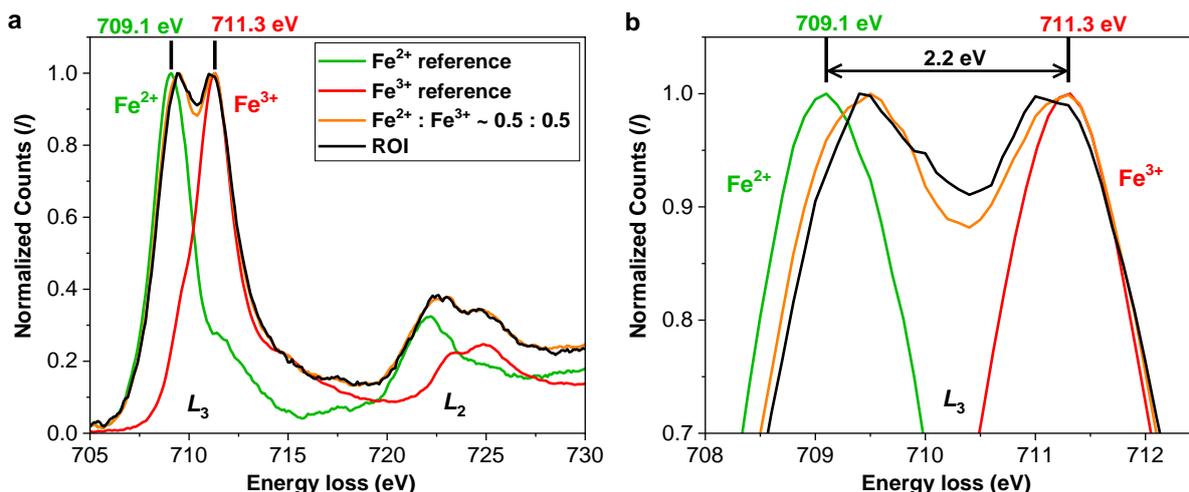

**Fig. S26** Comparison of the Fe-$L$ edge EELS spectrum obtained from the selected large area mapping of particle P2.05a from the STEM-EELS experiment with applied current stimuli (Fig. 1c) where the analysed area was chosen such (ROI in Fig. S25) that the maximal intensity of the two Fe-$L_3$ peaks was 1:1 (black line). The mathematical linear combination of the Fe-$L$ edge signals of the $Fe^{2+}$ (green line) and $Fe^{3+}$ (red line) references is shown for comparison (orange line). **a** whole Fe-$L$ edge, **b** zoomed Fe-$L_3$ peak.

A very good matching is shown in Fig. S26 between the mathematical linear combination of the signals of the references ($Fe^{2+}$ in green line and $Fe^{3+}$ in red line) where the $Fe^{2+}$ : $Fe^{3+}$ ratio of the Fe-$L_3$ peak-normalised reference signals is 1 : 1 (orange line) with the spectrum obtained from the selected LFP/FP area (Fig. S25) in particle P2.05a from the STEM-EELS experiment with applied current stimuli (Fig. 1c). The absolute ratio of the reference signals can be obtained by considering appropriate thickness-normalised reference spectra (see Fig. S31). The corresponding analysis shows that the signal with the 1 : 1 Fe-$L_3$ peak intensities corresponds to actual mixing of 56 % of $Fe^{2+}$ atoms and 44 % of $Fe^{3+}$ atoms. This is in agreement with the selected ROI (Fig. S25) where it can be seen that the part of the area with LFP phase was slightly larger than the area with the FP phase. Most importantly, the observed overall good matching between the ROI cumulative spectrum and the mathematical combination of the reference spectra supports the validity of the applied MLLS analysis method.

## 2.6 Sequence of lithiation in baseline -C/100 discharge experiment

In the baseline -C/100 discharge experiment stopped at DOD = 0.5 we found that the sequence of lithiation of particles seems to be strongly correlated with particle thickness. Fig. S27 shows a scanned region from the baseline -C/100 discharge with several particles (included in the area are particle B2.20a and particle B2.20b from Fig. 1 in the main text). We can clearly observe in the LFP/FP phase map as well as in the corresponding thickness map that the thinner particles (shown in area B2.20a) got completely lithiated, while at the same time the thicker particle B2.20b barely started with lithiation at the (thin) bottom left edge. We report a similar observation in all of the analysed areas from the baseline -C/100 discharge experiment. Thus, in experiments with very low rates of lithiation it appears that thin particles lithiate preferentially whereas thicker ones lithiate later during continuation of a (full) discharge. Further discussion of the details of the particles' aggregation and grain orientation directly related with the particles' thickness and consequential lithiation sequence is beyond the scope of the present paper.

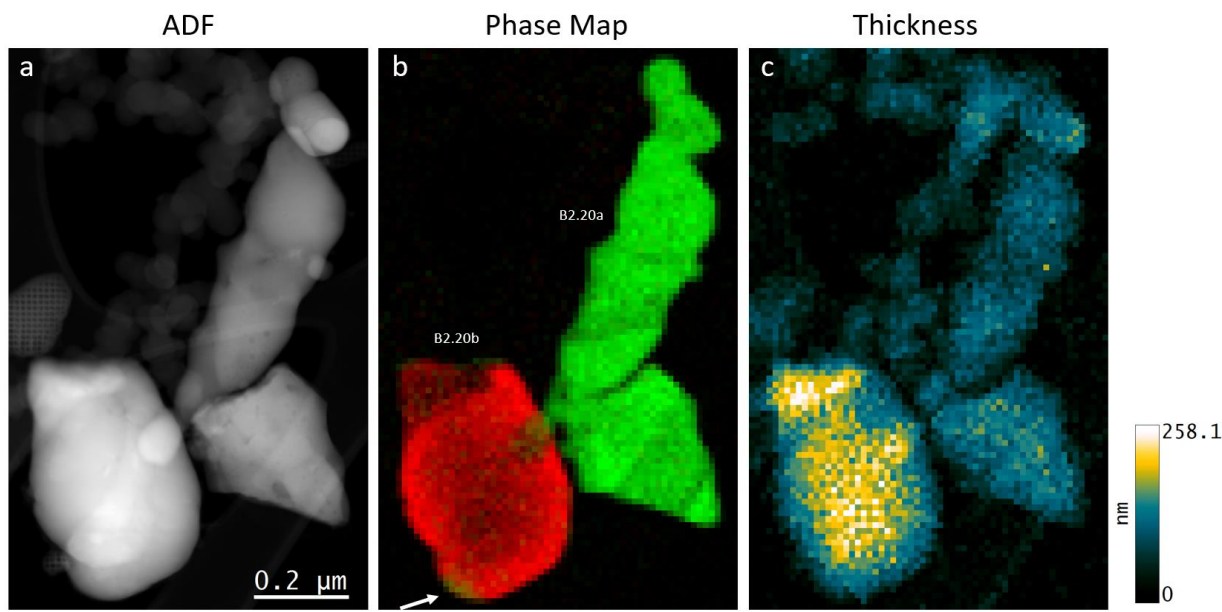

**Fig. S27** Scanned region from the baseline -C/100 discharge experiment with several particles included (in the areas with particle B2.20a and particle B2.20b from Fig. 1 in the main text). **a** ADF image of the mapped region (in addition to the FP and LFP particles of interest, some carbon particles can be observed as well), **b** LFP/FP phase map, and **c** the corresponding thickness map. Thinner particles (shown in area B2.20a) got completely lithiated, while the thicker particle B2.20b has barely started with lithiation, as can be seen at the (thin) bottom edge (marked by the arrow).

### 2.7 Line analysis of LFP/FP interface

We performed line analysis of phase composition in selected particles taken from the experiment with applied current stimuli that exhibited a well developed LFP/FP interface (Fig. 1c). Line profiles were taken from the already measured Fe-*L* edges EELS mappings. However, in this case, a deconvolution (MLLS)/quantification algorithm was employed using the measured standards for each phase (LFP and FP, as previously described) for cross-sections models. In addition, the effects of plural scattering were removed by deconvolving with the low loss signal. The line scan then illustrates the areal density of $Fe^{2+}$ and $Fe^{3+}$ atoms along the line, presented as normalised areal density (normalised by the sum of the intensities/densities for $Fe^{2+}$ and $Fe^{3+}$), and it is compared with the ADF intensity profile from the analyzed particle obtained along the same line to reveal the morphological features. Line profile Fe areal density analysis was preferably done on parts of the $Li_xFePO_4$ particles that were thin enough to avoid big effects of plural scattering. We also avoided regions where potential overlap of particles was observed. In Fig. S28 we show an example of particle P2.27 that started to undergo lithiation (approx. 1/6 of the particle being lithiated). Two line profile analyses were obtained: Line-1 is crossing the particle along the central part where it goes through a relatively wide LFP/FP interface transient region; Line-2, on the other hand, goes over a part of the particle where a relative narrow LFP/FP interface is found. The data collected from the line profile analysis corresponds to a 1x1 pixel step width (pixel size 5 nm).

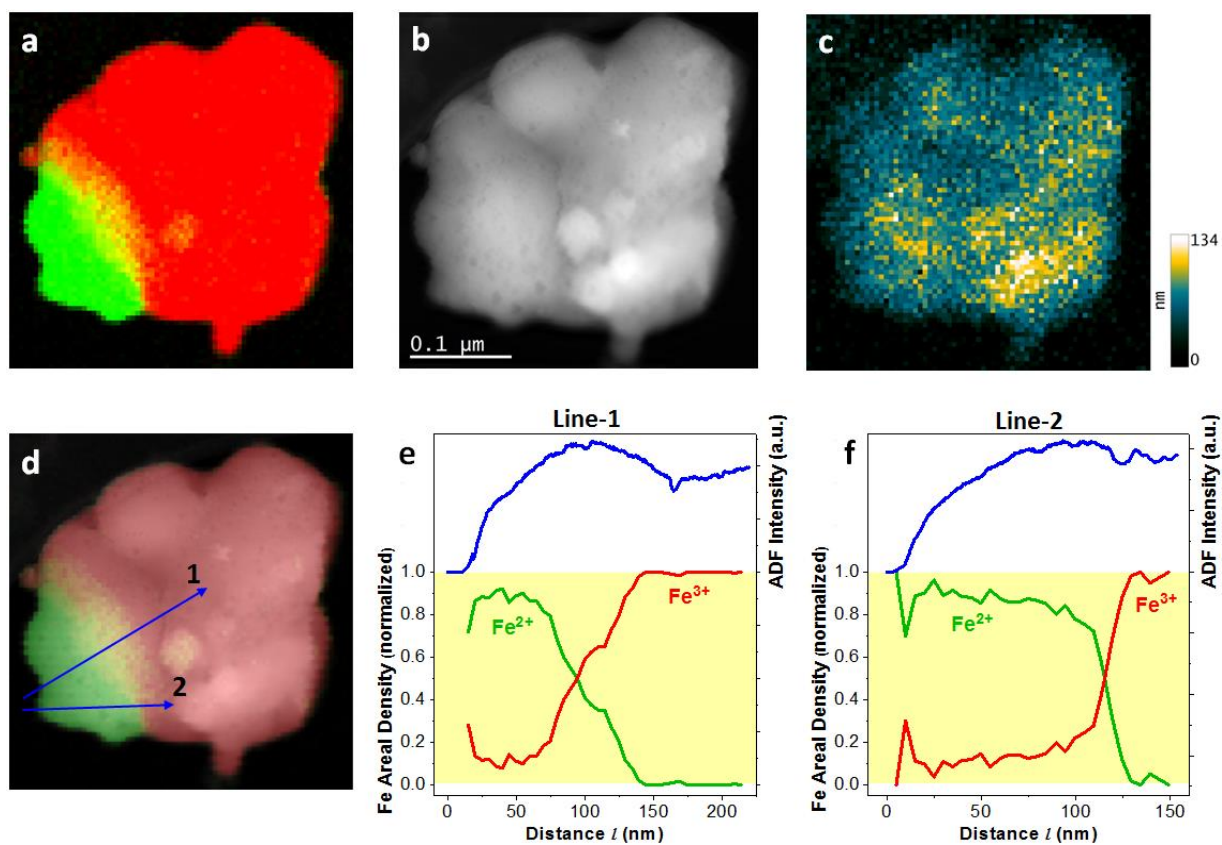

**Fig. S28** Particle P2.27 from Fig. 1c in the main text (experiment with applied current stimuli) that started to undergo lithiation and exhibiting defined LFP/FP phase boundary. **a** Phase map (color saturated), **b** ADF image of the particle, **c** the corresponding thickness map, **d** phase map overlaid with ADF image to display the positions of the line profile analysis along the LFP/FP interface marked with numbered arrows for: **e** Line-1 wide phase-boundary transition, and **f** Line-2 narrow phase-boundary transition.

Both line analyses of particle P2.27 start outside of the particle at the bottom left, and continue over the lithiated region, further crossing the LFP/FP phase boundary, and finish in the bulk of FP phase. Figs. S28 e-f show that the lithiated phase includes about 10 atomic % of $Fe^{3+}$, while on the contrary the FP phase exhibits practically pure $Fe^{3+}$. In other words, the LFP phase exhibits some small degree of "mixing" while the FP phase appears to be pure. This observation is suggesting that in case of particle P2.27 probably there was not enough time passed from the start of lithiation that the LFP phase would completely lithiate. Line-1 exhibits a "wide" LFP/FP phase boundary with the corresponding width of about 70 nm, and in case of Line-2 the LFP/FP phase boundary width appears to be about 20 nm. In both the cases at the phase boundary the areal density of $Fe^{3+}$ increases and that of the $Fe^{2+}$ decreases approximately linearly with distance. This observation is in accordance with data from literature where this trend is correlated with Li concentration across a phase boundary [13, 14].

Moreover, particle P2.27 is clearly demonstrating that the inter-particle LFP/FP phase boundary is not aligned with the internal grain boundary. This strongly indicates that the phenomenon of current stimuli induced intraparticle phase separation is not (at least not in the sense of a first order effect) dominantly governed by the intra-particle boundaries between grains within an aggregate. This observation justifies the suitability of the use of simulation model topology where aggregates are represented as non-assembled individual particles.

In Fig. S29 are shown acquired Fe-*L* edge EELS spectra obtained from line analysis of particle P2.27 shown in Fig. S28. The individual spectra were obtained at a step size of $\Delta l = 5$ nm whereby oversampling was used (adding up two neighbouring spectra) in order to reduce noise level. The spectra are plotted in the bottom panels of Fig. S29, while in the top panels the corresponding contour plots for the variation of the spectra with distance ($l$) along the selected line where the brightness corresponds to the intensity of signal are presented.

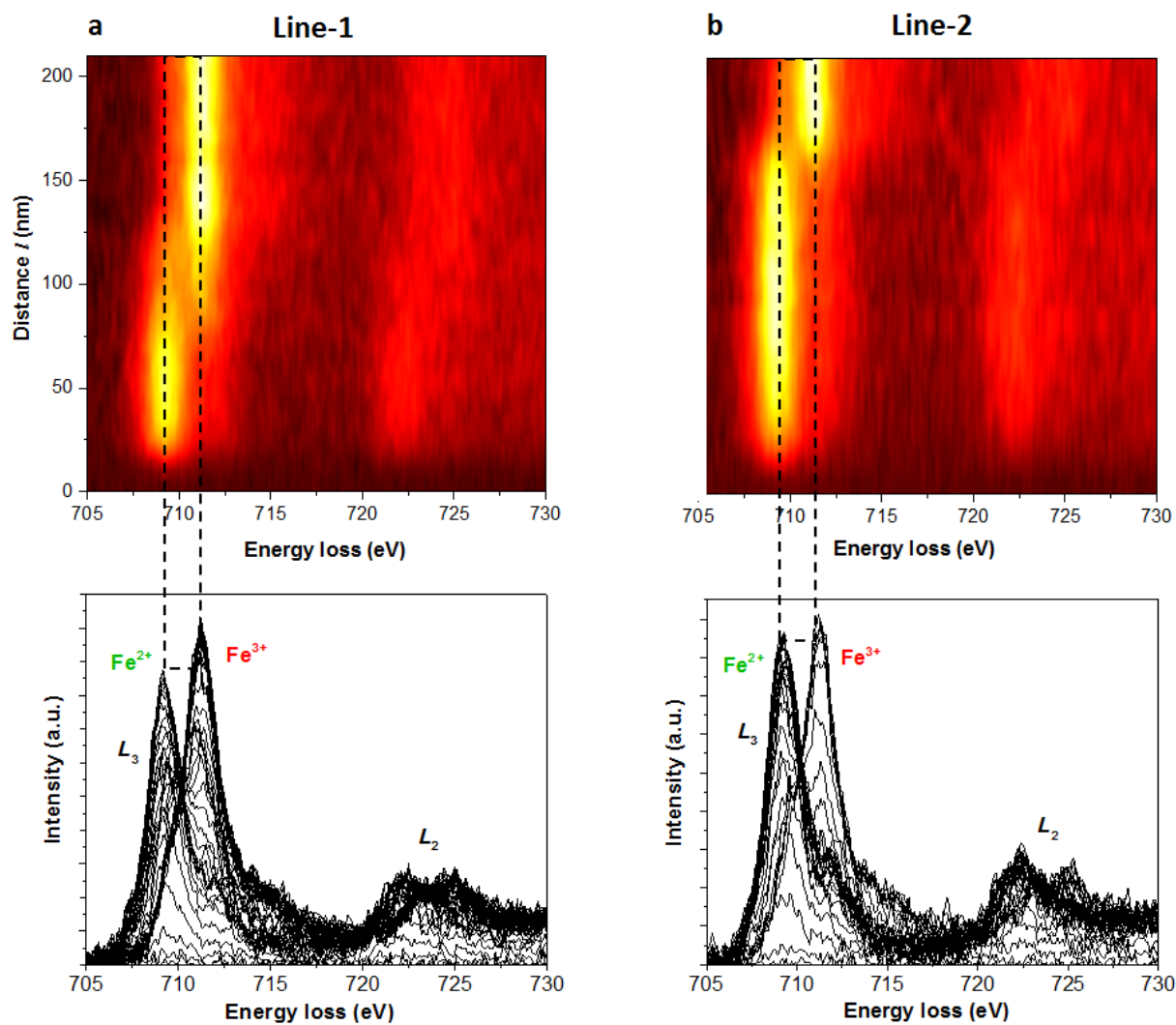

**Fig. S29** Acquired Fe-*L* edge EELS spectra obtained from line analysis along: **a** Line-1, **b** Line-2 of particle P2.27 shown in Fig. S28. Spectra is plotted in the bottom panels, while in the top panels the corresponding intensity contour plots for the variation of the spectra with distance ($l$) along the selected line are shown.

In Fig. S29 it is clearly seen that along the analysed lines at the LFP/FP phase boundary the intensity of the most prominent Fe-$L_3$ peak gradually decreases for $Fe^{2+}$ and simultaneously increases for $Fe^{3+}$ while we do not observe some "bridging" effect where the peak value of the Fe-$L_3$ energy loss would be observed somewhere in between those two corresponding to the LFP and FP phases.

In Fig. S30 a large platelet particle P2.18 is shown. It is partially lithiated whereby about 30% of the particle is lithiated at the upper thinner part that has an average thickness of around 100 nm. The lower half of the particle is thicker with a thickness in the range from about 110 nm to 140 nm. Line analysis was performed in the upper thinner half of the particle with the line starting on the left hand side outside the particle, going over the FP region, further crossing the FP/LFP phase boundary, and finally going over the LFP region to eventually reach the particle's edge on the right hand side. The direction of the line was chosen in such a way that the line perpendicularly crosses the FP/LFP phase boundary. The line width was 1 pixel (8 nm) as well as the step size.

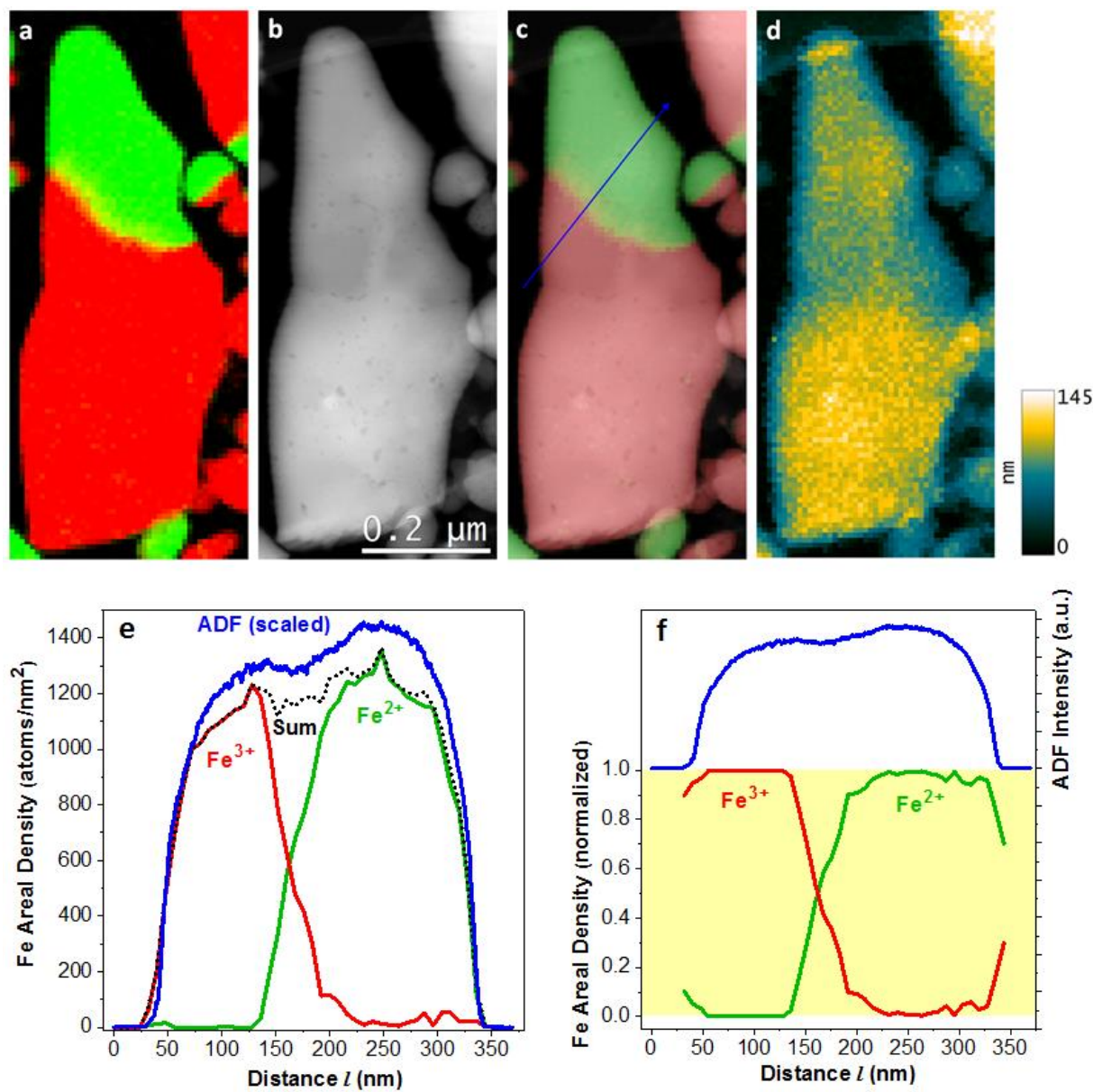

**Fig. S30** Particle P2.18 from Fig. 1c in the main text (experiment with applied current stimuli) that is in the lithiation process with about 30% of (areal fraction of) particle being lithiated and exhibiting clearly defined FP/LFP phase boundary. **a** Phase map (colour saturated), **b** ADF image of the particle, **c** phase map overlaid with ADF image with shown arrow where the line analysis was performed, and **d** the corresponding thickness map. **e** Obtained areal density for $Fe^{3+}$ and $Fe^{2+}$ and comparison of their sum with the (scaled) intensity obtained from the corresponding ADF image that in this case can be related with the particle morphology. **f** Normalised Fe areal density.

In Fig. S30f one can clearly see that particle P2.18 exhibits bulk FP phase with practically pure $Fe^{3+}$ composition. When crossing the FP/LFP phase boundary that is about 70 nm wide, the corresponding areal density of $Fe^{2+}$ increases and that of the $Fe^{3+}$ decreases approximately linearly with distance (position along the line). The lithiated LFP phase includes a small fraction of $Fe^{3+}$ when approaching the particle's edge on the right hand side with observed scatter in composition suggesting in average about 3% of $Fe^{3+}$. Thus, overall the large platelet particle P2.18 in the bulk of the two phases exhibits a very low degree of $Fe^{2+}/Fe^{3+}$ "mixing", which is in agreement with the above-shown good match between the large-area spectra of large platelet particles and the two reference spectra. It is worth mentioning that in these particular cases we have tried to perform the semi-quantitative line analysis on relatively thinner parts to avoid misinterpretations. Similar line analysis done on other suitable particles gave comparable results as shown in the two cases of particle P2.27 and particle P2.18. In summary, the detailed analysis of the measured STEM-EELS data provides a clear evidence that the application of current stimuli results in formation of a "large amount" of intra-particle LFP/FP phase boundaries. Since MLLS mappings suggest that in the most observed cases there is a single LFP/FP phase boundary in one particle the process results in a large increase of active particle population – the latter being true no matter whether individual particle is seemingly composed of a single primary particle or being an aggregate.

The set of ADF-intensity-normalised spectra obtained at the interface of the FP/LFP phase boundary of particle P2.18 (Fig. S30) is shown in Fig. S31 together with the corresponding mathematical combination of the signals for the $Fe^{2+}$ and $Fe^{3+}$ references. The ADF-normalisation of the spectra can be considered as a thickness-normalisation. It can be seen that the set of spectra exhibits the isosbestic points similarly as observed and strongly pointed out by Laffont et al. [9] in their pioneering work where they elegantly used HREELS for studying of the LFP/FP phase boundary. According to the authors the appearance of the isosbestic points on the overlaid EELS spectra should be considered as a direct experimental evidence that the interface at the LFP/FP phase boundaries can be treated as juxtaposition of the two end members: LFP ($LiFePO_4$) and FP ($FePO_4$).

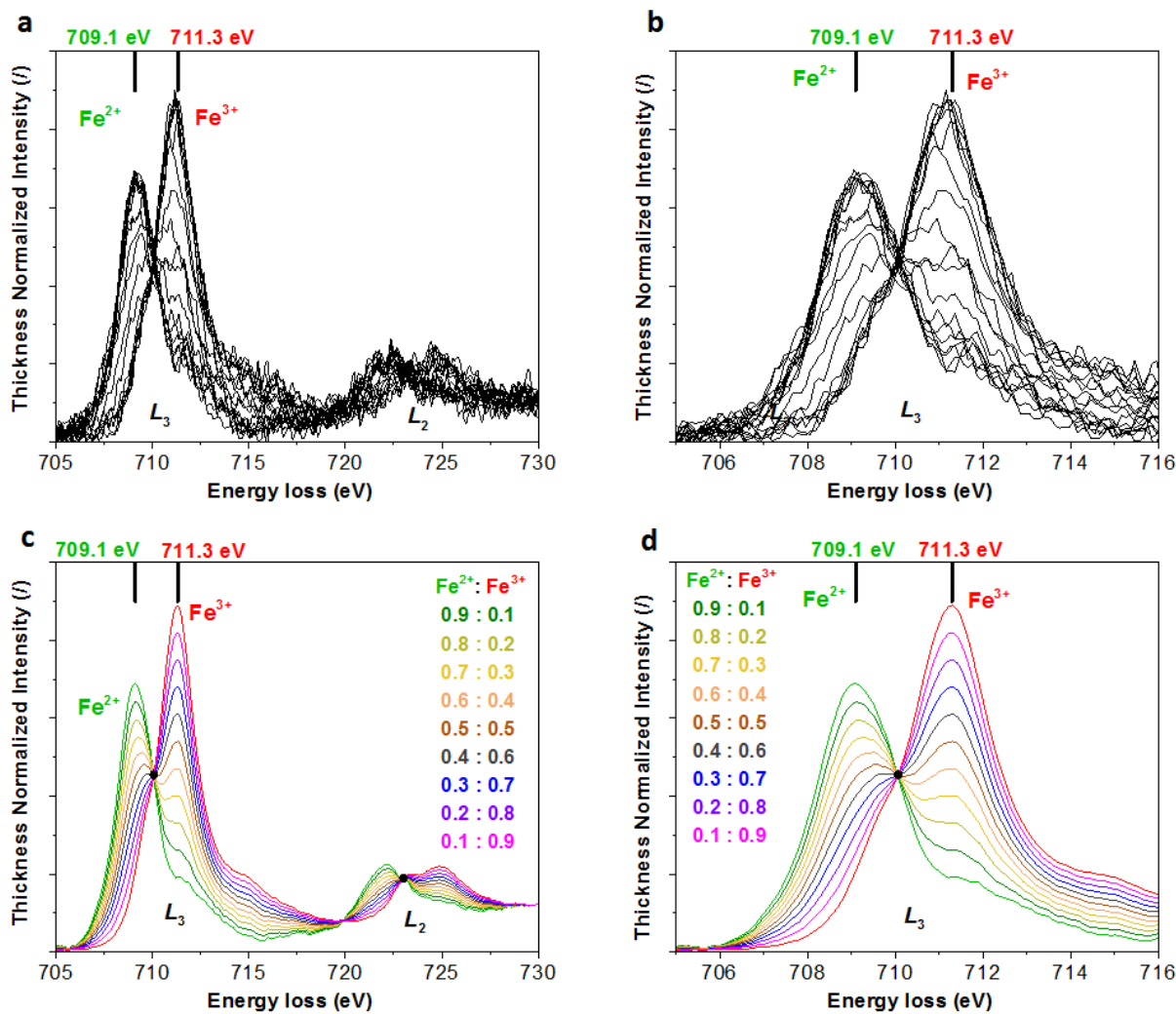

**Fig. S31 a-b** Set of thickness-normalised EEL spectra from the interfacial region at the FP/LFP phase boundary of particle P2.18 (Fig. S30) along the selected distance (100 nm $< l <$ 240 nm) of the line analysis (Fig. S30c). **c-d** The corresponding mathematical linear combination of the signals for the $Fe^{2+}$ and $Fe^{3+}$ references. The set of spectra shows distinct isosbestic points.

It is necessary to be noted that the presented results of the EELS analysis (MLLS mappings, evolution of spectra along analyzed line, as well as the Fe areal density line profiles) alone cannot provide information regarding the actual phase separation and/or solid solution formation in the observed $Fe^{2+}/Fe^{3+}$ "mixed" regions of the analyzed $Li_xFePO_4$ particles. For example, in the early work of Laffont et al. [9] the authors have considered that the presence of isosbestic points in the spatially-resolved EELS line analysis of LFP/FP interface alone provides direct evidence that the observed interface corresponds to the superposition of the two end members (e.g. $LiFePO_4$ and $FePO_4$) rather than a solid solution. More recently Ikuhara et al. [13] and Kobayashi et al. [14] have suggested that the gradual EELS $Fe^{2+}/Fe^{3+}$ compositional transient of LFP/FP interface region of a thin sample (in their case from ~50 to about 80 nm) can be interpreted as a Li concentration profile.

Most importantly, in terms of the present paper the actual determination of the nature of LFP/FP phase boundaries is not of crucial importance. Whether the latter is due to a gradual (continuous) change in Fe oxidation state corresponding to a Li concentration gradient in solid-solution compositions or it is actually a sharp boundary and the "mixed" regions are corresponding to the pixels embracing tilted boundaries and consequently simultaneously showing both $Fe^{2+}$ and $Fe^{3+}$ due to overlay of the LiFePO$_4$ and FePO$_4$ phases, this information is not needed to support the main findings of the present work. Regardless of the exact scenario (might be even some kind of "mixed" regime) it is clear that the application of current stimuli induces formation of a large number of additional intra-particle LFP/FP phase boundaries that push the voltage during the subsequent low-rate lithiation within the intrinsic voltage hysteresis loop of a LFP material.

## 3. Differentiation between the newly observed electrode state and previously reported "Memory effect" and ''Group-by-group'' multi-particle Li intercalation

The phenomenon presented in Fig. 1a in the main text could superficially be mixed up with the so-called "memory effect" presented by Novak [15] and later on mechanistically explained by Kondo et al. [16] and Zelič et al. [17]. Despite the fact that both phenomena are influenced by active particle population, the two effects do not feature the same causality of phenomena nor the same directionality of its impact. As already explained in the recent literature [16, 17], the memory effect originates from kinetically inhomogeneous reactions of particles featuring a non-monotone potential profile. This results in a so-called shim shaped distribution of particle lithiation as a function of particle size [16, 17] after not being fully discharged prior to the memory releasing charging cycle. On the other hand, as indicated above, the entrance into voltage hysteresis is a fundamentally different process that is directly associated with an intraparticle phase separated electrode state, where the majority of the particles constituting the electrode are in the intraparticle phase separated state. In addition, the intraparticle phase separated electrode state is associated with entrance into the voltage hysteresis, while the memory effect increases the voltage hysteresis [15, 14].

Likewise, the presented intraparticle phase separated electrode state and associated entrance into the voltage hysteresis is also fundamentally different to the ''group-by-group'' multi-particle Li intercalation in a battery system that undergoes Li phase separation yielding electrochemical oscillations [18]. Namely, the ''group-by-group'' multi-particle Li intercalation is a consequence of collective phase-separation dynamics, where one particle that reaches the critical Li concentration first initiates the phase separation, accompanied by a sudden rise in voltage [18]. Consequently, the rapid Li insertion into this particle suppresses the lithiation of other particles, even leading to release of Li ions from those ''inactive'' particles, until the phase transition is completed in the ''active'' particle [18]. This process is repeated until all particles become fully lithiated and the sudden rises in voltage are caused by discrete phase-separation events yielding observed voltage oscillations [18]. It is thus obvious that the newly reported entrance into the voltage hysteresis being associated with the intraparticle phase separated state is fundamentally different compared to the ''group-by-group'' multi-particle Li intercalation, as it is associated with a large share of active particle population being simultaneously in the so called phase separation regime, while the ''group-by-group'' multi-particle Li intercalation elucidates collective phase-separation dynamics, with just a very few particles in the so called phase separation regime. This also elucidates why the newly reported entrance in the voltage hysteresis features a larger monotonous deviation of the cell voltage, which has also an impact of cell power and heat generation.

# 4. Path dependency of known electrode states

This section provides additional analyses of different electrode states to further support the analysis of Fig. 2 in the main text.

### 4.1 Small current densities – an interparticle phase separated state

It is well known that discharging with small currents is characterised by small overpotentials (Fig. S32a) and a low specific free energy, as indicated in Fig. S32b. In this case of small overpotentials that barely exceed the initial energy barrier imposed by the convex part of the free energy density of a particle (Fig. S32c and S32d), the particles undergo the intraparticle phase separation. However, in contrast to the interparticle phase separated electrode state, which is a precondition for entering into the voltage hysteresis, only a small fraction of particles is active simultaneously, as reported in Fig. 1b and Fig. 3b in the main text as well as in Supplementary section 2. This regime is, therefore, characterised by a particle-by-particle mode of lithiation forming the so-called mosaic pattern [19].

This phenomenon is driven by two facts. Firstly, the lithiation level of the electrode is equal to the sum of Li mole fractions of the many-particle, i.e. eq. (1) in the main text. According to reference [19], this constraint imposes that in a many-particle system it is possible to control only the total amount of lithium in all particles, whereas the amount of Li inside individual particles is not uniquely determined by the total amount of lithium in all particles. Another driving force for this behaviour arises from free energy minimisation of the many-particle system, i.e. eq. (2) in the main text, which implies that the amount of lithium inside individual particle and the exchange of lithium among active particles is driven by minimisation of the energy of the whole ensemble.

The verification of this mechanism has already been carried out by several particle resolved X-ray analyses, e.g. [20], whereas it has also been thoroughly supported by innovative experimental results presented in Fig. 1b of the main text and Supplementary section 2. Since at small currents only a small share of particle population is active (Fig. 3b in the main text), the electrode's chemical potential remains at the higher level of the first spinodal during lithiation (Fig. S32c). This results in the well know voltage hysteresis, which is associated with interparticle phase separated state when subjecting a battery to a charge and discharge cycle (Fig. S32a). This explanation of the small current phenomena in phase separating materials follows directly the reasoning published in the original theory proposed in reference [19].

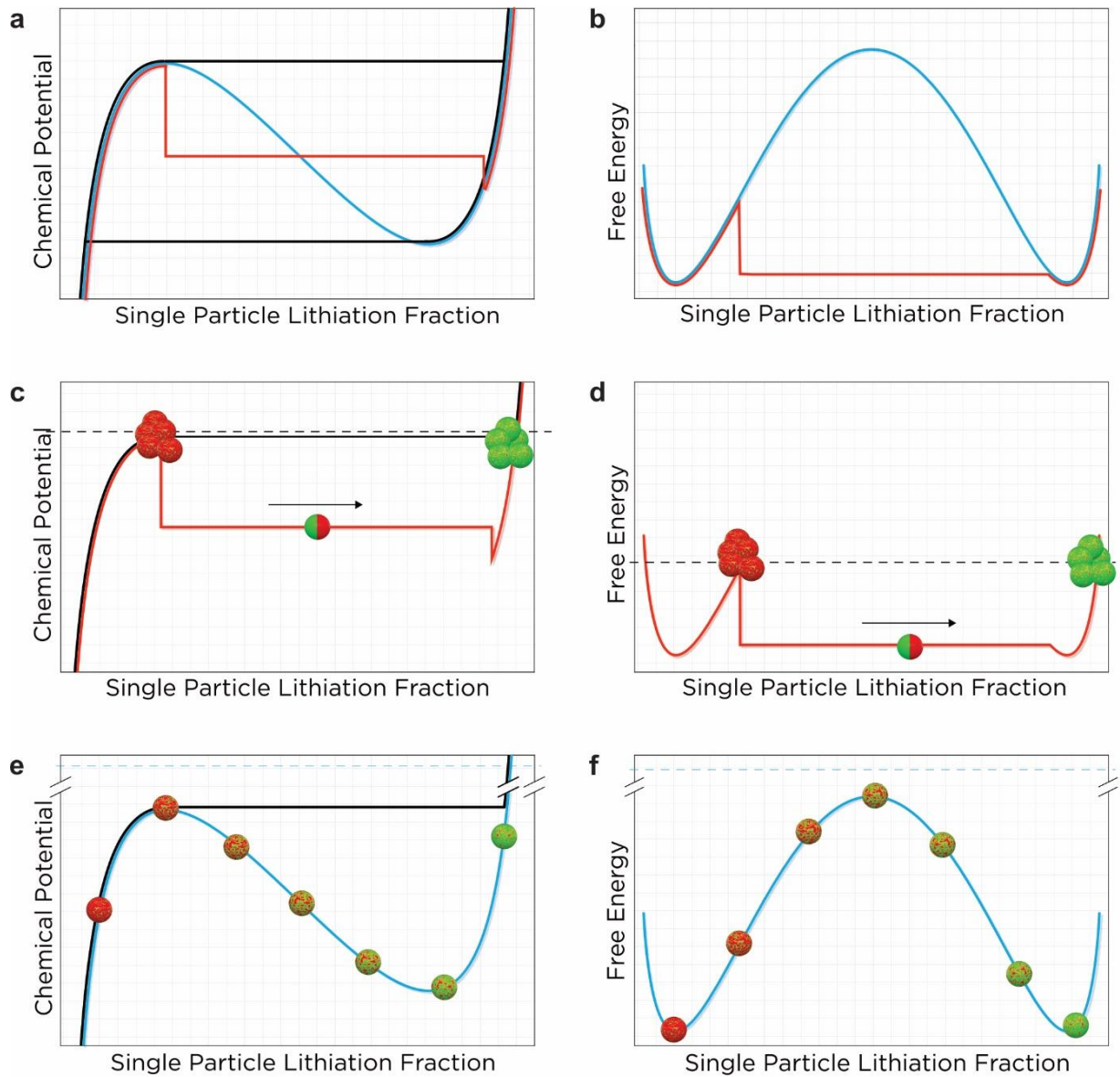

**Fig. S32** Schematic representation assuming identical particles and thus potential and free energy characteristics of particles, depicting: **a** potentials associated with high overpotential transient and interparticle as well as intraparticle phase separated state; and **b** corresponding representation of the high overpotential transient and intraparticle phase separated state in the specific free energy diagram, where free energy of the interparticle phase separated state is not presented, since it is a collective electrode parameter and not a particle related parameter; **c** a constant small current discharge, without a previous large current stimulus, corresponding to interparticle phase separated state and being characterised by particle-by-particle lithiation process and **d** corresponding representation in the free energy diagrams showing insufficiently high energy for particles to transverse the entire lithiation range in the high overpotential transient regime; **e** a large current discharge where large share of active particle population undergoes a high overpotential transient

transition and **f** corresponding representation in the free energy diagram showing sufficiently high energy for particles to transverse the entire lithiation range in the high overpotential transient regime.

### 4.2 High current densities – a high overpotential transient state

Another (limiting) (dis)charging regime that has also been extensively studied in the literature is observed at large currents. In this regime, active particles in the electrode are far from the thermodynamic equilibrium characterised by high over potentials [21, 22, 23]. There is a consensus that in this high current density regime, active particle population in the particle ensemble is significantly higher in comparison to the small current regime, whereas references report different specific values [23, 24, 25, 26, 27]. However, in the high current density regime and for lithiation levels within the miscibility gap, different possible (de)lithiation modes are reported in the literature, e.g. [21, 28, 29]. Due to the complexity of the intraparticle and interparticle phenomena of $LiFePO_4$, the detailed temporal and spatial dependence of this state is not yet unambiguously mechanistically understood and described with governing equations. Moreover, experimental studies, using different experimental techniques, further indicate that lithium distributions depend on active particle size [27, 4], its shape [4] and its ion and electron wiring [27, 30].

In the high current density regime active particles are far from the thermodynamic equilibrium motivating the naming "nonequilibrium solid solution" or "metastable solid solution", which was proposed in [21, 31]. This is related to the fact that lithiation of particles in the high overpotential limit is so rapid that characteristic time for particle lithiation is in the range of intraparticle transport phenomena [32, 33] and thus characteristic time needed for particle to phase separate [21, 34]. Reasoning of suppression of phase separation is proposed in [21]. Zhang et al. [24] and Liu et al. [25] present experimental evidence of the existence of nonequilibrium solid solution phase inside active particles in the high current regime. The measured relaxation times of this non-equilibrium state and the fraction of active particles in this non-equilibrium state reported in references [24] and [25] do not feature the same values, which can be attributed to the differences in the material and the electrode topology including ion and electron wiring. Chueh et al. [26] also provided comprehensive experimental studies on the interplay between nonequilibrium solid solution and phase separated lithium distribution in the particle ensemble. They have published a series of papers analysing the impact of different sizes and shapes of active particle [23, 27, 34, 35, 36]. In comparison to references [24, 25], they show higher fractions of active particles in the nonequilibrium solid solution mode.

Recent findings using operando TXM-XANEX method on the so called microrod shape active particles [28] show that inside active particles phase boundary can also be established in the direction perpendicular to the fast diffusion channels during rapid (de)lithiation [28]. Unlike in the case of slow lithiation (Supplementary Section 4.1), phase boundaries are in this fast lithiation case parallel to the fast diffusion direction [28]. The significance of this experimental evidence was also reinforced by the simulation results that show the same results as experiment [28]. Similar result, clearly showing existence of pure lithiated and delithiated phases inside single particle during rapid (de)lithiation, was also published by Yu et al. [37].

Since there is no ultimate consensus about the interplay between the nonequilibrium solid solution and the two phase particle nature in the large current limit, notation *high overpotential transient state* will be used in this article to denote electrode state at high overpotentials and lithiation level within the miscibility gap. Regardless of the previously mentioned missing consensus, aforementioned studies advocate two common facts, which are crucial for analyses in this article: i) the active particle population in the large current limit is significantly higher in comparison to the small current limit and

ii) active particles in the large current limit are far from the thermodynamic equilibrium during (de)lithiation (in the high overpotential transient state) and subsequently relax to more stable two-phase state in case of current termination (elaborated in more detail in Supplementary Section 5).

Despite these common facts, exact functional dependency of the total free energy of the representative LFP active particle or corresponding chemical potential is, as indicated previously, not yet elaborated. This challenge is further aggravated by the fact that, as also analysed above, exact values of the total free energy of the particle depend on the (de)lithiation rates, on the active particle size [27, 4], its shape [4] and its ion and electron wiring [27, 30] while, in addition, internal heterogeneities of the particle and defects also have an impact on the total free energy. To alleviate these challenges and to support theoretical analyses presented in this paper, regular solution theory will be applied as the theoretical basis. The core principle of the regular solution theory is an assumption of double well shaped free energy density of the system [31]. In this paper the most frequently used formulation of the free energy of LFP particle [19, 21, 22, 23, 4, 31, 38, 39, 40] is used. Such a presentation serves well to explain many basic phenomena in the LFP material [39] and it serves well as a plausible qualitative theory, while it features certain deficiencies in the quantitative prediction capability. However, due to clarity of presentation, this large current (and thus overpotential) operation in the lithiation range of the particles that corresponds to the miscibility gap will be graphically represented by regular solution inspired functional dependency of the total free energy as a function of particle lithiation fraction (presented in Fig. 2 of the main text and Fig. S32 and denoted as high overpotential transient state), which satisfies items i) and ii) from the previous paragraph. Note that minima in the free energy in Fig. 2 for LFP and FP phase are at the same value. This is due to the fact that this curve represents the free energy of mixing that is described by regular solution theory. As a consequence, the chemical potential curves in Fig. 2 are centred around zero, meaning that offset due to the chemical potential at standard condition is omitted. This is standard approach in regular solution theory since differences in chemical potential drive fluxes and the offset due to value of chemical potential at standard conditions does not involve in to the transport processes.

### 4.3 Intermediate current densities - interparticle phase separated state

Intermediate current densities have been analysed less frequently compared to both limiting cases. By intermediate current densities we refer to currents of around e.g. 0.5 C, which are frequently encountered in vehicles and consumer electronic applications [20]. This case is similar to the small current case in the sense that the portion of particles that simultaneously undergo the phase separation is still rather small - or moderate at slightly higher current values [20]. In this regime the overpotential is not high enough for the particles to transverse the entire lithiation range in the high overpotential transient regime as reported in [20, 21]. Briefly, this regime is also characterised by a relatively low active particle population and thus the electrode chemical potential remains above the higher level of the first spinodal point.

## 5. Intraparticle vs. interparticle phase separation

The concepts of intraparticle and interparticle phase separation have been frequently discussed in the literature [38, 39, 41]. In addition to the predominantly theoretical investigations, these states have recently also been clearly demonstrated experimentally, using X-ray diffraction and microscopy experiments [34] and they are further credibly supported by results of STEM-EELS experiments presented in this work. During low discharge rates the electrode will tend to relax directly to the lowest energy state, that is, it will undergo the interparticle separation (also referred to as mosaic pattern).

Conversely, the relaxation to intraparticle phase separated state is energetically less favourable as it is additionally associated with strain and interfacial energy due to internal phase boundary [4, 42]. Still, as also shown in this paper, both states are much closer to the equilibrium state than the high overpotential transient state.

For the purpose of present paper, it is highly important to have a rough estimation of the typical times needed to relax from the high overpotential transient state into intraparticle and interparticle separated states. In the aforementioned experimental paper [34] it was shown that a large portion of particles transformed from high overpotential transient state to intraparticle separated state in about 2 h. However, note that that experiment was carried out in absence of external current excitation (at the so-called open circuit voltage conditions). In the present case, where the electrodes are discharged/charged at certain finite rate and the particle size is much smaller, this transition might be even (much) faster. Based on the voltage response after the first relaxation time, which roughly corresponds to the time needed from the end of large-current stimulus to the peak of resulting voltage relaxation, the characteristic relaxation time, $\Delta t_1$, can be determined. The results in Fig. S33 indicate that $\Delta t_1$ is on the order from several minutes to a couple of hours, depending on the magnitude of charge-discharge current value following the high-current stimulus. Interestingly, this relaxation time is found to be roughly inversely proportional to the charge/discharge current, as can be inferred from Fig. S33.

As reported in [34], after the relaxation from the high overpotential transient state typically found at higher rates, individual particles initially separate into Li-rich and Li-poor domains confirming that lithium is mostly confined to its original particle in the first few hours. As there is not enough time for interparticle transport, the particles constituting the electrode can thus only relax by creating a new phase inside themselves.

In this paper, however, we demonstrate for the first time the actual effect of intraparticle phase separation on the electrochemical output. More specifically, Fig. 1 in the main paper shows a direct correlation between intraparticle phase separation and entrance into the hysteretic loop normally observed during charge and discharge of phase separating materials. Based on this finding, one might intuitively anticipate that the delayed voltage increase from the end of large-current stimulus to the peak of resulting voltage is thus only the consequence of intraparticle phase separation. However, Fig. 3d in the main text and figures in Supplementary section 6.2 provide another insightful explanation of this delayed voltage rise towards the equilibrium potential after termination of the current stimulus. When closely inspecting the lithiation levels of particles, which are lithiated far beyond the right spinodal during the current stimulus, i.e. the particles with the highest DOD values, it is discernible that these lithiation levels correspond to potentials that are far below the lower hysteresis potential. This lower hysteresis potential is namely represented by delithiated particles at the minimum of the spinodal, as explained along Fig. 2a of the main text and Fig. S32a. This very low potential of the most lithiated particles is a consequence of the very high overpotential during the current stimulus. After the end of the current stimulus, the overpotential decreases significantly at C/100, as indicated in Fig. 2a of the main text and in Fig. S32a. Consequently, these most lithiated particles tend to delithiate via the interparticle exchange of Li, which - in accordance with previous explanation – is a relatively slow process [34]. This phenomenon, which takes place during a delayed voltage increase in the time interval $\Delta t_1$, is insightfully demonstrated by decrease of DOD values of these most lithiated particles as discernible in Fig. 3d in the main text and figures in Supplementary section 6.2. This interaction takes place in the time interval $\Delta t_1$ and represents a complementary process to the intraparticle phase separation of the other particles.

On the other hand, the transition into the interparticle phase separated state was found to be much slower; clearly transformed particles with no internal boundary were only observed after 500 hours at OCV conditions. This second relaxation time ($\Delta t_2$) is much longer and lasts from the above mentioned peak voltage to more or less the end of ongoing charge/discharge or even longer (see the roughly indicated time lapse $\Delta t_2$ in Fig. S33. Such clearly separated two relaxation times are also observed in this study, see Fig. S33.

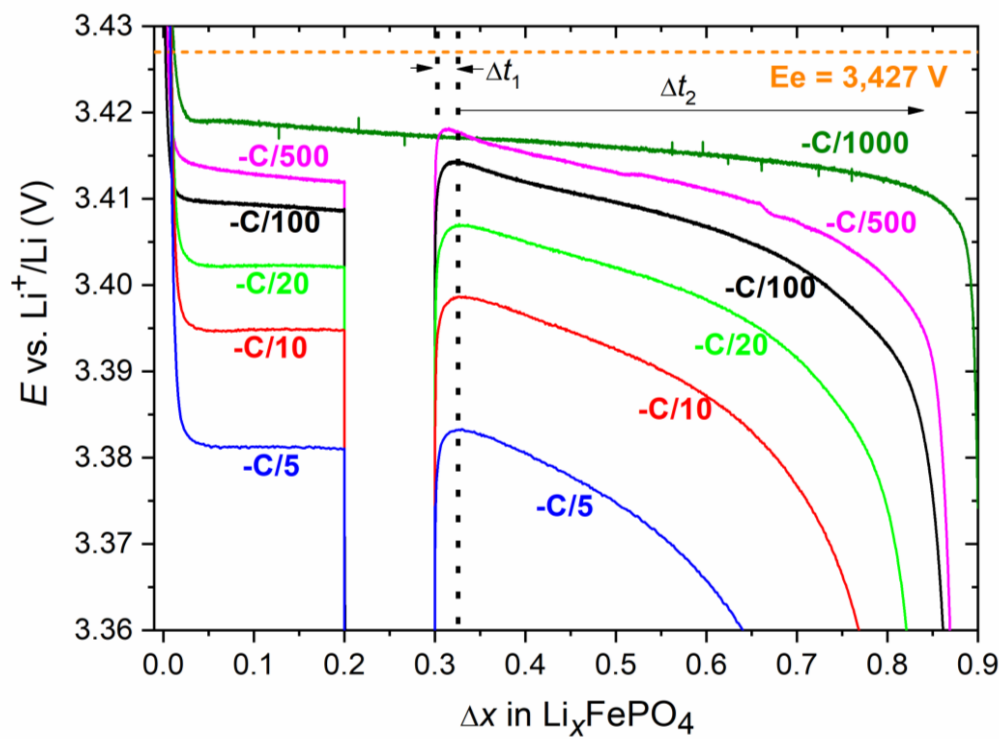

**Fig. S33** Effect of excitation of LFP with a current stimulus (the magnitude was always -1C) at different discharge rates (from -C/500 to -C/5) measured on cell #15. Note that the peak that follows the excitation occurs almost at the same $\Delta x$ value, regardless of the magnitude of discharge rate. This means that the corresponding relaxation time (which is proportional to $\Delta t_1$) is roughly inversely proportional to the discharge rate. The second relaxation time ($\Delta t_2$) is much longer, i.e. on the same order as the total discharge time.

The experiments in Fig. S34 further confirm the previous conclusions, conclusion from the main text (in particular Fig. 1) and also the consistency with the findings published in ref [34]. Using cell #12, two discharge protocols were investigated after applying a -5C current stimulus until half of cell capacity: i) a variant with baseline completion of discharge at -C/100 (red line) and ii) a variant where a long (240 h) open-circuit relaxation period was inserted between the -5C current stimulus and the baseline discharge at -C/100. It is discernible from Fig. S34 that the first variant is characterised by a higher electrode potential after the -5C current stimulus. The lower electrode potential during the -C/100 discharge in the case of variant ii) confirms the previous conclusions and conclusion from the main text, while it is found consistent also with the findings published in ref [34]. Firstly, a sharp peak in the voltage before the rest period indicates the entrance into the intraparticle phase separated state characterised by a lower chemical potential and thus a higher electrode potential. Secondly, the decrease of the electrode potential during the rest period together with the lower electrode potential during the -C/100 discharge compared to the variant i), is consistent with the assumption that the transition from intraparticle to interparticle state is a slow - but not extremely slow - process. This is reasoned by the fact that voltage of variant ii) remains between the voltage of variant i) and voltage of the interparticle phase separated state until the spinodal.

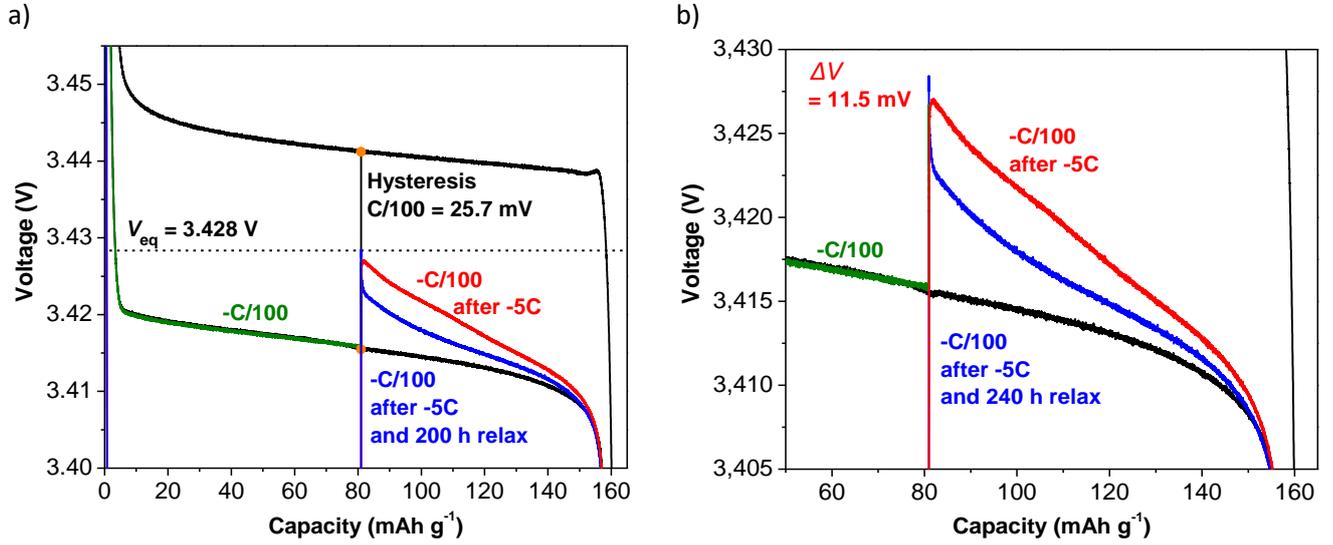

**Fig. S34** Two different variants of tests with long current stimuli ($\Delta X = 0.5$ in $Li_XFePO_4$): i) the -5C current stimulus is immediately followed by a baseline completion of discharge at -C/100 (red line), and ii) a variant where long (240 h) open-circuit relaxation period was inserted between the -5C current stimulus and baseline discharge at -C/100 (blue line). The measurements were done on LFP-Li cell #12 with regular mass loading (4.5 mg/cm²).

# 6. Numerical simulations

## 6.1 Simulation model

Simulations were conducted using an extended continuum level porous electrode model inspired by porous electrode theory [43] [44], which is upgraded to more consistently virtually represent the electrode topology and material characteristics. The previously published governing equations (ref. [12]) will be briefly summarised to ensure completeness of the article, while preserving the brevity of model descriptions and to indicate specifics of the model applied in this analysis. The concentration of ionic species in a binary electrolyte solution is governed by the following material balance equation

$$\frac{\partial(\epsilon_e c_e)}{\partial t} = \nabla \cdot \left(D_e^{\text{eff}} \nabla c_e\right) + a^{\text{e}}(1 - t_+)j^{\text{e}}, \quad (S1)$$

where $\epsilon_e$ represents local electrode porosity, $D_e^{\text{eff}} \nabla c_e$ represents the flux of ionic species with the effective diffusion constant evaluated from the Bruggeman relation $D_e^{\text{eff}} = D_e \epsilon_e^{\text{brugg}}$ [45, 46], whereas $t_+$ and $a^{\text{e}}$ represent the transference number and specific active surface exposed to the electrolyte, respectively.

The porous electrode theory is based on the assumption of electro-neutrality [47, 48] modelled as

$$\nabla \cdot i_e + \nabla \cdot i_s = 0, \quad (S2)$$

where $i_e$ and $i_s$ represent liquid-phase current density and solid phase current density, respectively.

Liquid-phase potential $\Phi_e$ is governed by the charge leaving or entering the liquid phase, and can be written as

$$\nabla \cdot i_e = \nabla \cdot \left( \kappa_e^{\text{eff}} \nabla \Phi_e - 2 \frac{\kappa_e^{\text{eff}} RT}{F} (1 - t_+) \left(1 + \frac{\partial \ln f_\pm}{\partial \ln c_e}\right) \nabla \ln c_e \right) = a^e F j^e, \tag{S3}$$

where $\kappa^{\text{eff}}$ represents the effective liquid-phase conductivity evaluated from the Bruggeman relation $\kappa_e^{\text{eff}} = \kappa_e \epsilon^{\text{brugg}}$ and $f_\pm$ represents the activity coefficient [49].

As in Equation S3, the solid phase potential $\Phi_s$ is governed by the charge leaving or entering the solid phase, and can be written in the form of Ohmic law as

$$\nabla \cdot i_s = \nabla \cdot \left(\sigma^{\text{eff}} \nabla \Phi_s\right) = -a^e F j^e. \tag{S4}$$

The parameters for the electrochemical and transport model are presented in Table A1 and Table A2.

For the charge transfer molar flux $j^e$, the widely adopted Butler-Volmer equation is used [50, 47, 51]

$$j^e = \frac{i_0}{F} \left[ \exp\left(-\frac{\alpha F}{RT} \eta\right) - \exp\left(\frac{(1-\alpha)F}{RT} \eta\right) \right], \tag{S5}$$

where $i_0$ represents the exchange current density, $\alpha$ represents the charge transfer coefficient, and $\eta$ represents overpotential. The exchange current density, $i_0$, in Equation S5 is derived from the regular solution theory [52] and can be written as

$$i_0 = i'_0 \sqrt{\frac{c_e}{c_e^{\text{ref}}}} x(1-x) \exp\left(\frac{\Omega}{RT}(1-2x)\right), \tag{S6}$$

where $c_e^{\text{ref}}$ represents the reference concentration of Li-salt in the electrolyte and $\Omega$ represents the regular solution parameter [40].

Overpotential $\eta$ is calculated from the following equation

$$\eta = (\Phi_s - \Phi_e) - U^{\text{EQ}}, \tag{S7}$$

where $U^{\text{EQ}}$ represents the equilibrium solid potential defined as

$$U^{\text{EQ}}(x) = U^0 - \frac{\mu(x)}{F}. \tag{S8}$$

$U^0$ is the standard equilibrium potential of the active material and $\mu$ represents the equilibrium chemical potential of an individual particle dependent on the filling fraction $x = \frac{\tilde{c}_s}{c_s^{\text{max}}}$.

Another merit of the applied modelling framework is a plausible determination of $\mu$ of active particles via a multi-scaling approach that preserves a high level of consistency with lower scales and thus increases the modelling fidelity of the cathode (elaborated in [12, 40, 53]). Due to its relatively small size and short characteristic times for diffusion [21], the particles were simulated as 0D particles [53] with chemical potential derived through a consistent reduction of the detailed spatially resolved model relying on the regular solution theory for phase separating materials [40]. The use of 0D particle model is inevitable to ensure finite computational times when performing electrode level simulations that are needed to virtually represent the electric performance response of the battery, which comprises a large number of particles. A detailed spatially resolved phase field model [40] that considers gradient penalty, phase boundary thickness between Li-rich and Li-poor phases, and surface free energy at the interface between the surface and the bulk of a particle is namely characterised by very long computational times. Despite this fact, the application of this detailed spatially resolved phase field model [40] is very important to plausibly map the material specific chemical potential of the phase separating material, which is inherently non-monotone, to the chemical potential of the particle.

Consequently, a 0D particle model was derived by applying a thermodynamically consistent analytical procedure, which allows for preserving the main physicochemical relevance during reduction of the detailed phase field model based on Cahn-Hilliard dynamics [40] to a 0D particle model and reducing the computational times by up to six orders of magnitude [53]. The derived 0D model thus considers the dependency of the particle surface reactivity on the particle state and effects of nucleation barrier on particle dynamics [53]. However, due to the nature of the 0D model it does not directly resolve the impact of the gradient penalty, the strain and surface free energy at the interface between surface and bulk of the particle with a spatial resolution. In a 0D model these effects are considered in the offset of the free energy, as presented in Fig. 2b and Figs. S32b and S32d. Consequently, this yields a gradient free chemical potential for interparticle phase separated particles, which is analogous to [23]. In general, the interparticle phase separated chemical potential is not necessarily fully gradient free, however, as discernible from [54], this challenge is not yet uniquely resolved, while sensitivity to this assumption is analysed in Section 6.2.

As a result, the thermodynamically consistent reduction to 0D enables mapping the material specific spinodal chemical potential of the phase separating material to the chemical potential of the particle for high and low overpotential limits. At high overpotentials, the chemical potential of 0D particles follows the spinodal-curve-shape (blue curve in Fig. 2a in the main text and Fig. S32a). At low currents, the chemical potential of 0D particles was represented by its low overpotential limit, i.e. the red curve in Fig. 2a in the main text and Fig. S32a denoting the intraparticle phase separated state. The fact that both limits are characterised by small spatial variations in chemical potential [40] ensures that averaged (volume integrated) 0D chemical potential is very close to the chemical potential of each point of spatially resolved particle (more details are given in reference [53]). This way several important aspects of detailed single particle model (i.e. nucleation barrier, dependence of particle surface reactivity on particle lithiation level, gradient penalty, strain) are introduced in the multi particle porous electrode model of the full cell via averaged 0D chemical potential.

Further, it was reported in [4] that the regular solution parameter, $\Omega$, features particle size dependency. Due to a limited number of active particles, which is significantly smaller that the number in real electrodes, the maximum value of $\Omega$ was in this analysis set below $4.5 k_B T$, as proposed in [4]. This is reasoned by two facts. Firstly, systematically gathered measurements of the potential value at the point of phase transition initiation in LFP indicate that they fall below the theoretical limit for perfect crystal [4]. Secondly, the presence of large particles with a value of $\Omega = 4.5 k_B T$ yields a hysteresis in the range of 70 mV, which is considerably more than measured in the present experiment (Fig. 1 in the main text) and also in the pioneering work on particle-by-particle (de)lithiation [19]. This approach was directly applied to the virtual representation of the electrode topology, where aggregates were modelled as ensemble of interconnected primary particles, while size dependency of this phenomenon was plausible scaled for the virtual representation of the electrode topology with experimentally measured topology of aggregates, which is presented in the subsequent text.

The equilibrium chemical potential is inherently linked to the lithiation level of a particle and thus to the mass conservation in the active particle, which can, in its most general form, be written as

$$\frac{\partial c_s}{\partial t} = \nabla \cdot \left( \frac{D_s c_s}{k_B T} \nabla \mu \right), \tag{S9}$$

where $c_s$ represents the concentration of Li in solid and $D_s$ represents the diffusion constant in the solid.

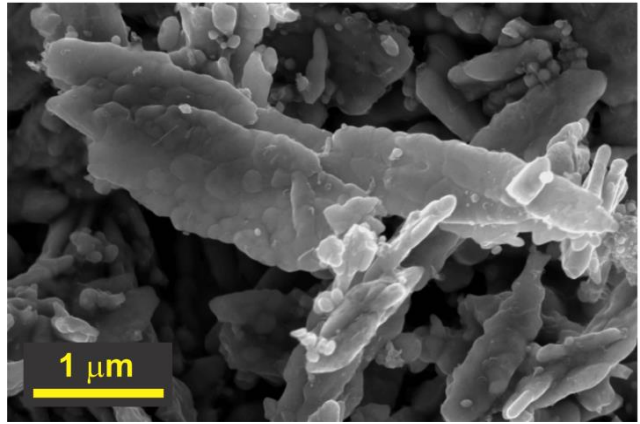

**Fig. S35** SEM image of the commercial LFP cathode material [12].

Electrode topology, i.e. topological characteristics of aggregates (Fig. S35) and size distribution of particles (Supplementary section 2), also influence the rate of transport and electrochemical phenomena. To address this important aspect and to further confirm that virtual replication of entering and remaining within the voltage hysteresis under current bias in phase separating materials is not depending on the virtual representation of the electrode topology, results are presented using two levels of detail of virtual representation of the electrode topology.

First, as already presented in the main text, experimentally measured topology of the active particles, i.e. aggregates (as indicatively shown in Supplementary section 2.4), was used as an input of the model, where aggregates were modelled as 0D particles.

Second, to provide a simulation proof that observed phenomenon of entering and remaining within the hysteresis under bias in phase separating materials can be virtually replicated also with significantly different virtual representation of the electrode topology, topology of the active particles was virtually represented by modelling representation featuring an additional level of detail. In this representation, aggregates were modelled as ensemble of interconnected primary particles, modelled as 0D particles, as schematically represented in Fig. S36. In this case, experimentally measured size distribution of primary particles [12] was used in the model (Fig. S19b), while primary particles were interconnected into aggregates considering experimental determined topologies of aggregates. In such a framework, primary particles share areas of direct contact with other particles in aggregates, which are position and size dependent, while the particle connectivity and its position in the aggregate determine the areas of the primary particle being in the direct contact with the electrolyte. This so-called connectivity scheme (Fig. S36) allows for the decoupling of a single particle's total surface between the surface in contact with the electrolyte and the surface of direct connections to neighbouring particles, and can be written as [12]

$$A_k^{\text{tot}} = A_k^{\text{e}} + \sum_l A_{k,l}^{\text{DC}}, \tag{S10}$$

where indexes $k$ and $l$ denote the index of the primary particle and the index of the connection, respectively.

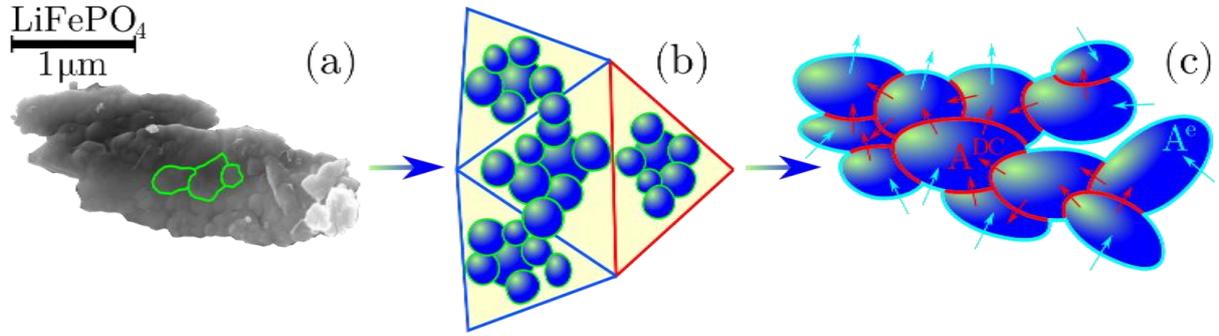

**Fig. S36 a** SEM image of a commercial cathode material; **b** a schematic representation of the decoupling of the particle size and the size of the computational cell and a representation of particle connectivity; **c** schematic representation of the secondary LFP particle (aggregate) built from the primary particles. Surface of the aggregate exposed to the electrolyte $A^\text{e}$ is depicted with cyan colour whereas direct contact surface $A^\text{DC}$ is depicted with red colour. Adapted from [12].

The presented extension of the continuum level porous electrode model enables redistribution of Li between active particles via electrolyte or via direct contact by considering the electrode topology and materials properties [12]. The molar flux at the interface between two particles in the direct contact is derived from equation of mass conservation in active particle. It is inspired by the equation proposed by Orvananos et al. [55], but used in its more general form [12].

$$\hat{n}_{k_1,k_2} \cdot j^\text{DC} = \frac{\bar{c}_{s,k_1} + \bar{c}_{s,k_2}}{2} \frac{P}{RT} (\mu_{k_2} - \mu_{k_1}). \tag{S11}$$

The rate of Li exchange between particles with indexes $k_1$ and $k_2$ is determined by their difference in chemical potential. $\bar{c}_s$ represents the average concentration of Li in active particles, whereas $P$ represents the permeability of the direct contact surface.

To preserve as high consistency with experimentally determined topological representation of the electrode, both levels of detail of virtual representation of the electrode topology feature also limited inter-aggregate connectivity. Inter-aggregate connectivity is based on the same modelling framework as connectivity between primary particles into aggregates, while considering (consistent with microscopic observations Fig. S35) much less area of direct inter-aggregate contact.

For the simulations of the half-cell, a 2D unstructured mesh with total of 273 control volumes was used. In the case where aggregates were modelled as 0D particles, 20 representative aggregates comprise each control volume, whereas in the case where aggregates were modelled as ensemble of interconnected primary particles, 60 inter-connected representative primary particles comprise in each control volume. Size distribution of the aggregates and primary particles was extracted from the experimental TEM imaging of the electrode material (Fig. 1 main text & Fig. S35). Approach with 2D computational mesh allows to investigate the impact of electrode-scale inhomogeneities and examine the impact of spatial variations in the discharge. This is presented with the case, where the porosity in each control volume was randomly distributed between 0.3 and 0.7 with the aim to intentionally analyse the results under such a large range of variation. The variation in porosity has a direct impact on the amount of active material in the control volume. Thus, the porosity variation correlates with the size of aggregates particles, which further correlates with the number of representative particles.

This extended continuum level porous electrode model features quite high level of sophistication, which enables adequately addressing recent experimental observations such as C-rate dependence of active particle population [20, 23, 25, 12] and the resulting homogeneous electrode charging at high rates [21, 23, 56] as well as particle-by-particle charging at low rates [19, 20, 23]. Despite these merits, we are fully aware of the limitations of the presented modelling framework, which, in addition to certain remaining deficiencies related to the porous electrode model [12], does not yet directly model the carbon-binder domains [47, 57], the fluid-enhanced surface diffusion [34] and the defects in crystal lattices. In addition, as addressed previously in this section, the applied 0D model considers the dependency of the particle surface reactivity on the particle state and effects of nucleation barrier on particle dynamics [53]. However, due to the nature of the 0D model, it does not resolve the intra particle phenomena with a spatial resolution. In addition, exact modelling of yet unambiguously unexplored intraparticle Li redistribution at high overpotentials (Sections 4.2) represents another potential limitation of the applied 0D particle model. Consequently, the finite rate of intraparticle diffusion and the rate of establishing the phase boundaries is not considered in the implemented 0D approach. Despite these generally persisting challenges, the present extended continuum level porous electrode model offers an adequate basis for interpreting and qualitatively validating the intriguing experimental phenomena observed in this work. Thus, the main aim of the simulating model is to provide a bridge between the thermodynamic reasoning presented in Fig. 2 and experimental evidence presented in Fig. 1 of the main text.

### 6.2 Supplementary simulation results

This section provides simulation results that further support analyses and statements in the main text.

One of the crucial preconditions that allow entering and remaining within the hysteresis under current bias is the preference of intraparticle phase separation over the interparticle phase separation due to longer characteristic times for interparticle phase separation, as elaborated in Section 5 of this Supplementary. Thus, the results in Fig. S37 are calculated using the same model setting as used to predict the results in Fig. 3 in the main text, with the following virtually imposed differences: a) assuming that dynamics is always, at all overpotentials, governed by the high overpotential transient state particle potential (Fig. 2 of the main text and Fig. S32) and b) assuming that the contribution of phase boundary within intraparticle phase separated particles to the free energy [54, 58] are such that chemical potential is slightly tilted in the plateau region after the point A in Fig. 2a of the main text (denoted as "tilted potential"). These analyses were performed to further examine the plausibility of the key conditions that need to be satisfied in order to enter into the hysteretic loop under current bias, as presented in Fig. 2 and accompanying text in the main text.

Although, when considering only the chemical potential representation, it might appear feasible to use high overpotential transient state particle potential to model particle dynamics at small currents and after current stimulus, the free energy representation opposes such dynamics (Fig. 2). Despite this fact the virtual experiment allows assessing also such a scenario. Fig. S37a clearly shows that the use of high overpotential transient state potential significantly increases the rate of decay of the active particle population, which is further visualised in Fig. S37c. The decay time of the active particle population is thus only approximately 200 minutes, which is significantly shorter than reported in [34] and also significantly shorter as indicated in Fig. S34 These results thus confirm thermodynamic reasoning presented in the main text, while they are also in-line with previous findings, e.g. [23], thereby underpinning the statement that preferential occurrence of intraparticle phase separation resulting in lower intraparticle phase separated chemical potential (Fig. 2) is a crucial precondition that allows entering and remaining within the hysteresis under current bias (in continuation "bias" related to hysteresis is always meant as current bias).

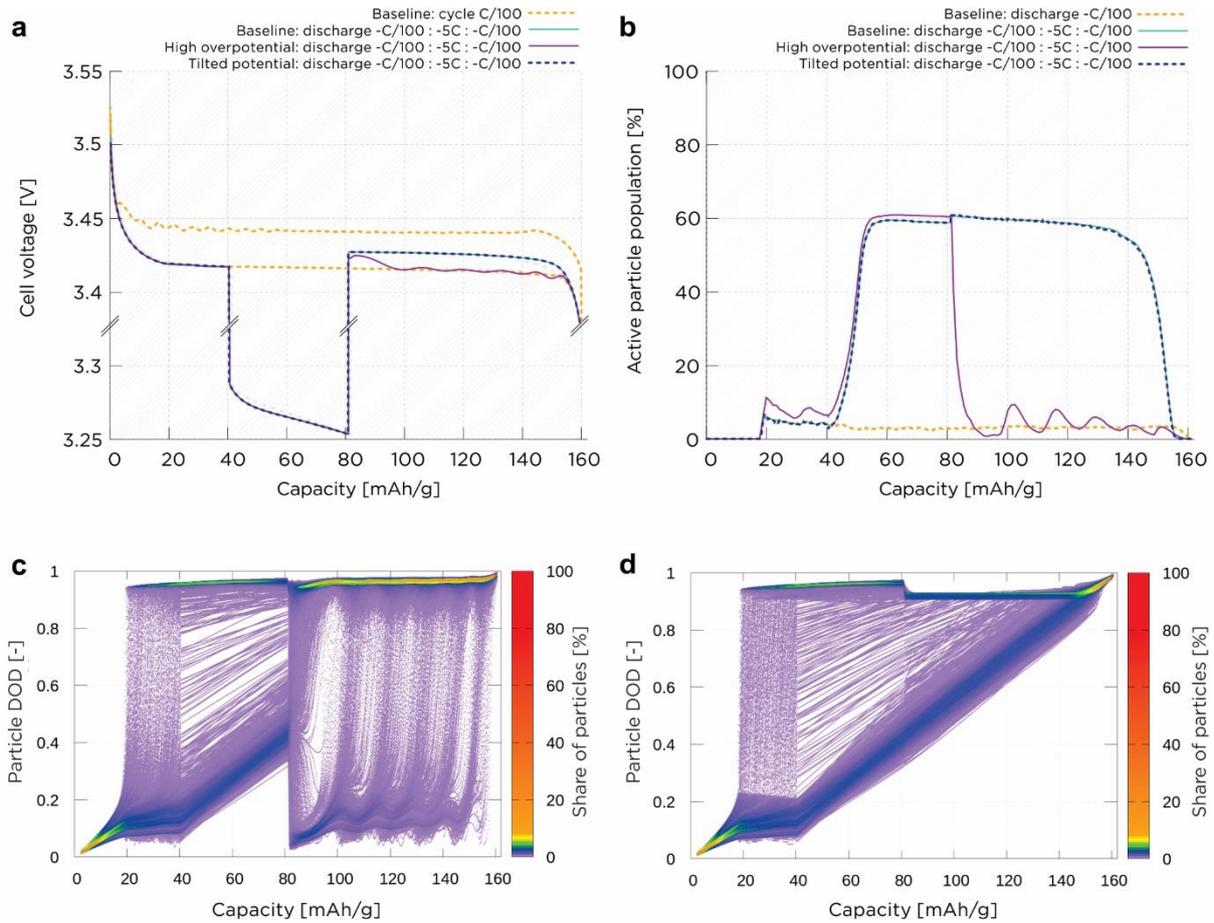

**Fig. S37** Comparison between **a** voltage and **b** active particle population of the baseline simulation results presented in Fig. 3 of the main text (denoted as baseline), which was calculated using intraparticle phase separated particle potential at small currents and thus overpotentials, and the results calculated with the same model and applying high overpotential transient state particle potential (Fig. 2) at all currents and thus overpotentials as well as the results calculated with the model assuming that the contribution of phase boundary within intraparticle phase separated particles to the free energy [58, 54] are such that chemical potential is slightly tilted in the plateau region after the point A in Fig. 2a of the main text (denoted as "tilted potential"). **c** and **d** evolution of the particle's DODs as a function of cell capacity during discharge clearly showing: **c** significantly increased rate of decay of the active particle population when applying high overpotential transient state particle potential and **d** very similar trend as shown in Fig. 3d of the main text when applying slightly tilted potential.

Fig. S37 takes analyses on the sensitivity of the chemical potential variation further by analysing the impact of virtually imposing tilted chemical potential in the region after the point A in Fig. 2a of the main text. As addressed in Supplementary Section 6.1, in general, the phase separated particle chemical potential in phase separating electrode materials is not fully gradient free in the region after the point A in Fig. 2a of the main text, as discernible from the free energy trends published in references [54, 58]. Both references offer comprehensive comparison between different models of the free energy trends in phase separating electrode particles in this region and a thorough parametric study. However, spread of results of [54, 58] indicates that the challenge of unambiguously determining the gradient of the chemical potential in the region after the point A in Fig. 2a of the main text is not yet resolved. Therefore, to analyse the impact of this gradient of the chemical potential, a moderate and constant gradient of the chemical potential, which is within the range of results reported in [54], was imposed in the region after the point A in Fig. 2a of the main text. Fig. S37b and S37d indicate that particles tend to separate only a bit faster in the case of the tilted chemical potential compared to the baseline results (presented also in Fig. 3 in the main text). Resultantly, active particle population and battery voltage are very similar. In general, these results indicate that entering into the interparticle phase separated state, which is virtually represented by the interparticle phase separated chemical potential (Fig. 2 in the main text), is crucial for ensuring sufficiently longevity of the intraparticle phase separated state to enter and remain within the hysteresis under bias. It further indicates that the exact value of the gradient of the chemical potential in the region after the point A in Fig. 2a of the main text does not influence virtual reproduction of the first order phenomena related to entering and remaining within the hysteresis under bias until values are in reasonable ranges.

To further strengthen findings presented in the main text, an additional analysis confirming that virtual replication of the entering and remaining within the hysteresis under bias in phase separating materials is not depending on the virtual representation of the electrode topology was performed. As outlined in Supplementary Section 6.1, this was performed by comparing results calculated with the two levels of detail of virtual representation of the electrode topology (Fig. S38), while using the same model setting as used to predict the results in Fig. 3 in the main text. Thus, baseline results (Fig. 3) calculated using experimentally measured topology of the active particles, i.e. aggregates (as indicatively shown in Fig. 1b and 1c), which were modelled as 0D particles, are first compared to the results where aggregates were modelled as ensemble of interconnected primary particles, modelled as 0D particles (denoted as "0D primary"), as schematically represented in Fig. S36. These results (Fig. S38) clearly indicate that observed phenomenon of entering and remaining within the hysteresis under bias in phase separating materials can be virtually replicated using both modelling approaches featuring significantly different level of detail of virtual representation of the electrode. In addition, both approaches feature qualitatively similar trends of active particle population (Fig. S38b), while specific differences in active particle population and evolution of the particle's DODs as a function of the cell capacity (Fig. S38c and Fig. 3d of the main text) arise from the modelling depth. When aggregates are modelled as ensemble of interconnected primary particles, some primary particles feature very small or even negligible area in contact with the electrolyte areas of direct contact, which impedes their lithiation. If such particles are also larger and thus feature smaller area to volume ratio this additionally impedes their lithiation, whereas furthermore size dependent $\Omega$ prefers lithiation of smaller particles under assumption of particle size independent OCV. Consequently, in this case of "0D primary" particles some particles did into reach lithiaton level corresponding to the spinodal point at 50% of the cell DOD, i.e. approx. 81 mAh/g. This is higher level of aggregate resolution thus mainly contributes to the lower active population (Fig. S38b), whereas particles did into reach lithiaton level corresponding to the spinodal point at 50% of the cell DOD tend to delithiate afterwards when the current and overpotential are reduced as seen in Fig. S38c. Despite these minor modelling depth specific differences, the main trends are virtuall reproduced with both modelling approaches, i.e.: a) particle-by-particle regime characterised by small active particle population and high rates of lithiation of active particles indicated by a few step lines until the cell capacity that corresponds to 25% of the cell DOD, i.e. approx. 40.5 mAh/g; b) increase of active particle population during the current stimulus phase and c) retention of the high share of active particle population in the interparticle phase separated state after the current and overpotential are reduced at 50% of the cell DOD, i.e. approx. 81 mAh/g. Likewise, comparing -C/100 discharge before and after current stimulus, it can be observed in both modelling representations that after current stimulus, much more particles are active and that they are also lithiated at lower rates being indicated by much more lines with lower gradient compared to the particle-by-particle lithation regime. Both modelling approaches thus expectedly yield very similar results in terms of voltage response and qualitative evolution of the active particle population, which further supports the fact that entering into the interparticle phase separated state, which is virtually represented by the interparticle phase separated chemical potential, is crucial to enter and remain within the hysteresis under bias.

A further step in analysing generality of the observed phenomena comprises sensitivity on the size dependent variation of the regular solution parameter, $\Omega$, as addressed in Section 6.1. This analysis was performed the modelling approach where aggregates were modelled as ensemble of interconnected primary particles, modelled as 0D particles, while neglecting particle size dependent variation of the regular solution parameter (denoted as "0D primary + const. $\Omega$"). When additionally assuming a constant $\Omega$, the voltage trend remains nearly identical to the baseline case, whereas the active particle population is higher (Fig. S38b and comparison of Fig. S38c and Fig. S38d). Assuming particle size independent OCV, this phenomenon originates from the fact that size dependent $\Omega$ prefers lithiation of smaller particles in the initial phase of discharge. These particles are namely characterised by lower values of $\Omega$ and thus lower spinodal value of the chemical potential (point A in Fig. 2a of the main text), while in addition they feature a larger area to volume ratio which, in general and by disregarding their position in secondary particles, further favours their faster lithiation. Unlike, in the case of constant $\Omega$ value, much less of the smaller particles are already lithiated at the onset of the current stimulus thus yielding a much larger active particle population (Fig. S38b and Fig. S38d). However, as discernible from Fig. S38, this has a negligible influence on battery voltage. It can thus again be summarised that presented significant variation of model parameters does have a notable impact on key results, further supporting the fact that applied variation do not influence virtual reproduction of the first order phenomena related to entering and remaining within the hysteresis under bias.

Fig. S39 investigates the impact of electrode-scale inhomogeneities by examining spatial variation of the electrode porosity. For this purpose, the results of the baseline simulation results presented in Fig. 3 of the main text were compared to the results calculated with a 2D mesh where the porosity in each control volume was randomly distributed between 0.3 and 0.7, whereas all other model settings were identical to the model setting used to predict the results in Fig. 3 of the main text. As indicated in Supplementary Section 6.1, the variation in porosity has a direct impact on the amount of active material in the control volume. The results indicate some, though minor, differences in the overpotential during a 5C current stimulus, which can be associated with the variation of transport losses. However, in general, it can be observed that voltage trends (Fig. S39a) and, in particular, the active particle population (Fig. S39b) and also evolution of the particle's DODs as a function of the cell capacity (comparison of Fig. S39c and Fig. 3d of the main text) feature nearly identical results, irrespective of the spatial inhomogeneities.

This sensitivity analysis on model parameters proves that virtual reproduction of the key conditions that need to be satisfied to enter into the hysteretic loop under bias (listed in the main text), is possible within a wide range of plausible model parameters further supporting the claim on the generality of the observed phenomenon. However, the sensitivity analysis also shows that simulation results clearly deviate if non-plausible model parameters or model inputs, as for example application of high overpotential transient state particle potential, are imposed. As already addressed, this underpins the statement that preferential occurrence of intraparticle phase separation resulting in lower intraparticle phase separated chemical potential (Fig. 2) is a crucial precondition that allows entering and remaining within the hysteresis under bias.

Fig. 3b and Fig. 3d of the main text and several figures in this section expose an additional phenomenon that influences the active particle population at the end of current stimulus, i.e. at 50 % DOD or at approx. 81 mAh/g in the analysed case. This phenomenon was already discussed in Section 5 of this Supplementary. The increase in active particle population at the end of current stimulus, i.e. at 50 % DOD in Fig. 3 of the main text and in Fig. S37, Fig. S39 and, in particular, in Fig. S38, is namely associated with delithiation of most lithiated particles, which were lithiated far beyond the right spinodal during the current stimulus. Namely, upon delithiation some of these particles enter, once again, the lithiaton level that characterises them as active. As discussed in Section 5 of this Supplementary, this finite rate of delithiation of such particles significantly influences the characteristic time needed from the end of large-current stimulus to the peak of resulting voltage relaxation, i.e. time interval $\Delta t_1$.

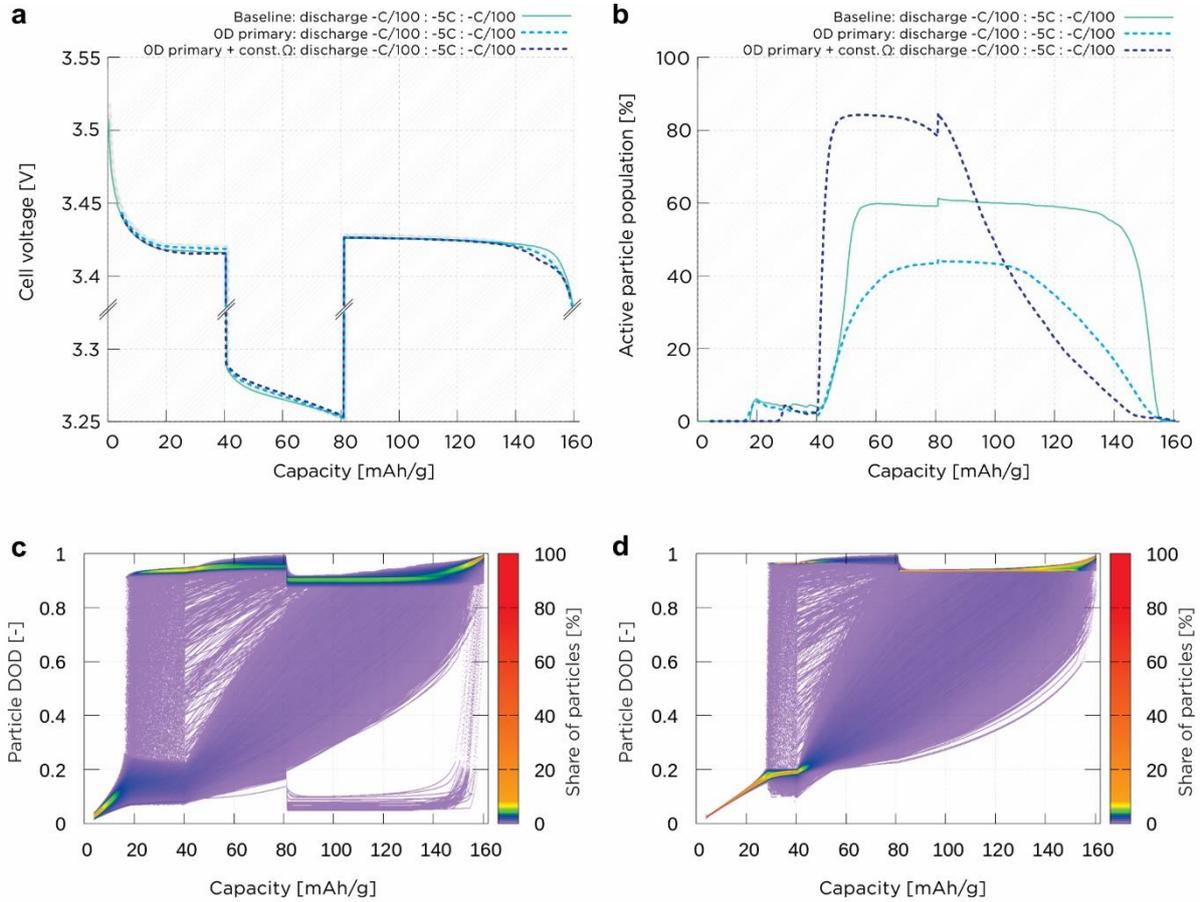

**Fig. S38** Comparison between **a** voltage and **b** active particle population of the baseline simulation results presented in Fig. 3 of the main text (denoted as baseline) and results calculated with the model where aggregates were modelled as ensemble of interconnected primary particles, modelled as 0D particles (denoted as "0D primary"), and results calculated with the model where aggregates were modelled as ensemble of interconnected primary particles, modelled as 0D particles, while neglecting particle size dependent variation of the regular solution parameter (denoted as "0D primary + const. $\Omega$"). **c** and **d** evolution of the particle's DODs as a function of cell capacity during discharge for: **c** "0D primary" and **d** "0D primary + const. $\Omega$" model.

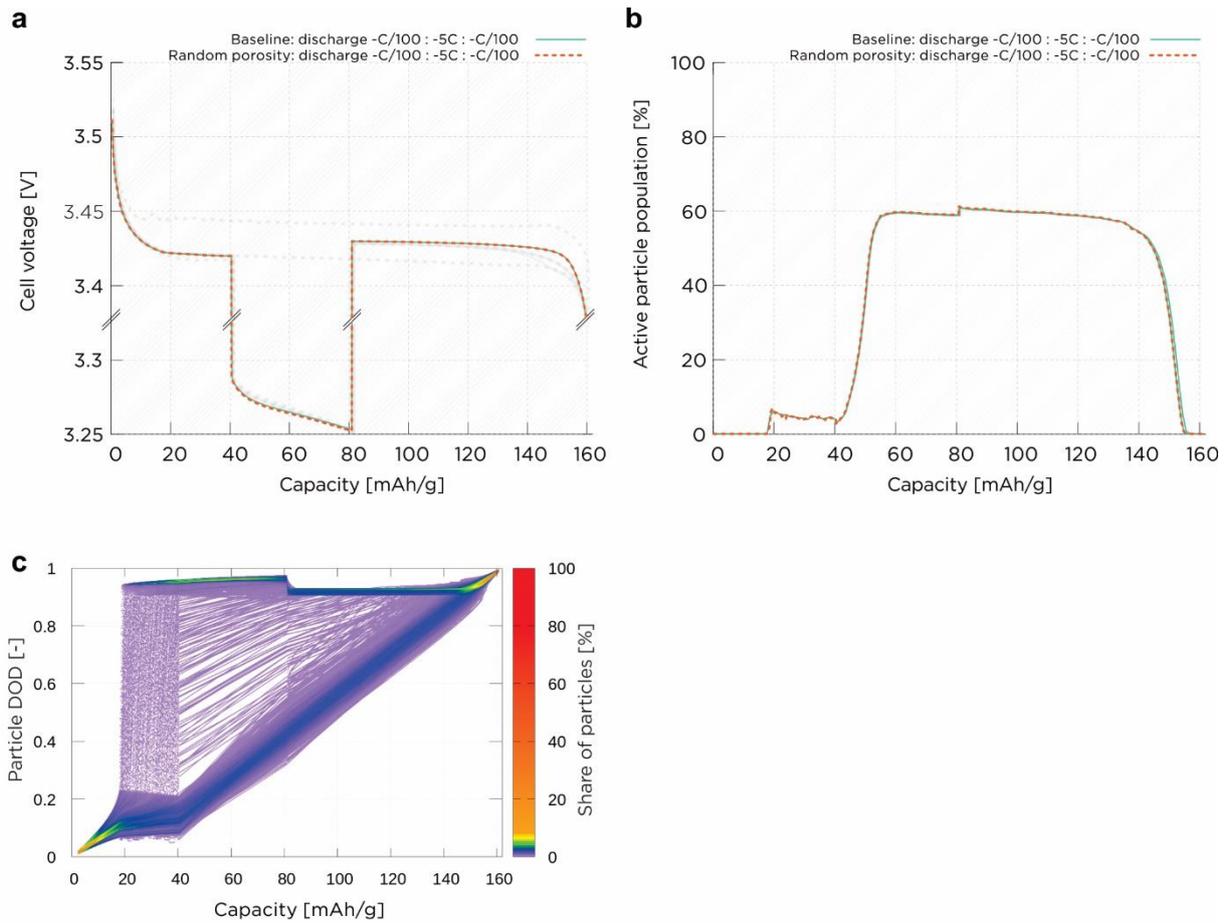

**Fig. S39** Comparison between **a** voltage and **b** active particle population of the baseline simulation results presented in Fig. 3 of the main text (denoted as baseline) and results calculated with a 2D mesh where porosity in each control volume was randomly distributed between 0.3 and 0.7. **c** evolution of the particle's DODs as a function of cell capacity during discharge for the case where porosity in each control volume was randomly distributed between 0.3 and 0.7.

### 6.3 Model parameters

This section presents model parameters used in the modelling framework. The parameters are divided into two groups. The parameters related to the electrolyte are listed in Table A1. The parameters related to the electrode materials and separator are listed in Table A2.

Table A1: Model parameters for the electrolyte. Remarks: *assumed. References [59, 60, 61].

| Parameter | Unit | Value |
|---|---|---|
| $c_e^{ref}$ | mol/m$^3$ | 1000 |

| | | | |
|---|---|---|---|
| $D_e$ | m²/s | 4 · 10⁻¹⁰ | |
| $\kappa_e$ | A/Vm | Functional dependency [61] | |
| $T$ | K | 293.0 | |
| $t_+$ | - | 0.363 | |

Table A2: Model parameters for electrode material and separator. Remarks: †laboratory-built half-cell (measured), *carbon coated LFP, **aggregate size, ‡assumed. References [46, 62, 61].

| Parameter | Unit | Cathode | Separator |
|---|---|---|---|
| **brugg** | - | 1.5‡ | 1.5‡ |
| $c_{p,max}$ | mol/m³ | 22800 | / |
| $D_s$ | m²/s | 5.0 · 10⁻¹³ | / |
| $i'_0$ | A/m² | 0.02 | / |
| $L$ | µm | 25† | 250† |
| $P$ | m/s | 1 · 10⁻¹⁰‡ | / |
| $\overline{R_p}$ | m | **793 · 10⁻⁹ | / |
| $T_{ref}$ | K | 293.0 | 293.0 |
| $U^0$ | V | 3.428 | / |
| $\alpha$ | - | 0.50 | / |
| $\epsilon$ | - | 0.49† | 0.92† |
| $\sigma$ | A/Vm | 1.0* | / |